\documentclass[12pt]{JHEP3} 
\input colordvi

\title{\textRed{Robust signatures
of solar neutrino oscillation
solutions}\textBlack}

\title{\textRed{Robust signatures of solar neutrino oscillation
solutions}\textBlack}
\author{John N. Bahcall\\
  School of Natural Sciences, Institute for Advanced Study, Princeton,
  NJ 08540\\
    E-mail: \email{jnb@ias.edu}}
\author{M. C. Gonzalez-Garcia \\
  Theory Division, CERN, CH-1211, Geneva 23, Switzerland,\\
  Y.I.T.P., SUNY at Stony Brook, Stony Brook,NY 11794-3840\\
and IFIC, Universitat de Val\`encia -- C.S.I.C., Apt 22085, 46071
  Val\`encia, Spain\\
    E-mail: \email{concepcion.gonzalez-garcia@cern.ch}}
\author{Carlos Pe\~na-Garay\\
        IFIC, Universitat de Val\`encia -- C.S.I.C., Apt 22085, 46071
  Val\`encia, Spain\\
       E-mail: \email{penya@ific.uv.es}}

\abstract{\textBlue With the goal of identifying signatures that
select specific neutrino oscillation parameters, we test the
robustness of global oscillation solutions that fit all the
available solar and reactor experimental data.  We use three
global analysis strategies previously applied by different authors
and also determine the sensitivity of the oscillation solutions to
the critical nuclear fusion cross section, $S_{17}(0)$, for the
production of ${\rm ^8B}$. Our standard results make use of the
precise new measurement of $S_{17}(0)$ by Junghans et al. The
globally favored solutions are, in order of goodness of fit:
LMA(the only solution at $2\sigma$), LOW, and VAC. The  NC to
CC ratio for SNO is predicted by the standard global analysis to
be $3.45^{+0.70}_{-0.54}(1\sigma)$ which is separated from the
no-oscillation value of $1.0$ by much more than the expected
experimental error. The predicted range of the day-night
difference in CC rates is $8.3^{+5.0}_{-5.6}(1\sigma)$\%. A
measurement by SNO of either a NC to CC ratio $> 3.3$ or a
day-night difference $> 10$\%, would favor a small region of the
currently allowed LMA neutrino parameter space. The global
oscillation solution predicts a ${\rm ^7Be}$ neutrino-electron
scattering rate in BOREXINO and KamLAND in the range
$0.65^{+0.04}_{-0.03}(1\sigma)$ of the BP00 standard solar model
rate, a prediction which can be used to test both the solar model
and the neutrino oscillation theory. Only the LOW solution
predicts a large day-night effect($\leq 42$\%, $3\sigma$) in
BOREXINO and KamLAND. For the reactor KamLAND experiment, the LMA
solution predicts a charged current rate relative to the standard
model of $0.44^{+0.22}_{-0.07}(1\sigma),~~E_{\rm threshold} =
1.22{\rm ~MeV}$. We have also evaluated the effects of including
preliminary Super-Kamiokande data for $1496$ days of observations.
}

\keywords{solar and atmospheric neutrinos, neutrino and gamma
astronomy, neutrino physics}

\begin{document}
\input psfig
\textBlack

\section{Introduction}

How robust are the predictions  of the allowed neutrino
oscillation solutions that are obtained by fitting the currently
available solar neutrino data with different neutrino oscillation
parameters? How can we determine experimentally the neutrino
oscillation parameters? We address these questions in the present
paper.

To determine the robustness of the oscillation solutions and their
predicted experimental consequences, we compute the currently
allowed regions in neutrino oscillation space using three
different analysis strategies, each strategy previously advocated
by a different set of authors. We include all the available solar
neutrino and reactor oscillation
data~\cite{chlorine,gallex,gno,sage,superk,chooz,sno2001}. We
also present ``Before and After'' comparisons of the globally
allowed solutions using both the previous(1998) standard value of
$S_{17}(0)$,the low energy cross section factor for the
production of ${\rm ^8B}$, and a precise new
measurement~\cite{junghans01} for this important quantity.

In order to identify which quantities are most likely to lead to
an experimental determination of the neutrino  oscillation
parameters, we use the allowed regions in neutrino parameter
space  to predict the expected range of the most promising
quantities that can be measured accurately in solar neutrino
experiments. For the Sudbury Neutrino Observatory
(SNO)~\cite{sno2001,sno}, we report the predictions of the currently
allowed regions for the neutral current to charged current double
ratio, [NC]/[CC], the day-night effect, and the first and second
moments of the electron recoil energy spectrum. For experiments
like BOREXINO ~\cite{borexino} and KamLAND~\cite{kamland} that
can detect ${\rm ^7Be}$ solar neutrinos, we evaluate the
predicted range of the neutrino-electron scattering rate and the
difference between the night and the day event rates.

It pays to be lucky. We show in figure~\ref{fig:lucky} and
figure~\ref{fig:be7daynight} that if SNO measures (as predicted
by some LMA oscillation parameters) a large value for either the
neutral current to charged current ratio or the day-night
difference in rates, or if BOREXINO or KamLAND measures(as
predicted by some LOW oscillation parameters) a large day-night
difference, then this measurement will uniquely select the
correct neutrino oscillation solution and fix the neutrino
oscillation parameters to within rather small uncertainties. On
the other hand, a small value of the neutral current to charged
current ratio would favor the SMA solution.

How should this paper be read? Since you got this far, we recommend
that you now turn to the final section, the summary and discussion
section, and see what results seem most interesting to you. The
summary and discussion section refers to what we believe are the most
useful figures and tables, which the reader may wish to glance at
before deciding whether to invest the time to read the detailed text
in the main sections.

This paper is organized as follows. In section~\ref{sec:newbp} we
present the predicted solar neutrino fluxes and their uncertainties
that are derived from the BP00 standard solar model~\cite{bp2000} with
and without including the recent precise measured value for
$S_{17}(0)$ determined by Junghans et al.~\cite{junghans01}. We
determine in section~\ref{sec:global} the regions in neutrino
oscillation parameter space that are allowed using three different
analysis scenarios; we present results obtained using both the new
value for $S_{17}(0)$ and the previous(1998) standard value.  We use
the three sets of global oscillation solutions in
section~\ref{sec:sno} to determine for SNO the $3\sigma$ (and
$1\sigma$) ranges of predicted values for the neutral current to
charged current ratio, the difference between the day and the night CC
event rates, and the first two moments of the energy spectrum of the
CC recoil electrons. In section~\ref{sec:be7}, we determine for the
BOREXINO and KamLAND solar neutrino experiments the $3\sigma$ (and
$1\sigma$) ranges of allowed values for the neutrino-electron
scattering rate and the day-night effect. For the reactor KamLAND
experiment, we present in section~\ref{sec:kamland} the predicted
$3\sigma$ (and $1\sigma$) ranges of the charged current (antineutrino
absorption) rate and the distortion of the visible energy spectrum. We
summarize and discuss our conclusions in section~\ref{sec:discuss}.

We use, unless stated otherwise, the techniques and parameters for the
analysis that we have described elsewhere~\cite{bks2001,cc2001,bks98,bgp}.

\section{BP00 + New ${\rm\bf ^8B}$}
\label{sec:newbp}

In this section, we present the predicted solar neutrino fluxes
(table~\ref{tab:rates}) and their uncertainties
(table~\ref{tab:uncertainties}) that were computed with the BP00 solar
model~\cite{bp2000}, with and without taking account of a recent
precise measurement of the production cross section that leads to the
emission of ${\rm ^8B}$ neutrinos. Both the solar neutrino fluxes and
their uncertainties are used in calculations of the allowed regions
for neutrino oscillation parameters.

There has been a lot of beautiful experimental work devoted to
measuring the rate of the fusion reaction ${\rm
^7Be(p,\gamma)^8B}$~\cite{junghans01,b8refs}. Recent experiments have
concentrated on reducing the systematic uncertainties, in going to
lower center-of-mass energies that are more relevant to the sun, and
in using new measurement techniques. The goal is to reduce the
uncertainty in the low energy cross section factor for the reaction
${\rm ^7Be(p,\gamma)^8B}$ to less than or equal to $5$\%~\cite{book},
so that it is no longer the dominant uncertainty in the prediction of
the ${\rm ^8B}$ solar neutrino flux. Very important experiments are
continuing to be performed on this reaction.

The recent Junghans et al. precision measurement of the low energy
cross section for the fusion reaction ${\rm ^7Be(p},\gamma){\rm ^8B}$
yields~\cite{junghans01}
\begin{equation}
S_{17}(0) = \left(22.3 \pm 0.7 \left({\rm expt}\right) \pm 0.5
\left({\rm theor}\right) \right){\rm eVb}.
\label{eq:b8}
\end{equation}
The quoted uncertainty for the Junghans et al. measurement, which
combines statistical and systematic uncertainties, is smaller than for
all previous low energy measurements. Other relatively recent
measurements~\cite{b8refs} have yielded values of $S_{17}(0)$ that are
smaller than the Junghans et al. value.

For illustrative purposes, we adopt in this paper $S_{17}(0) =
\left(22.3 \pm 0.9\right) {\rm eVb}$ instead of the previous(1998)
standard value of $S_{17}(0) = 19^{+4}_{-2} {\rm
eVb}$~\cite{adelberger} that was used in the original
calculations~\cite{bp2000} of the BP00 standard model neutrino fluxes.
The value of $S_{17}(0)$ given in eq.~(\ref{eq:b8}) is $0.8\sigma$
larger than the previously used value.

We will present throughout this paper the differences in the
calculated quantities that result from using $S_{17}(0) = \left(22.3
\pm 0.9\right) {\rm eVb}$ instead of $S_{17}(0) = 19^{+4}_{-2} {\rm
eVb}$. These two values of $S_{17}(0)$ span the set of recent
experimental results~\cite{junghans01,b8refs}.

\subsection{Fluxes}
\label{subsec:fluxes}

\TABLE[!t]{ \centering \caption{{\bf Standard Model Predictions (BP00
+ New ${\rm ^8B}$)}: solar neutrino fluxes and neutrino capture rates,
with $1\sigma$ uncertainties from all sources (combined
quadratically). The neutrino fluxes are the same as in the original
BP00 model~\cite{bp2000} except for the ${\rm ^8B}$ flux, which is
increased because of the larger adopted value of $S_{17}(0)$, see
eq.~(\ref{eq:b8}).  Using the 1998 standard value $S_{17}(0) =
19^{+4}_{-2} {\rm eVb}$~\cite{adelberger}, the ${\rm ^8B}$ neutrino
flux was calculated previously to be $\phi\left({\rm ^8B}\right) =
5.05\left(1.00^{+0.20}_{-0.16}\right)$. The total rates were
calculated using the neutrino absorption cross sections and their
uncertainties that are given in ref.~\cite{nuabs} }
\protect\label{tab:rates}
\begin{tabular}{llcc}
\noalign{\bigskip}
\hline
\noalign{\smallskip}
Source&\multicolumn{1}{c}{Flux}&Cl&Ga\\
&\multicolumn{1}{c}{$\left(10^{10}\ {\rm cm^{-2}s^{-1}}\right)$}&(SNU)&(SNU)\\
\noalign{\smallskip}
\hline
\noalign{\smallskip}
pp&$5.95 \times 10^{0}~~\left(1.00^{+0.01}_{-0.01}\right)$&0.00&69.7\\
pep&$1.40 \times 10^{-2}\left(1.00^{+0.015}_{-0.015}\right)$&0.22&2.8\\
hep&$9.3 \times 10^{-7}$&0.04&0.1\\
${\rm ^7Be}$&$4.77 \times 10^{-1}\left(1.00^{+0.10}_{-0.10}\right)$&1.15&34.2\\
${\rm ^8B}$&$5.93 \times 10^{-4}\left(1.00^{+0.14}_{-0.15}\right)$&6.76&14.2\\
${\rm ^{13}N}$&$5.48 \times
10^{-2}\left(1.00^{+0.21}_{-0.17}\right)$&0.09&3.4\\
${\rm ^{15}O}$&$4.80 \times
10^{-2}\left(1.00^{+0.25}_{-0.19}\right)$&0.33&5.5\\
${\rm ^{17}F}$&$5.63 \times
10^{-4}\left(1.00^{+0.25}_{-0.25}\right)$&0.00&0.1\\
\noalign{\medskip}
&&\hrulefill&\hrulefill\\
Total&&$8.59^{+1.1}_{-1.2}$&$130^{+9}_{-7}$\\
\noalign{\smallskip}
\hline
\end{tabular}
}

Any plausible  change in the value of $S_{17}(0)$ has no
discernible numerical effect on the standard solar
model~\cite{book}. The reaction ${\rm ^7Be(p},\gamma){\rm ^8B}$
is so rare (it occurs of order twice in every $10^4$ completions
of the $p-p$ chain) that even a factor of ten change in the cross
section does not affect, to the required accuracy, the calculated
solar model temperatures and densities. The $15$\% increase in the
best-estimate value of $S_{17}(0)$ represented by
eq.~(\ref{eq:b8}) does not change significantly any of the
computed physical variables in the standard solar model except
the ${\rm ^8B}$ neutrino flux.

For a given solar model with specified distributions of temperature,
density, and composition, the calculated ${\rm ^8B}$ neutrino flux
depends~\cite{book} linearly upon the adopted value of $S_{17}(0)$,
i.e., $\phi({\rm ^8B}) \propto S_{17}(0)$.

Table~\ref{tab:rates} presents the predicted neutrino fluxes, and
the predicted chlorine and gallium capture rates, for the BP00
standard solar model and the ${\rm ^7Be(p},\gamma){\rm ^8B}$
cross section factor of eq.~(\ref{eq:b8}).  Except for $\phi({\rm
^8B})$, all of the fluxes in table~\ref{tab:rates} are the same
as in table~7 of the original BP00 paper~\cite{bp2000}.

\subsection{Uncertainties}
\label{subsec:uncertainties}

Table~\ref{tab:uncertainties} summarizes the uncertainties in the
most important solar neutrino fluxes and in the Cl and Ga event
rates that are due to different nuclear fusion reactions (the
first four entries), the heavy element to hydrogen mass ratio
(Z/X), the radiative opacity, the solar luminosity, the assumed
solar age, and the helium and heavy element diffusion
coefficients.  In addition, the ${\rm ^{14}N} + p$ reaction
causes a $0.2$\% uncertainty in the predicted pp flux,  a $0.1$
SNU uncertainty in the Cl  event rate, and a $1$ SNU uncertainty
in the Ga event rate.

The uncertainty in the laboratory measurement of the low energy cross
section factor for the ${\rm ^3He}$(${\rm ^4He}$,$\gamma$)${\rm ^7Be}$
reaction is the dominant uncertainty in the calculation of the
${\rm ^7Be}$ neutrino flux and one of the two largest uncertainties in the
calculation of the ${\rm ^8B}$ neutrino flux.

For the original BP00 solar model predictions~\cite{bp2000}, the
uncertainties in the individual parameters and in the fluxes are, with
the exception of the two non-zero entries in the column under ${\rm
^7Be + p}$, the same as in table~\ref{tab:uncertainties}. Instead of
the entries $0.040$ under ${\rm ^7Be + p}$, the original BP00 model
predictions had a much larger uncertainty, $0.105$. The original BP00
uncertainties were used, for example, in the analyses of solar
neutrino oscillations described in refs.~
\cite{bks2001,bgp,krastev01,foglipostsno}.

The input data that were used to calculate the uncertainties
listed in table~\ref{tab:uncertainties} are given in the papers
in which the BP00 model~\cite{bp2000} and the BP98
model~\cite{bp98} were originally presented.  The principal
change from the uncertainties that were calculated for the BP98
model (cf. table~2 in ref.~\cite{bp98}), and which were used in many
previous analyses of solar neutrino oscillations, is that, assuming the
validity of eq.~(\ref{eq:b8}), the uncertainty in the cross
section for the ${\rm ^7Be(p},\gamma){\rm ^8B}$ reaction is
reduced by a factor of more than three and the estimated
uncertainty in the heavy element to hydrogen ratio, $Z/X$, is
increased by almost a factor of two( cf. discussion in section
5.1.2 of ref.~\cite{bp2000}).

The predicted event rates for the chlorine and gallium experiments
make use of neutrino absorption cross sections from
refs.~\cite{nuabs}. The uncertainty in the prediction for the standard
(no oscillation) gallium rate is dominated by uncertainties in the
neutrino absorption cross sections, $7$ SNU ($5$\% of the predicted
rate). The uncertainties in the chlorine absorption cross sections
cause an error, $\pm 0.22$ SNU ($2.6$\% of the predicted rate), that
is relatively small compared to other uncertainties in predicting the
rate for this experiment.  For calculations that involve non-standard
neutrino energy spectra that result from neutrino oscillations or
other new neutrino physics, the uncertainties in the predictions for
currently favored solutions (which reduce the contributions from the
least well-determined ${\rm ^8B}$ neutrinos) will in general be less
than the values quoted here for standard spectra and must be
calculated using the appropriate cross section uncertainty for each
neutrino energy~\cite{nuabs}.

\TABLE[!t]{
\centering
\caption{{\bf For the standard solar model (BP00 + New ${\rm\bf ^8B}$), the
average uncertainties in neutrino fluxes and event rates due to
different input data.}  The flux uncertainties are expressed in
fractions of the total flux and the event rate uncertainties are
expressed in SNU.  The ${\rm ^7Be}$ electron capture rate causes an
uncertainty of $\pm 2\%$ \cite{be7paper} that affects only the ${\rm
^7Be}$ neutrino flux.  The average fractional uncertainties for
individual parameters are shown. \label{tab:uncertainties}}
\begin{tabular}{@{\extracolsep{-5pt}}lccccccccc}
\noalign{\bigskip}
\hline
\noalign{\smallskip}
$<$Fractional&pp&${\rm ^3He ^3He}$&${\rm ^3He ^4He}$&${\rm ^7Be} +
p$&$Z/X$&opac&lum&age&diffuse\\
uncertainty$>$&0.017&0.060&0.094&0.040&0.061&&0.004&0.004&0.15\\
\noalign{\smallskip}
\hline
\noalign{\smallskip}
Flux\\ \cline{1-1}
\noalign{\smallskip}
pp&0.002&0.002&0.005&0.000&0.004&0.003&0.003&0.0&0.003\\
${\rm ^7Be}$&0.016&0.023&0.080&0.000&0.034&0.028&0.014&0.003&0.018\\
${\rm ^8B}$&0.040&0.021&0.075&0.040&0.079&0.052&0.028&0.006&0.040\\
\noalign{\medskip}
SNUs\\ \cline{1-1}
\noalign{\smallskip}
Cl&0.3&0.2&0.6&0.3&0.6&0.4&0.2&0.04&0.3\\
Ga&1.3&1.0&3.3&0.6&3.1&1.8&1.3&0.20&1.5\\
\noalign{\smallskip}
\hline
\end{tabular}
}

\section{Global neutrino oscillation solutions with different analysis
prescriptions}
\label{sec:global}

We present in this section the global solutions for the allowed
neutrino oscillation parameters that are derived using the neutrino
fluxes and uncertainties given in table~\ref{tab:rates} and
table~\ref{tab:uncertainties}. We describe in
section~\ref{subsec:three} the oscillation solutions that are allowed
with three different analysis prescriptions (see
figure~\ref{fig:global3}) and show how adopting the new Junghans et
al. value of $S_{17}(0)$ reduces the allowed regions in neutrino
parameter space (see figure~\ref{fig:beforeafter}). We have at
different times used all three of the analysis strategies and various
colleagues have advocated strongly one or the other of the strategies
described here. In section~\ref{subsec:methods}, we describe in more
detail the three different analysis
prescriptions. Section~\ref{subsec:three} is intended for a general
readership, while section~\ref{subsec:methods} is intended primarily
for aficionados of neutrino oscillations.

The interested reader may wish to consult also a number of recent
papers,
refs.~\cite{bgp,krastev01,foglipostsno,barger2001,bayesian,goswami},
that have determined the allowed solar neutrino oscillation solutions
including the CC data from SNO. We will spare the reader erudite
comparisons, in the cases where there is overlap in the calculated
quantities, between our detailed results and those of the other
authors referred to above. In general, when the same analysis
procedures and input data are used, all authors obtain consistent
results.

In this paper, we study the sensitivity of the calculated
neutrino predictions to the adopted analysis strategy by comparing the
numerical results obtained with three different analysis
procedures. We also compare, for global neutrino solutions and for
predicted neutrino measurable quantities, the results that are
obtained using the previous standard value for $S_{17}(0)$ with the
results that are obtained using the recent precise measurement by
Junghans et al.~\cite{junghans01} of $S_{17}(0)$. In our previous paper on solar
neutrino oscillations~\cite{bgp}, we used only the analysis strategy (a)
(cf. Section 3.3 for a description of all three analysis strategies)
and the previously standard value for $S_{17}(0)$.  We apply the
sensitivity studies described in section 3 to the calculation of a
variety of solar neutrino measurables in sections 4-6, providing for
the first time a quantitative evaluation of the robustness of the
predictions to diverse analysis strategies.

\subsection{Global oscillation solution: three different analyses}
\label{subsec:three}

\FIGURE[!t]{
\centerline{\psfig{figure=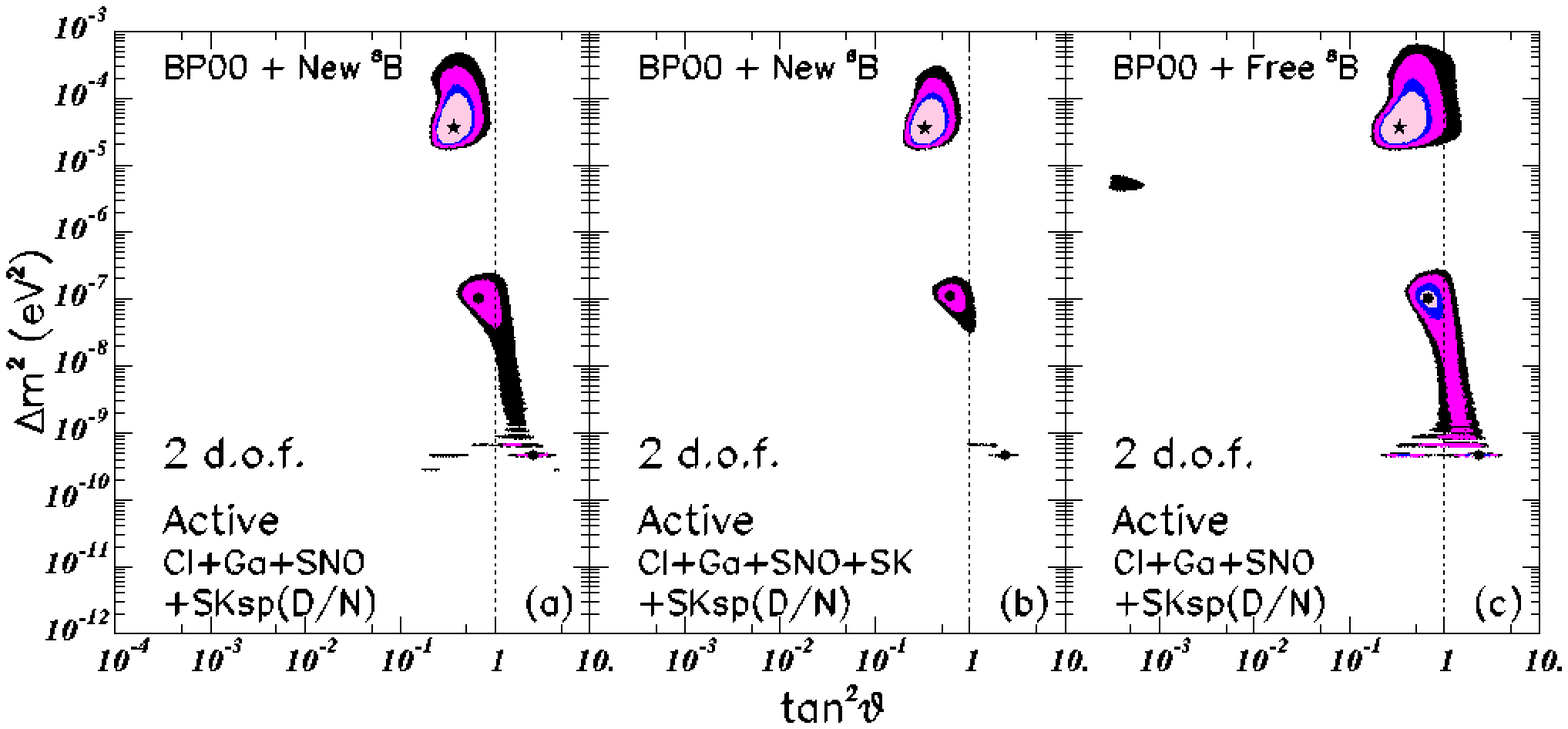,width=5.5in}}
\caption{{\bf Global neutrino oscillation solutions for three
different analysis strategies.} The three different analysis
strategies are described in section~\ref{subsec:methods}. In all
cases, the value of $S_{17}(0)$ reported in ref.~
~\cite{junghans01} and shown in eq.~(\ref{eq:b8}) was used.  The
input data include the neutrino fluxes and uncertainties
predicted by the BP00 solar model~\cite{bp2000} and the total
measured event rates from the SNO~\cite{sno2001},
Chlorine~\cite{chlorine}, and Gallium
(averaged)~\cite{gallex,gno,sage} experiments, as well as the
recoil electron energy spectrum measured by
Super-Kamiokande~\cite{superk} during the day and separately the
energy spectrum measured at night. The C.L. contours shown in the
figure are $90$\%, $95$\%, $99$\%, and $99.73$\% ($3\sigma$). The
allowed regions are limited by the Chooz reactor
measurements~\cite{chooz} to mass values below $\sim$ (7--8)
$\times 10^{-4} {\rm eV^2}$. The local best-fit points are marked
by dark circles. \label{fig:global3}}}

We first discuss in section~\ref{subsubsec:newglobal} the new global
solutions for each of three different analysis strategies and then compare in
section~\ref{subsubsec:beforeafter} the global solutions that are
obtained with the older and newer (cf. eq.~\ref{eq:b8}) values of
$S_{17}(0)$.

\subsubsection{New global solution}
\label{subsubsec:newglobal}

Figure~\ref{fig:global3} shows the allowed ranges of the neutrino
oscillation parameters, $\Delta m^2$ and $\tan^2\theta$, that were
computed following three different analysis approaches that have been
used previously in the literature. These approaches are described in
more detail in section~\ref{subsec:methods}. We follow
refs.~\cite{lisitan,GFM} in using $\tan^2\theta$ (rather than $\sin^2
2\theta$) in order to display conveniently solutions on both sides of
$\theta = \pi/4$.

Figure~\ref{fig:global3}a presents the result for our standard
analysis, which includes the BP00 predicted fluxes and uncertainties
and all the experimental data except the Super-Kamiokande total event
rate. The measured day and night electron recoil energy spectra
together represent the Super-Kamiokande total rate in this
approach. Figure~\ref{fig:global3}b was constructed using an analysis
similar to that used to construct figure~\ref{fig:global3}a except
that for figure~\ref{fig:global3}b the total Super-Kamiokande rate is
included explicitly together with a free normalization factor for the
day and night recoil energy spectra.  Figure~\ref{fig:global3}c shows
the results for the free ${\rm ^8B}$ analysis, in which the calculation is
the same as for the standard case, figure~\ref{fig:global3}a, except
that the ${\rm ^8B}$ neutrino flux is not constrained by the solar model
predictions.

The allowed regions obtained in the three panels of
figure~\ref{fig:global3}, although different, do not depend
dramatically on the adopted strategy.  The allowed regions are reduced
in size for strategy (a) and strategy (b) as compared to the regions
derived using the 1998 standard value of
$S_{17}(0)$~\cite{adelberger}, cf. the results presented in
refs.~\cite{bgp} and ~\cite{foglipostsno}, respectively.

The most restricted allowed regions are obtained for strategy (b),
which has been used by the Bari group~\cite{foglipostsno}.  As
discussed in ref.\cite{bgp}, the smallness of the allowed regions for
this strategy results from the absence in the analysis of the ${\rm
^8B}$ theoretical error for the Super-Kamiokande spectrum
normalization.  We will discuss this point more in the next section.

The largest allowed regions in figure~\ref{fig:global3}
correspond to strategy (c), used among others by Krastev and
Smirnov~\cite{krastev01}, in which the
${\rm ^8B}$ neutrino flux is treated as a free parameter. This was not
the case when using the 1998 standard value of
$S_{17}(0~\cite{adelberger})$ (see fig.9 in ref~\cite{bgp}). The
only changes in the quantities derived from
BP00 and    BP00 + New ${\rm^8B}$ are the ${\rm^8B}$
neutrino flux and its associated uncertainty.
Therefore the allowed regions are unchanged when going from BP00 to
BP00 + New ${\rm^8B}$ in strategy (c), for which the  ${\rm^8B}$
neutrino flux is treated as a free parameter.
Thus Fig.~\ref{fig:global3}c is equal to the right panel of
Fig.9 in Ref.~\cite{bgp} and it is also in excellent agreement with the
results from the free $^8$B analysis of ref.~\cite{krastev01}. The
change in the relative size of the allowed regions between strategy
(a) and (c) is due to the reduction in size, discussed above, of the
allowed regions for strategy (a).

In constructing figure~\ref{fig:global3}, we assumed that only active
neutrinos exist.  We derive therefore the allowed regions in $\chi^2$
using only two free parameters: $\Delta m^2$ and $\tan^2\theta$.  As
discussed below, oscillations into pure sterile neutrinos are highly
disfavored (see table~\ref{tab:globalbestbp00}) with the larger value
of $S_{17}(0)$ given in eq.~(\ref{eq:b8}), and are not included in our
standard demarcation of the allowed regions.  The allowed regions for
a given C.L. are defined in this paper as the set of points satisfying
the condition
\begin{equation}
    \chi^2 (\Delta m^2,\theta)
    -\chi^2_{\rm min}\leq \Delta\chi^2 \mbox{(C.L., 2~d.o.f.)} ,
\label{eqn:chidiff}
\end{equation}
with $\Delta\chi^2(\mbox{C.L., 2~d.o.f.}) = 4.61$, $5.99$, $9.21$, and
$11.83$ for C.L.~= 90\%, 95\%, 99\% and 99.73\% ($3\sigma$).
We use the standard least-square analysis approximation
for the definition of the allowed regions with a given confidence level.
As shown in ref.~\cite{bayesian} the allowed regions obtained in this
way are very similar to those obtained by a bayesian analysis.

\TABLE[!t]{ \centering \caption{\label{tab:globalbestbp00} {\bf Best-fit
global oscillation parameters with all solar neutrino data.}  The
table gives for the global solution illustrated in
figure~\ref{fig:global3}a the best-fit values for $\Delta m^2$,
$\tan^2 \theta$, $\chi^2_{\rm min}$, and g.o.f. for all the neutrino
oscillation solutions that were discussed in our previous analysis in
ref.~\cite{bgp}.  The differences of the squared masses are given in
${\rm eV^2}$.  The number of degrees of freedom is 39 [38(19 night
spectrum, 19 day spectrum) + 3(rates) $-$2(parameters: $\Delta {\rm
m}^2$, $\theta$)].  The BP00 best-fit fluxes and their estimated
errors have been included in the analysis (see table~\ref{tab:rates}
and table~\ref{tab:uncertainties}).  The rates from the GALLEX/GNO and
SAGE experiments have been averaged to provide a unique data point.
The goodness-of-fit given in the last column is calculated from the
value of $\chi^2/{\rm d.o.f}$ at each local minimum (i.e., for LMA,
SMA, VAC). Solutions that have $\chi_{\rm min}^2 \geq 34.5 + 11.83$ are
not allowed at the $3\sigma$ C.L.}
\begin{tabular}{lcccc}
\noalign{\bigskip}
\hline
\noalign{\smallskip}
Solution&$\Delta m^2$&$\tan^2(\theta)$& $\chi^2_{\rm min}$ &g.o.f. \\
\noalign{\smallskip}
\hline
\noalign{\smallskip}
LMA& $3.7\times10^{-5}$  &$3.7\times10^{-1}$ & 34.5 &$67$\% \\
LOW& $1.0\times10^{-7}$  &$6.7\times10^{-1}$ & 40.8 &$39$\%\\
VAC& $4.6\times10^{-10}$ &$2.5\times10^{0}$ & 42.3 &$33$\%\\
Sterile VAC & $4.7\times10^{-10}$ & $3.0\times10^{0}$ & 49.1 & $13$\%\\
SMA& $5.2\times10^{-6}$  &$1.8\times10^{-3}$ & 49.9 &$11$\%\\
Just So$^2$ & $5.5\times10^{-12}$
&\hskip -6pt$0.61 (1.6) \times10^{0} $ & 52.1 &$7.8$\%\\
 Sterile Just So$^2$
& $5.5\times10^{-12}$  &\hskip -6pt$0.61 (1.6) \times10^{0} $ &
52.1 &$7.8$\%\\
Sterile SMA & $4.6\times10^{-6}$ & $3.4\times10^{-4}$ & 52.3 & $7.5$\%\\
\noalign{\smallskip}
\hline
\end{tabular}
}

Table~\ref{tab:globalbestbp00} gives for our standard analysis strategy
(cf. figure~\ref{fig:global3}a) the best-fit values for $\Delta m^2$
and $\tan^2 \theta$ for all the neutrino oscillation solutions that
were discussed in our previous analysis in ref.~\cite{bgp}. The table
also lists the values of $\chi_{\rm min}^2$ for each solution. The
regions for which the local value of $\chi_{\rm min}^2$ exceeds the
global minimum  by more than $11.83$ are not allowed at $3\sigma$
C.L. Thus, all solutions with $\chi_{\rm min}^2$ as bad or worse than
the VAC Sterile solution are not allowed at $3\sigma$ for the standard
analysis.

The set of allowed solutions at $3\sigma$ C. L. would not change if we
considered active-sterile admixtures and defined the allowed regions
as in figure~1 of ref.~\cite{bgp} in terms of
$\Delta\chi^2(\mbox{C.L., 3~d.o.f.})$ instead of
$\Delta\chi^2(\mbox{C.L., 2~d.o.f.})$.  Table~\ref{tab:globalbestbp00}
shows that all regions with $\Delta\chi^2 > 11.83$ (corresponding to
$3\sigma$ C.L. for $2$ d.o.f.) also satisfy $\Delta\chi^2 > 14.16$
(corresponding to $3\sigma$ C.L. for $3$ d.o.f.).

Many authors(see, e. g., ref.~\cite{bimaximal} and references quoted
therein) have discussed the possibility of bi-maximal neutrino
oscillations, which in the present context implies $\tan^2 \theta
= 1$. Figure~\ref{fig:global3} shows that bi-maximal mixing is disfavored
for the preferred LMA solution. Quantitatively, we find that there
there are no solutions with $\tan^2 \theta = 1$ at the $99.87$\%
C.L. for our standard analysis strategy, corresponding to panel (a) of
figure~\ref{fig:global3}. For the strategies corresponding to the
panels (b) and (c), respectively, there are no solutions with $\tan^2
\theta = 1$ at the $99.95$\% C.L. and $98.95$\% confidence
level. Maximal mixing is allowed for the less favored LOW and VAC
solutions at a C.L. that varies between $95$\% and $3\sigma$, depending
upon the analysis strategy.  In summary, maximal mixing is not favored
but it is not rigorously ruled out.

The upper limit on on the allowed value of $\Delta m^2$ is important
for neutrino oscillation experiments, as stressed in
ref.~\cite{krastev01}. In units of $10^{-4}~{\rm eV^2}$. we find the
following $3\sigma$ upper limits on $\Delta m^2$ (values in parenthesis are
obtained with the older value of $S_{17}(0)$, see ref.~\cite{adelberger}):
case a: 4.7 (7.5); case b: 3.0 (6.8); and case c: 6.5 (6.5). The upper
limit therefore lies between $3.0$ and $7.5$, in units of
$10^{-4}~{\rm eV^2}$, depending upon what is assumed about $S_{17}(0)$
and the analysis strategy.

\subsubsection{Global ``Before and After''}
\label{subsubsec:beforeafter}
\FIGURE[!ht]{
\centerline{\psfig{figure=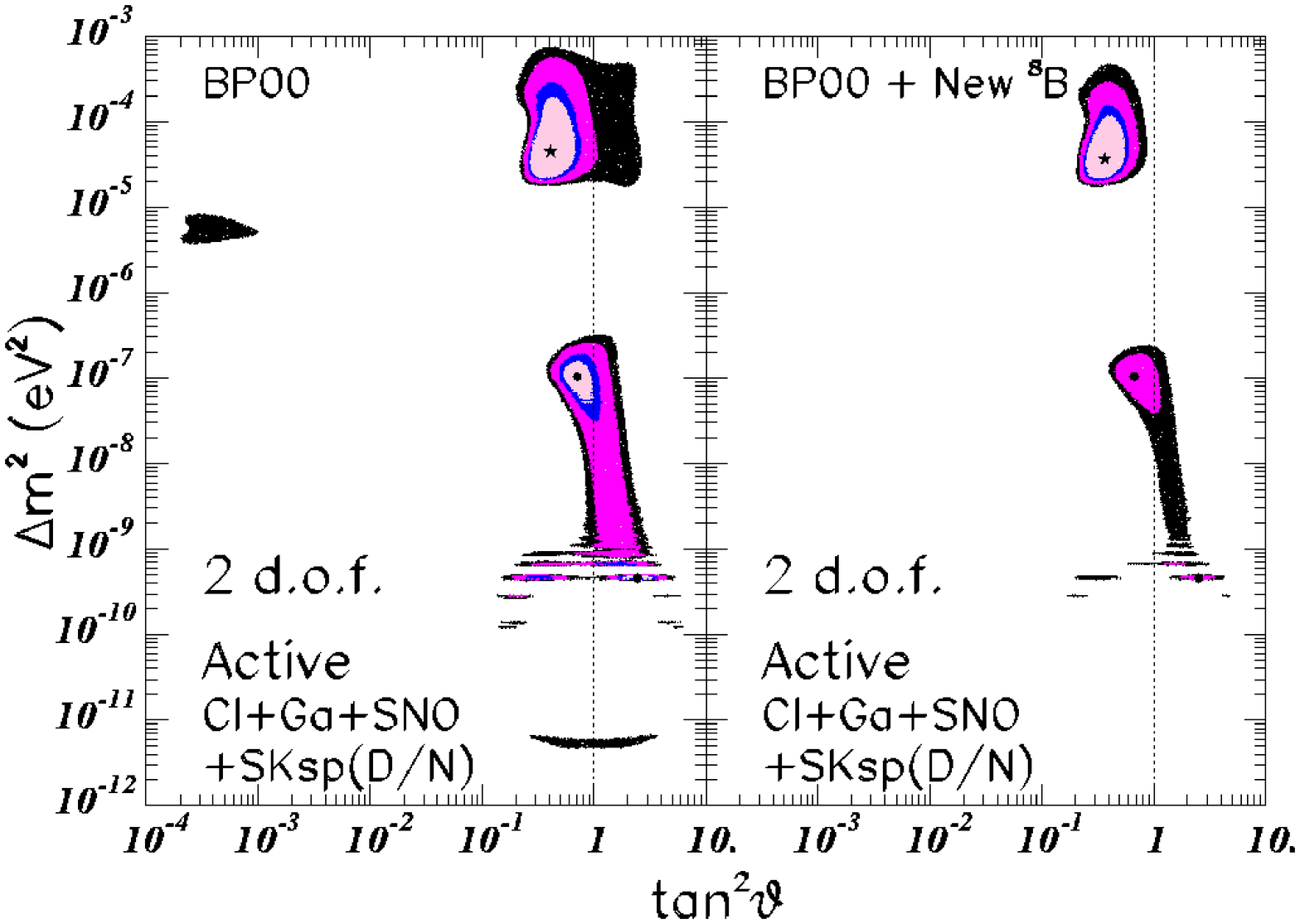,width=4.5in}}
\caption{{\bf Global ``Before and After.''} The left panel shows the allowed
regions for the neutrino oscillations computed in ref.~\cite{bgp} (see
figure 9a of that paper) with
$S_{17}(0) = 19^{+4}_{-2} {\rm eVb}$~\cite{adelberger} and the
standard analysis procedure used here(cf. figure~\ref{fig:global3})
and in ref.~\cite{bgp}. The right panel, which is the same as
figure~\ref{fig:global3}a, shows the allowed regions computed with the
same procedure but with the Junghans et al. value of $S_{17}(0) =
(22.3 \pm 0.9) {\rm eVb}$~\cite{junghans01}.
\label{fig:beforeafter}}}

Figure~\ref{fig:beforeafter} compares the allowed regions found with
the new ${\rm ^8B}$ theoretical neutrino flux and uncertainties
(table~\ref{tab:rates} and table~\ref{tab:uncertainties}) with our
previously published results(see figure 9a of , ref.~\cite{bgp}).  We like to
refer to this comparison as our global ``Before-After'' figure.

The modification of the ${\rm ^8B}$ flux and uncertainties
considered in this paper results in a reduction of the allowed
neutrino oscillation parameter space.  Since no pure sterile
oscillation solutions are found at the $3\sigma$ level, we focus
on oscillations involving only active neutrinos. We find allowed
regions for the LMA, LOW, and VAC solutions with our standard
analysis strategy. The allowed regions for the preferred LMA and
LOW solutions both become smaller with the larger adopted
$S_{17}(0)$, and the fit to the LOW solution becomes relatively
less good. The SMA and Just-so$^2$ solutions, which are
marginally allowed in the standard 'Before' panel of
figure~\ref{fig:beforeafter}, are not present in the 'After'
panel at the 3-$\sigma$ level.

The differences between the Before and After panels are mainly due to
two effects, which are described below.
\begin{description}
\item[(a)] The increase of the ${\rm ^8B}$ normalization results in a
smaller value for $\chi^2_{\rm min}$, which is found for the LMA
best-fit point.  With the new value of $S_{17}(0)$, we find
$\chi^2({\rm LMA})_{\rm min} = 34.5$ while before we had $\chi^2({\rm
LMA})_{\rm min} = 35.3$. Although the theoretical errors are reduced
in the new solution, the global minimum (which lies within the
boundaries of the LMA solution) is ``deeper.'' This effect can be
confirmed by comparing the results of figure~\ref{fig:global3}a and
figure~\ref{fig:global3}c, which were computed with and without the
BP00 constraint on the ${\rm ^8B}$ neutrino flux.  The global minimum
points are much closer together in neutrino parameter space than they
were when the older value of $S_{17}(0)$ was used (cf.  results given
in ref.~\cite{bgp}).  With the new ${\rm ^8B}$ normalization the LMA
best-fit point is not only the best-fit point in oscillation parameter
space but it also corresponds very closely to the predicted ${\rm
^8B}$ neutrino flux given in table~\ref{tab:rates}. We find that
$\phi({\rm ^8B};{\rm LMA~ best-fit}) = 0.999\phi({\rm ^8B};{\rm BP00 +
New ^8B})$, where the coefficient $0.999$ applies for case (c) (free
${\rm ^8B}$ analysis). For our standard analysis, case (a), the
coefficient is $0.96$\footnote{There is a simple reason why the
coefficient, defined in eq.~(\ref{eq:fbdefn}) as ${\rm f_B}$, deviates
slightly more from unity when $\phi({\rm ^8B})$ is constrained by the
standard solar model than when it is not constrained. The reason is
that the theoretical uncertainty is given as a fraction of the total
${\rm ^8B}$ neutrino flux. For higher fluxes, the theoretical error is
larger and the value of $\chi^2$ is reduced somewhat more. As can be
seen from eq.~(\ref{eq:fbdefn}), ${\rm f_B}$ depends inversely upon
the theoretical flux. Therefore, higher values of the flux, i. e.,
lower values of ${\rm f_B}$, are slightly favored.}.  Presumably, this
precise agreement between the best-fit LMA neutrino flux and the BP00
+ New ${\rm ^8B}$ flux is accidental, since the allowed $3\sigma$
range for the ${\rm ^8B}$ flux, within the domain of the LMA
oscillation parameters, is $(0.54-1.36)\phi({\rm ^8B};{\rm BP00 + New
^8B})$.

\item[(b)] The theoretical error for the ${\rm ^8B}$ flux is reduced
according to eq.~(\ref{eq:b8}) from an average value of $19$\% with
the Adelberger et al. value of $S_{17}(0)$~\cite{adelberger} to
$14.5$\% with the Junghans et al. value~\cite{junghans01} of
$S_{17}(0)$. The reduction of this theoretical error pushes the
previously marginally allowed SMA and Just So$^2$ solutions (for these
somewhat unfamiliar solutions see the original discovery
papers~\cite{raghavan} and ref.~\cite{bks2001}) over the edge (beyond the
$3\sigma$ allowed region).  As discussed in section 5.3 of
ref.~\cite{bgp}, the error on the predicted ${\rm ^8B}$ neutrino flux
played a crucial role in previously allowing, with the standard
analysis prescription, the marginal SMA and \hbox{Just So$^2$
solutions}.
\end{description}

We have performed a series of numerical experiments to determine which
effect, either the increase in the predicted ${\rm ^8B}$ flux or the
decrease in the uncertainty in the predicted flux, is the dominant
effect in reducing the allowed parameter space. We find that the
increase in the predicted flux contributes the most to worsening the
fit for the LOW solution while the decrease in the uncertainty in the
predicted flux is dominant reason why the SMA and Just So$^2$ regions
no longer appear at $3\sigma$. The increase in the predicted flux and
the decrease in the theoretical uncertainty contribute comparably to
reducing the size of the LMA allowed region.

\subsection{The allowed range for the ${\rm ^8B}$ neutrino flux}
\label{subsec:fb8}

The analysis described above yields best-estimates and $3\sigma$
allowed ranges for the total ${\rm ^8B}$ neutrino flux.  Let $f_{\rm
B}$ be the inferred ${\rm ^8B}$ neutrino flux produced in the sun in
units of the best-estimate predicted BP00 neutrino flux,
\begin{equation}
f_{\rm B} ~ =~
\frac{\phi({\rm ^8B})}{\phi({\rm ^8B})_{\rm BP00 + New ^8B}}.
\label{eq:fbdefn}
\end{equation}

Table~\ref{tab:fb8} gives the best-estimate and allowed range of the
solar ${\rm ^8B}$ neutrino flux. The values in brackets in
table~\ref{tab:fb8} were obtained by our standard analysis, case (a),
while the values without brackets were derived allowing the ${\rm
^8B}$ neutrino flux to be a free parameter, i. e., case (c).  The
bracketed range is always slightly smaller than the unbracketed range,
which just reflects the fact that for strategy (a) the $\chi^2$
penalizes large differences from the standard solar model value of the
$^8$B flux while the $^8$B flux is unconstrained for strategy
(c). However, the two strategies, (a) and (c), yield similar allowed
ranges for $f_{\rm B}$\footnote{For strategy (a), we first obtain the
allowed region of oscillation parameters using the BP00 values for the
$^8$B neutrino flux and uncertainties and then for each pair of
($\Delta m^2$ , $\tan^2\theta$) within the allowed region we obtain
the ``optimum'' value of $f_{\rm B}$ that best fits the data. The
range given in Table~\ref{tab:fb8} is the range of the optimum
$f_{\rm B}$ values obtained with this procedure for all the
oscillation parameters within the $3\sigma$ allowed region in
Fig.1(a).  For strategy (c), we defined the allowed regions allowing
the $^8$B neutrino flux to be a free parameter, unconstrained by the
BP00 predictions. Thus the range given in Table~\ref{tab:fb8} is the
range of the optimum $f_{\rm B}$ obtained for all the oscillation
parameters within the $3\sigma$ allowed region in Fig.1(c).}.

In all cases, the inferred range of $f_{\rm B}$ allowed by
the global oscillation analysis using data from all of the experiments
is smaller than the allowed range obtained by the SNO
collaboration~\cite{sno2001} from comparing the SNO and
Super-Kamiokande total rate measurements. In our notation, the SNO
collaboration found $f_{\rm B} = 0.92 \pm 0.50$, where the result is
expressed in terms of the BP00 flux calculated with the Junghans et
al. value of $S_{17}(0)$ (see table~\ref{tab:rates} and
ref.~\cite{junghans01}) and we quote the $3\sigma$ allowed range. Our
result is $f_{\rm B} = 0.88 \pm 0.48$[or, for strategy (a), $f_{\rm B}
= 0.86 \pm 0.38$]. Of course, the SNO determination is more direct. The
procedure described here assumes the validity of the two-neutrino
oscillation analysis and uses solar model fluxes and uncertainties
(see especially discussion in section~\ref{subsec:discussmodel}).

Even when the ${\rm ^8B}$ neutrino flux is unconstrained in the
oscillation analysis, almost all of the allowed range of $f_{\rm B}$
is within the $3\sigma$ uncertainty of the BP00 solar model
predictions. Only for the SMA solution does the allowed range of
$f_{\rm B}$ not overlap with the $3\sigma$ predicted range of the
BP00 solar model.

The procedure described above assumes the validity of the two-neutrino
oscillation analysis for pure active oscillations. As shown in
ref.~\cite{barger2001}, the allowed range of $f_{\rm B}$ can be
modified by allowing for the possibility of an admixture of sterile
neutrinos in the oscillations within the adiabatic regime.  In order
to test this possibility we have repeated our global analysis (c) with
an arbitrary active-sterile admixture within the region $10^{-3} <
\Delta m^2/{\rm eV}^2 < 10^{-5}$ , $0.05<\tan^2\theta<5$.  We find
that the maximum allowed ${\rm ^8B}$ neutrino flux is increased to
$f_{\rm B} \leq 2.9$ (at 3$\sigma$ for 3 d.o.f.) and corresponds to
oscillations into a $25$\% active and $75$\% sterile state. However,
these mixed scenarios including sterile neutrinos give a worse fit to
the data than pure active oscillations. A large component of sterile
neutrinos does better if one includes in the analysis only the total
event rates, as in ref.~\cite{barger2001}. For completeness, we
briefly describe how this is possible.  In order to fit the appearance
of active neutrinos at Super-Kamiokande versus SNO, one needs to
decrease $P_{ee}$ for ${\rm ^8B}$ neutrinos \cite{sno4} (which is
compensated by the increase of $f_{\rm B}$ ), while in order to keep
the agreement with gallium and chlorine rates the survival probability
at lower energies must not be substantially modified. Barger et
al. have pointed out in ref. ~\cite{barger2001} that this tuning can
be achieved within the adiabatic regime. However, as discussed in
ref. ~\cite{bgp}, one has less freedom for tuning in the global
analysis. Within the adiabatic regime, the tuning of the probability
can only be achieved by lowering both the value of $\tan^2\theta$ and
$\Delta m^2$, intruding into the region of a predicted large day-night
variation in the Super-Kamiokande experiment.  \TABLE[!t] { \centering
\caption{{\bf Allowed values for the ${\rm ^8B}$ neutrino flux.} The
table gives for each allowed oscillation solution the best-fit values
for $f_{\rm B}$ and the $3\sigma$ range of allowed values. The
quantity $f_{\rm B}$ is the ${\rm ^8B}$ neutrino flux expressed in
units of the best-estimate BP00 value for the ${\rm ^8B}$ neutrino
flux, see eq.~(\ref{eq:fbdefn}). The values of $f_{\rm B}$ given in
brackets were obtained by our standard analysis procedure, (a); the
values presented without brackets were obtained by analysis procedure
(c), in which the $^8$B flux is not constrained by solar model
predictions.}  \protect\label{tab:fb8}
\begin{tabular}{lccc}
\noalign{\bigskip}
\hline
\noalign{\smallskip}
Solution&$f_{\rm B}({\rm Best})$&$f_{\rm B} (3\sigma {\rm~ range})$\\
\noalign{\smallskip}
\hline
\noalign{\smallskip}
LMA&$1.00$&$0.54-1.36\;[0.57-1.24]$\\
LOW&$0.73$&$0.54-0.93\;[0.55-0.91]$\\
VAC&$0.59$&$0.48-0.68\;[0.48-0.66]$\\
SMA&$0.44$&$0.40-0.50\;$[{ no ~ solution}]\\
\noalign{\medskip}
&&\hrulefill&\hrulefill\\
\noalign{\smallskip}
\hline
\end{tabular}
}

\subsection{Methods of analysis}
\label{subsec:methods}

In this subsection, we describe the implementations of the three
different analysis strategies that were used to construct
figure~\ref{fig:global3} and figure~\ref{fig:beforeafter}. For most
readers, the outline of the analysis strategies given in
section~\ref{subsubsec:newglobal} will be
sufficient. Section~\ref{subsec:methods} is intended only for
aficionados of solar neutrino oscillation studies.

The results shown in figure~\ref{fig:global3} and
figure~\ref{fig:beforeafter} were derived using the CC event rate
measured by SNO~\cite{sno2001}, the Chlorine~\cite{chlorine} and
Gallium event rates~\cite{gallex,sage,gno} (we use here the
weighted averaged of the GALLEX/GNO and SAGE rates), and the
$2\times 19$ bins of the (1258 day)
Super-Kamiokande~\cite{superk} electron recoil energy spectrum
measured separately during the day and night periods.  We
include, as described below, the predicted neutrino fluxes and
their uncertainties, table~\ref{tab:rates} and
table~\ref{tab:uncertainties} of the present paper, for the
standard solar model~\cite{bp2000} (BP00 + New ${\rm ^8B}$). We
use the distribution for neutrino production fractions and the
solar matter density given in ref.~\cite{bp2000} and tabulated in
http://www.sns.ias.edu/$\sim$jnb.

In order to explore the robustness of the inferred allowed regions in
neutrino parameter space, we obtain the permitted regions using
calculations based upon three different plausible prescriptions for
the statistical analysis that have been used in the published
literature and which we label as follows.

\begin{description}
\item[(a)] For our ``standard'' analysis,
we adopt the prescription described in refs.~\cite{bks2001,bgp}. We do not
include here the Super-Kamiokande total rate, since to a large extent
the total rate is represented by the flux in each of the spectral
energy bins.
We define the $\chi^2$ function for the global analysis
as:
\begin{equation}
\chi^2_{\rm global,a}=\sum_{i,j=1,41} (R^{th}_i- R^{\exp}_i)
\sigma_{G,ij}^{-2} (R^{th}_j- R^{\exp}_j) ,
\label{chi2a}
\end{equation}
where $\sigma_{G,ij}^2 = \sigma_{R,ij}^2 + \sigma_{SP,ij}^2$.
Here $\sigma_{R,ij}$ is the corresponding $41 \times 41$ error
matrix containing the theoretical as well as the experimental
statistical and systematic uncorrelated errors for the 41 rates
while $\sigma_{SP,ij}$ contains the assumed fully--correlated
systematic errors for the $38\times 38$ submatrix corresponding
to the Super-Kamiokande day--night spectrum data.  We include
here the energy independent systematic error which is usually
quoted as part of the systematic error of the total rate. The
error matrix $\sigma_{R,ij}$ includes important correlations
arising from the theoretical errors of the solar neutrino fluxes,
or equivalently of the solar model parameters. We follow the
principles described by the Bari group~\cite{fogli} in including
the solar model uncertainties, using the updated data presented
in table~\ref{tab:uncertainties}.

\item[(b)] For the ``all rates'' analysis, we follow the prescription
described in refs.~\cite{foglipostsno,fogli,bks99,cc2001,sno4}. In
this approach, we use all the total event rates in the analysis,
including the Super-Kamiokande total event rate with its corresponding
theoretical error.  We
define the $\chi^2$ function for this global analysis as:
\begin{equation}
\chi^2_{\rm global,b}=\chi^2_{\rm Rates,b}+\chi^2_{\rm SP,b} ,
\label{chi2b}
\end{equation}
where
\begin{eqnarray}
\chi^2_{\rm Rates,b}&=&\sum_{i,j=1,4} (R^{th}_i- R^{\exp}_i)
\tilde \sigma_{R,ij}^{-2} (R^{th}_j- R^{\exp}_j),\\
\chi^2_{\rm SP,b}&=& \sum_{i,j=1,38} (\alpha_{\rm SP} R^{th}_i- R^{\exp}_i)
\tilde \sigma_{SP,ij}^{-2} (\alpha_{\rm SP} R^{th}_j- R^{\exp}_j) .
\end{eqnarray}
In $\chi^2_{\rm Rates}$, we include the CC event rate measured at
SNO~\cite{sno2001}, the Chlorine~\cite{chlorine} and
Gallium~\cite{gallex,gno,sage} event rates and the
Super-Kamiokande~\cite{superk} total event rate. The matrix $\tilde
\sigma_{R,ij}$ contains the theoretical uncertainties, as well as the
experimental statistical and systematic errors, for the total rates.
In particular, $\tilde \sigma_{R,ij}$ includes the energy
independent systematic error for the Super-Kamiokande rate.  On the
other hand, the matrix $\tilde \sigma_{{\rm SP},ij}$ contains both the
statistical as well as the systematic errors (both those that are
correlated with energy and those that are uncorrelated) of the $38$
spectral energy data; $\tilde \sigma_{{\rm SP},ij}$ does not include the
theoretical error of the ${\rm ^8B}$ flux.

In order to avoid double counting, we allow a free normalization (for
which no theoretical error is included) for the Super-Kamiokande day
and night electron recoil energy spectra.  Following the usual
procedure for this approach, we introduce an arbitrary normalization
factor, $\alpha_{\rm SP}$, for the amplitude of the energy spectra.
For each value of the oscillation parameters $\chi^2_{{\rm global},b}$
is minimized with respect to $\alpha_{\rm SP}$.

\item[(c)] The ``free ${\rm ^8B}$'' analysis is identical to our standard
analysis, described in (a) above, except that the ${\rm ^8B}$
neutrino flux is not constrained by solar model predictions. This
freedom is implemented by multiplying the ${\rm ^8B}$ flux
contribution to the $R^{th}_i$ in eq.~(\ref{chi2a}) by the
normalization factor, $f_{\rm B}$,
 and removing the theoretical
${\rm ^8B}$ flux errors from $\sigma_{R,ij}$. In this case,  the
resulting $\chi^2$ function can also be written as
\begin{equation}
\chi^2_{\rm global,c}=\chi^2_{\rm Rates,c}+\chi^2_{\rm SP,c},
\label{chi2c}
\end{equation}
where
\begin{eqnarray}
\chi^2_{\rm Rates,c}&=&\sum_{i,j=1,3} (R^{th}_i(f_{\rm B})- R^{\exp}_i)
\hat \sigma_{R,ij}^{-2} (R^{th}_j(f_{\rm B})- R^{\exp}_j), \\
\chi^2_{\rm SP,c}&=& \sum_{i,j=1,38} (R^{th}_i(f_{\rm B})- R^{\exp}_i)
\hat \sigma_{SP,ij}^{-2} (R^{th}_j(f_{\rm B})- R^{\exp}_j) .
\end{eqnarray}
In $\chi^2_{\rm Rates, c}$, we include only the CC event rates measured
in the SNO, Chlorine, and Gallium experiments. For these three rates,
the error matrix $\hat \sigma_{R,ij}$ only differs from $\tilde
\sigma_{R,ij}$ due to the absence of the theoretical error for the
$^8$B neutrino flux.  The error matrix $\hat \sigma_{SP,ij}$ only
differs from $\tilde \sigma_{SP,ij}$ by the inclusion of the energy
independent systematic error, which is usually quoted as part of the
systematic error of the total rate.

For each value of the oscillation parameters, $\chi^2_{\rm global,b}$
is minimized with respect to $f_{\rm B}$.  Notice that the only
difference in the ``free ${\rm ^8B}$'' analysis when performed in
terms of BP00 or in terms of BP00 + New ${\rm^8B}$ is a shift in the
corresponding best value of $f_{\rm B}$ for each value of the
oscillation parameters $f_{\rm B}\rightarrow f_{\rm
B}\left(\frac{5.05}{5.93}\right)$; the allowed regions are left
unchanged as can be seen by comparing Fig.~\ref{fig:global3}c with the
right panel of Fig.9 in Ref.~\cite{bgp}.
\end{description}

\section{Predictions for SNO}
\label{sec:sno}

In this section, we present the predictions for the Sudbury Neutrino
Experiment (SNO), ref.~\cite{sno2001}, of the currently allowed
neutrino oscillation solutions (cf. figure~\ref{fig:global3}) for the
neutral current to charged current ratio
(section~\ref{subsec:double}), for the CC day-night effect
(section~\ref{subsec:ccdaynight}), for the $\nu_e + e$ day-night
effect (section~\ref{subsec:esdaynight}), and for the first and second
moments of the electron recoil energy spectrum
(section~\ref{subsec:moments}).  In Section~\ref{subsec:lucky}, we
show that the neutrino oscillation parameters will be well determined
if we are lucky and SNO measures a large value ($> 3.3$) of the
neutral current to charged current ratio or a large value ($> 0.1$)
for the day-night effect.

The SNO collaboration is studying  charged current
(CC) neutrino absorption  by deuterium,

\begin{equation}
\label{eq:cc}
\nu_e + d \to p + p + e^-\ ,
\end{equation}
as well as the
neutral current (NC) neutrino disassociation of
deuterium,
\begin{equation}
\label{eq:nc}
\nu_x + d \to n + p + \nu'_x  \ ,\quad (x=e,\mu,\,\tau).
\end{equation}
SNO is the only operating (or completed) solar neutrino experiment
that determines a CC rate for electrons whose energies are
measured. The SNO neutral current detection is also unique.  In
addition, SNO measures neutrino-electron scattering,
\begin{equation}
\label{eq:es}
\nu_x + e \to e + \nu'_x  \ ,\quad (x=e,\mu,\,\tau).
\end{equation}
The statistical uncertainties for neutrino-electron scattering in the
SNO detector are less favorable than those available from the
Super-Kamiokande experiment~\cite{superk}.

\FIGURE[!!hb]{
\centerline{\psfig{figure=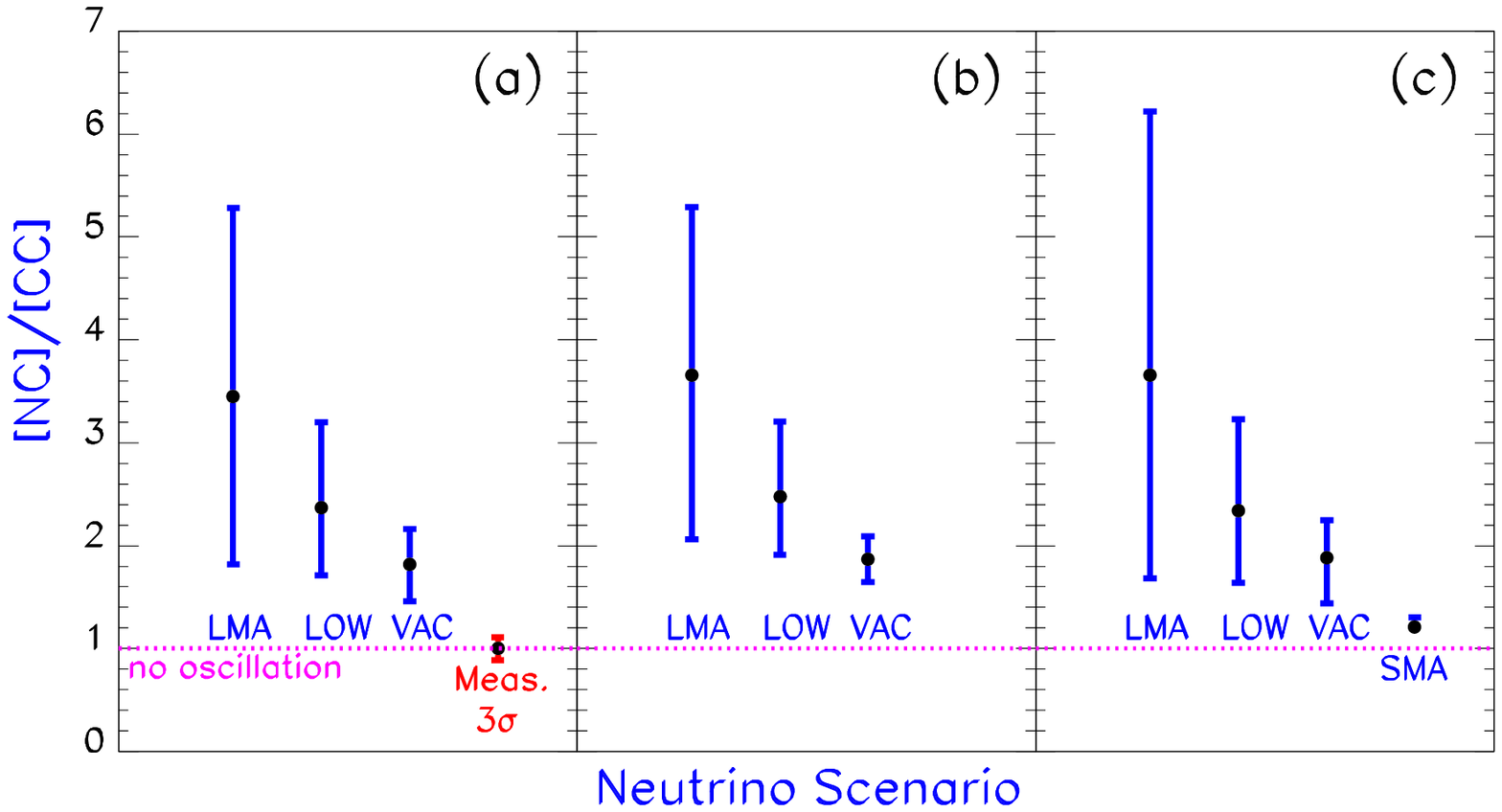,width=4.5in}}
\caption{{\bf The neutral current to charged current double ratio,
[NC]/[CC] .}  The double ratio, [NC]/[CC], is defined by
eq.~(\ref{eq:defnncovercc}).  The three panels of predictions in
figure~\ref{fig:nccc} were derived from the $3\sigma$ ranges shown in
the three panels of global solutions illustrated in
figure~\ref{fig:global3} and obtained using the three analysis
strategies described in section~\ref{sec:global}.
Figure~\ref{fig:nccc} was constructed using a $6.75$ MeV total
electron energy for the CC threshold. Table~\ref{tab:nccc} shows that,
within the allowed domain of neutrino oscillation solutions, the value
of [NC]/[CC] is not sensitive to the threshold energy adopted for the
CC reactions. The tiny $3\sigma$ estimated measuring error in the left
hand panel assumes the systematic uncertainties projected in
ref.~\cite{tencommand} and $3000$ CC events and $1200$ NC events. The
ultimate SNO measurement will probably be even more precise than this
estimate.
\label{fig:nccc}}}

The predicted measurable effects for SNO have been investigated in
detail in ref.~\cite{tencommand} for the larger range of neutrino
oscillation solutions that were viable prior to the
publication~\cite{sno2001} of the first SNO results on the rate of the
CC reaction. We follow here the notation and the techniques described
in ref.~\cite{tencommand}, concentrating on those effects that were
shown previously to be the most likely to be measurable.  We do not
repeat here calculations of the effects that were discussed in
ref.~\cite{tencommand} and found to be small, such as the neutral
current day-night effect or the seasonal difference (winter-summer
difference).

\TABLE[!t]{
\centering
\caption{\label{tab:nccc} {\bf SNO Neutral Current to Charged Current
Double Ratio and Day-Night CC Asymmetry.}
The table presents the double ratio, ${\rm [NC]/[CC]}$ and
$A_{\rm N-D}$ (in \%). The results are tabulated
for different neutrino oscillation scenarios and for two different
thresholds of the recoil  electron kinetic energy used in computing the
CC ratio, $4.5$ MeV (columns two through four) and $6.75$ MeV (columns five
through seven). The ranges are obtained for the 3$\sigma$ regions for
analysis (a).}
\begin{tabular}{@{\extracolsep{-5pt}}ccccccc}
\noalign{\bigskip}
\hline
\noalign{\smallskip}
 &\multicolumn{1}{c}{b.f.}&max&min&\multicolumn{1}{c}{b.f.}&max&min\\
Scenario & 4.5 MeV& 4.5 MeV& 4.5 MeV& 6.75 MeV& 6.75 MeV& 6.75 MeV\\
\noalign{\smallskip}
\hline
\noalign{\smallskip}
&[NC]/[CC]&[NC]/[CC]&[NC]/[CC]&[NC]/[CC]&[NC]/[CC]&[NC]/[CC]\\
\noalign{\smallskip}
\hline
\noalign{\smallskip}
 LMA& 3.44 & 5.28   & 1.79   &3.45    & 5.28 &  1.82\\
 LOW& 2.39 & 3.22   & 1.71   &2.37    & 3.20 &  1.71   \\
 VAC& 1.76 & 2.06   & 1.43   &1.82    & 2.17&  1.46 \\
\noalign{\smallskip}
\hline
\noalign{\smallskip}
&A$_{\rm N-D}$&A$_{\rm N-D}$&A$_{\rm N-D}$&A$_{\rm N-D}$&A$_{\rm N-D}$&A$_{\rm N-D}$\\
\noalign{\smallskip}
\hline
\noalign{\smallskip}
 LMA& 7.4  & 19.5& 0.0   & 8.3   &  21.4   & 0.0  \\
 LOW& 4.3  & 10.4& 0.0   & 3.7   &  9.5    & 0.0  \\
 VAC& 0.1  & 0.3 & -0.2 &0.2 &  0.5  & -0.3  \\
\noalign{\smallskip}
\hline
\end{tabular}
}

For the CC reaction, we adopt a $6.75$ MeV kinetic energy threshold
for the recoil electrons,
the threshold used by the SNO collaboration in their CC measurement
reported in ref.~\cite{sno2001}, as the standard value used in the
figures. In order to illustrate the dependence upon the CC threshold,
we also present results in the tables that refer to a $4.5$ MeV recoil
electron kinetic energy threshold.

\subsection{The Neutral Current to Charged Current Double Ratio}
\label{subsec:double}

In this subsection, we present  predictions for the ratio of neutral
current events (NC) to charged current events (CC) in SNO. The most
convenient form in which to discuss this quantity is~\cite{howwell}

\begin{equation}
{\rm  {[NC]} \over { [CC]} } =
{
{\left({\rm (NC)_{Obs}/(NC)_{SM} }\right) } \over
{\left({\rm (CC)_{Obs}/(CC)_{SM} }\right) }
}.
\label{eq:defnncovercc}
\end{equation}
The ratio ${\rm {[NC]}/ { [CC]} }$ is equal to unity if nothing
happens to the neutrinos after they are produced in the center of
the sun (no oscillations occur). Also, ${\rm {[NC]} / { [CC]} }$
is independent of all solar model considerations provided that
only one neutrino source, $^8$B, contributes significantly to the
measured rates. Inserting into eq.~(25) of ref.~\cite{tencommand}
the $3\sigma$ upper limit measured for the $hep$ flux by the
Super-Kamiokande collaboration~\cite{superk}, one finds that
$hep$ neutrinos affect the value of ${\rm {[NC]}/ { [CC]} }$ by
less than $0.05$\% for a $4.5$ MeV CC kinetic energy threshold
(and by less than $2$\% for a $6.75$ MeV CC threshold.) Moreover,
the calculational uncertainties due to the interaction cross
sections and to the shape of the $^8$B neutrino energy spectrum
are greatly reduced by forming the double
ratio~\cite{bl,yhh,nakamura,bc}.

Figure~\ref{fig:nccc} and table~\ref{tab:nccc} present the calculated
range of the double ratios for the oscillation solutions that are
currently allowed at $3\sigma$.  The table gives the best-fit values
for ${\rm {[NC]} / { [CC]} }$ as well as the maximum and minimum
allowed double ratios for a threshold recoil electron kinetic energy
for the CC reaction of $4.5$ MeV and separately for a CC threshold of
$6.75$ MeV. The calculated double ratio is insensitive to the CC
threshold within the range of thresholds that are considered.

The $1\sigma$ predicted range for our standard global analysis
strategy (a) is $3.45^{+0.70}_{-0.54}$ for a CC threshold of $6.75$
MeV($3.44^{+0.70}_{-0.55}$ for a $4.75$ MeV threshold). The predicted
range for analysis strategy (b) is slightly smaller and for analysis
strategy (c) is somewhat larger ( $3.65^{+0.97}_{-0.80}$).

\FIGURE[!!t]{
\centerline{\psfig{figure=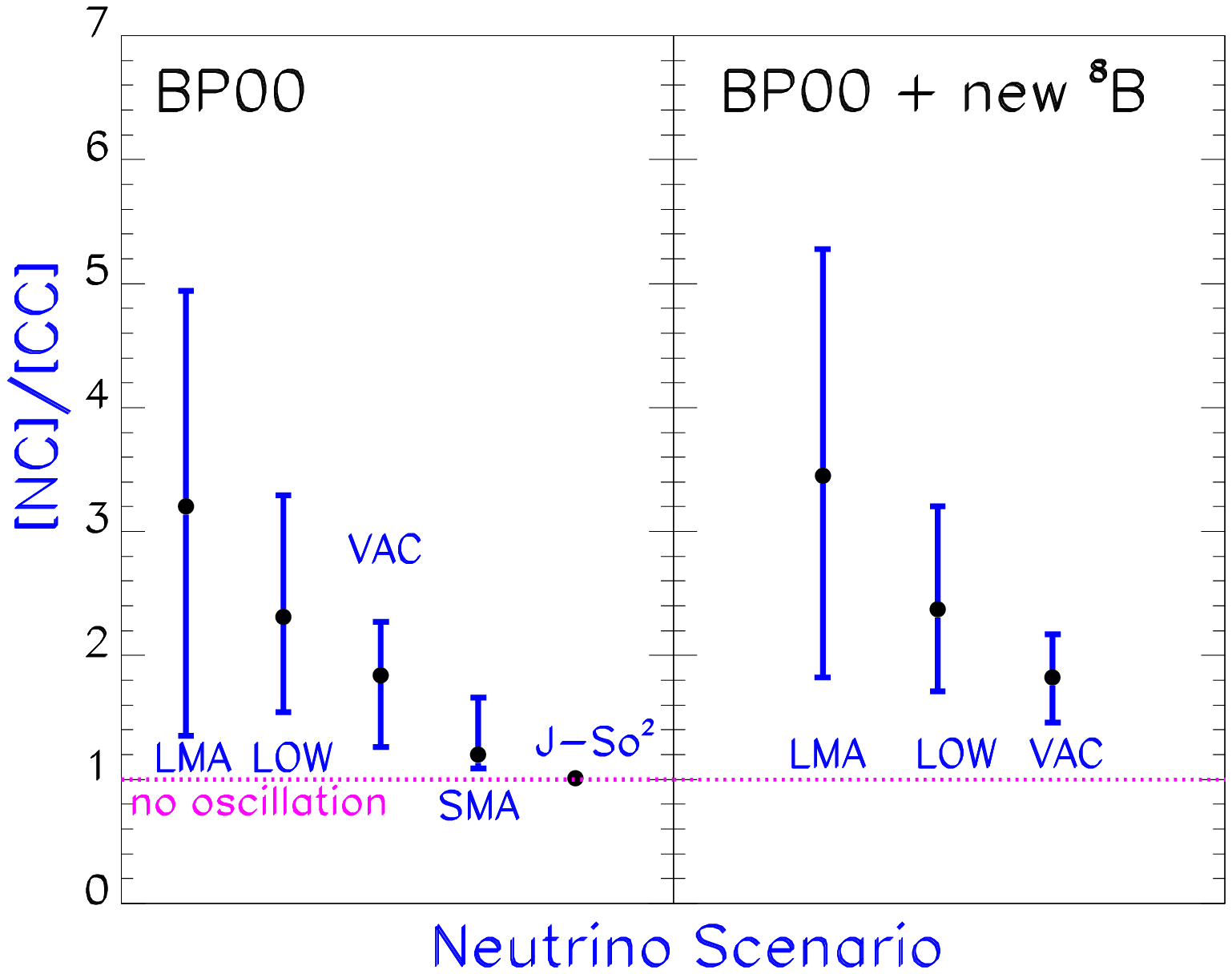,width=4.5in}}
\caption{{\bf [NC]/[CC] ``Before and After.''} The left panel shows
the $3\sigma$ allowed ranges for the neutral current to charged current double
ratio using the neutrino oscillations solutions computed in
ref.~\cite{bgp} with $S_{17}(0) = 19^{+4}_{-2} {\rm
eVb}$~\cite{adelberger} and the standard analysis procedure used
here(cf. figure~\ref{fig:global3}) and in ref.~\cite{bgp}. The right
panel, which is the same as figure~\ref{fig:nccc}a, shows the
allowed values for [NC]/[CC] computed with the same procedure but with
the Junghans et al. value of $S_{17}(0) = (22.3 \pm 0.9) {\rm
eVb}$~\cite{junghans01}.
\label{fig:ncccbeforeafter}}}

The best-fit values for the double ratio for oscillations into
active neutrinos range between $1.8$ and $3.4$ ($1.8$ and $3.5$)
for a $4.5$ MeV ($6.75$ MeV) CC threshold.  The maximum predicted
value for ${\rm {[NC]} / { [CC]} }$ is an enormous $5.3$ for an
extreme LMA solution. The minimum calculated value  for the
double ratio is $1.4$, which is implied by an extreme VAC solution.

Figure~\ref{fig:ncccbeforeafter} compares the ``Before'' and
``After'' predicted values for the neutral current to charged
current double ratio. For the favored three solutions, LMA, LOW,
and VAC, the predicted double ratios are not very sensitive to
the precise value of $S_{17}(0)$ that is adopted. We note however
that in all three cases adopting the Junghans et al. value of
$S_{17}(0)$ causes the values of [NC]/[CC] to become slightly
larger (more distant from the no oscillation value of $1.0$) as a
consequence of the reductions and shifts of the allowed regions.

\subsection{The Charged Current Day-Night Effect}
\label{subsec:ccdaynight}

\FIGURE[!!ht]{
\centerline{\psfig{figure=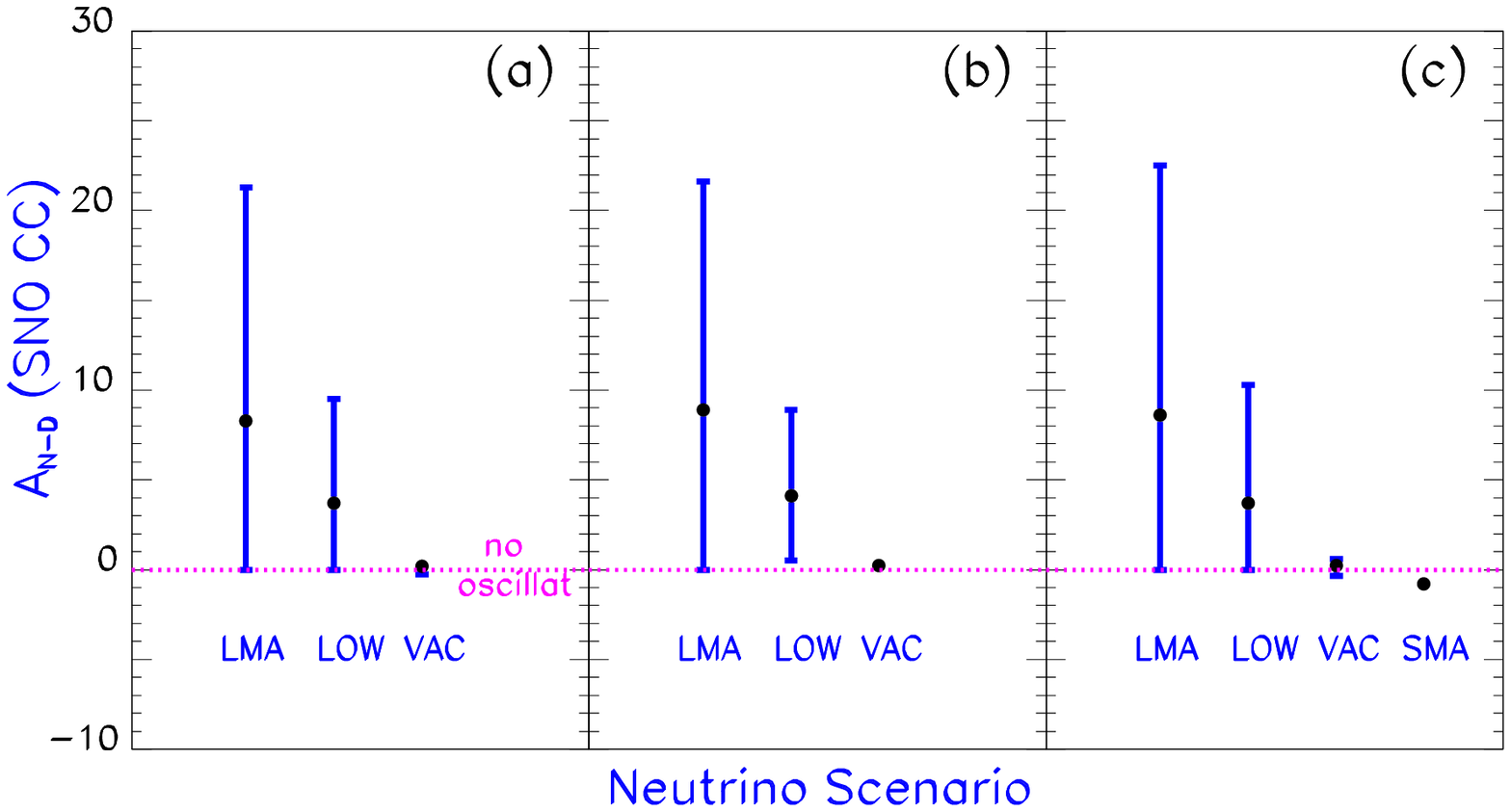,width=4.5in}}
\caption{{\bf The percentage difference between the night and the day
CC rates.} The figure shows the night-day percentage difference for
the charged current rate in SNO, i.e., $100 \times A_{\rm N-D}$
defined in eq.~(\ref{eq:daynightdefn}). Predictions are shown for the
solar neutrino oscillation scenarios allowed at $3\sigma$ and
illustrated in figure~\ref{fig:global3}. The three panels refer to
results for different analysis strategies described in
section~\ref{sec:global}.  Figure~\ref{fig:snodaynight} was
constructed using a recoil electron kinetic energy threshold of $6.75$
MeV for the CC events. Table~\ref{tab:nccc} gives numerical results
for $A_{\rm N-D}$ for kinetic energy thresholds of $4.5$ MeV and
$6.75$ MeV.
\label{fig:snodaynight}}}

\FIGURE[!!hbt]{
\centerline{\psfig{figure=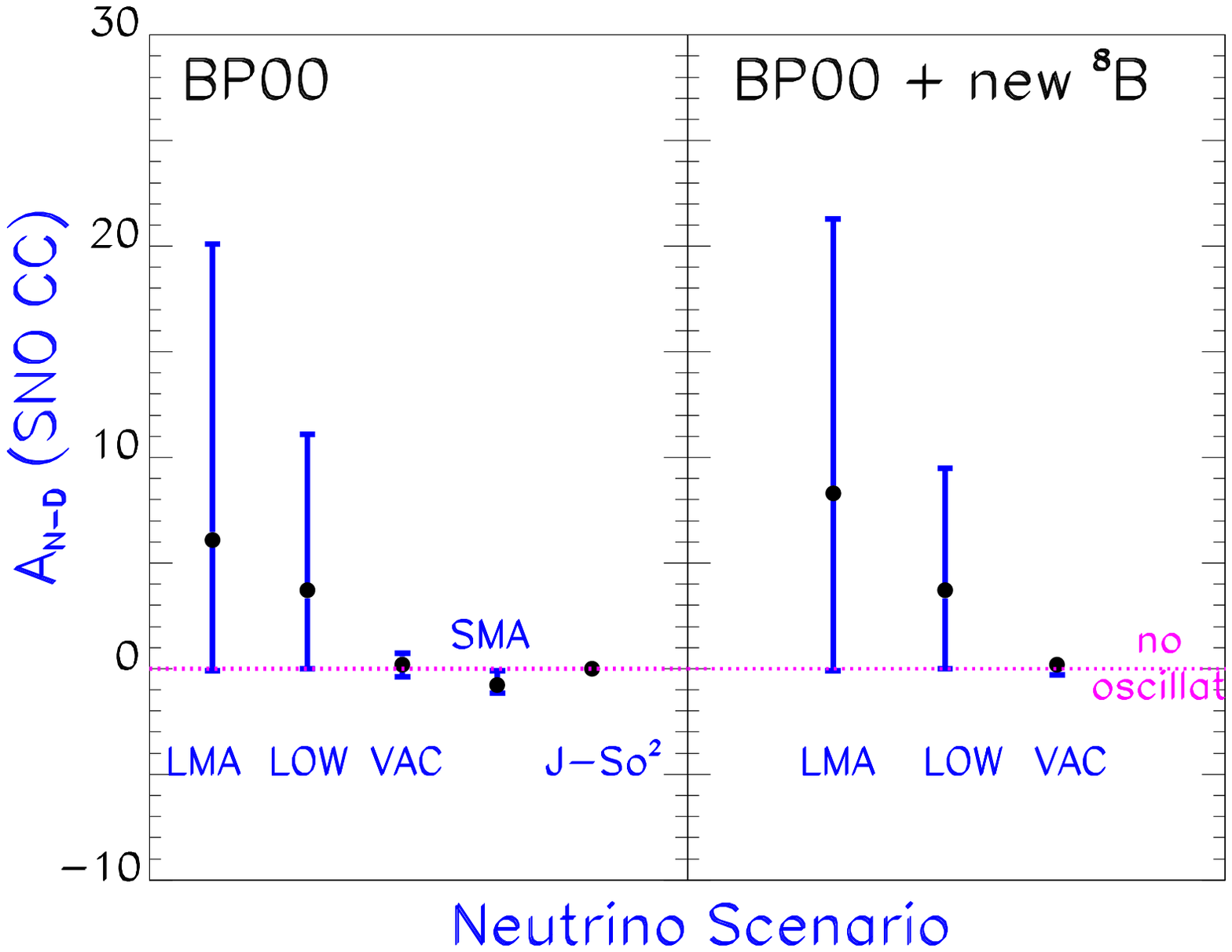,width=4.5in}}
\caption{{\bf Night-Day CC ``Before and After.''} The left panel shows
the $3\sigma$ allowed ranges in percent for the average CC night minus
the day event rates, $A_{\rm N-D}$, as defined in
eq.~(\ref{eq:daynightdefn}); the values in the left panel were
computed using the neutrino oscillations solutions in ref.~\cite{bgp}
obtained with $S_{17}(0) = 19^{+4}_{-2} {\rm eVb}$~\cite{adelberger}
and the standard analysis procedure used
here(cf. figure~\ref{fig:global3}) and in ref.~\cite{bgp}. The right
panel, which is the same as figure~\ref{fig:snodaynight}a, shows the
allowed values for $A_{\rm N-D}$ computed with the same procedure but
with the Junghans et al. value of $S_{17}(0) = (22.3 \pm 0.9) {\rm
eVb}$~\cite{junghans01}.
\label{fig:ccabeforeafter}}}

In this subsection, we discuss the difference between the charged
current event rate observed at night and the charged current event
rated observed during the day. This difference in event rates has been
evaluated previously for a variety of experiments by many authors,
including those listed in
refs~\cite{howwell,tencommand,daynight,krastev01,Gonzalez-Garcia:2001dj}.

We concentrate here on the difference, $A_{\rm N-D}$, between the
nighttime and the daytime CC rates for SNO, averaged over one year. The
definition of $A_{\rm N-D}$ is
\begin{equation}
A_{\rm N-D} ~=~2{\rm  {[Night - Day]} \over {[Night + Day]}}.
\label{eq:daynightdefn}
\end{equation}
The value of $A_{\rm N-D}$ is zero for neutral current detection of
oscillations into active neutrinos. (For neutrino oscillations into
sterile neutrinos, $A_{\rm N-D}$ has
been calculated in
ref.~\cite{tencommand} and shown to be small, less than $1$\% for the
range of neutrino parameters allowed before the SNO
experimental results were available.)


Table~\ref{tab:nccc} and figure~\ref{fig:snodaynight} present the range of
predicted differences between the average rate at night and
the average rate during the day [i.e., $100 \times
A_{\rm N-D}({\rm SNO \; CC})$ of
eq.~(\ref{eq:daynightdefn})].  The calculated predictions in
table~\ref{tab:nccc} are given for a $4.5$ MeV and an $6.75$ MeV CC
electron recoil kinetic energy threshold.

For most of the MSW oscillation solutions, the predicted
day-night differences are only of order a few percent. However,
for the LMA solution, the predicted difference can reach as high
as $20$\% for a $4.5$ MeV threshold ($21$\% for a $6.75$ MeV
threshold). For very special choices of the LOW parameters, $A_{\rm
N-D}({\rm SNO \; CC})$ can be as large as $9$\% or $10$\%. For vacuum
oscillations, there is a small day-night effect that is due to
the dependence of the survival probability upon the earth-sun
distance. This dependence has been calculated in
ref.~\cite{tencommand} and corresponds in all the allowed cases
to $|A_{\rm N-D}| \leq 1$\%.

The $1\sigma$ predicted range for our standard global analysis
strategy (a) is $A_{\rm N-D}\break ({\rm SNO \; CC}) = 8.3^{+5.0}_{-5.6}\;$\%
for a $6.75$ MeV CC threshold and $A_{\rm N-D}({\rm SNO \; CC}) =
7.4^{+4.7}_{-5.1}\;$\% for a $4.5$ MeV threshold. The predictions for
analysis strategies (b) and (c) are very similar.

Figure~\ref{fig:ccabeforeafter} shows the ``Before and After''
comparison between the range of predicted values for $A_{\rm N-D}({\rm
SNO \; CC})$ using the older $S_{17}(0)$ and the most recent and
accurate determination of $S_{17}(0)$. For the three favored
solutions, LMA, LOW, and VAC, the allowed range of $A_{\rm N-D}({\rm
SNO \; CC})$ is only changed slightly, the maximum value is increased
(decreased) a small amount for the LMA (LOW) solution.

\subsection{The $\nu + e$  Scattering Day-Night Effect}
\label{subsec:esdaynight}

We present in section~\ref{subsubsec:esdaynight} the calculated
day-night effects for neutrino-electron scattering in the SNO detector
that are predicted by the current globally-allowed oscillation
solutions. In section~\ref{subsubsec:correlation}, we present and
discuss the predicted correlation, which depends somewhat on the
particular oscillation solution, between the day-night effect in the
CC event rate and the day-night effect in the neutrino-electron
scattering event rate. This correlation constitutes an important
consistency check on the oscillation solution~\cite{bkscorr}.

\subsubsection{Predicted values for $A_{\rm N-D}({\rm SNO
~ ES})$}
\label{subsubsec:esdaynight}

\FIGURE[!!ht]{
\centerline{\psfig{figure=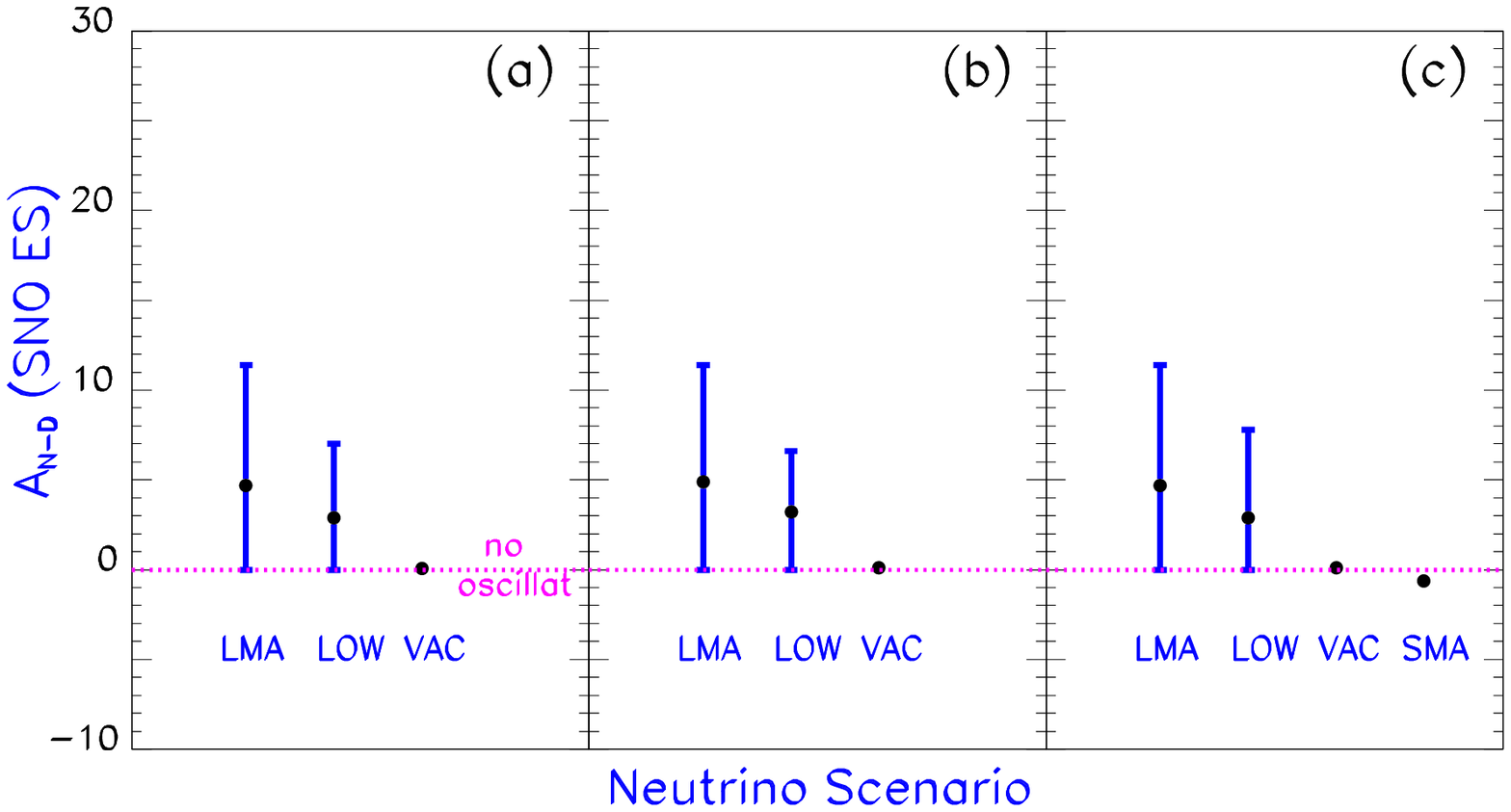,width=4.5in}}
\caption{{\bf The percentage difference between the night and the day
$\nu + e$ scattering rates.} The figure shows the night-day percentage
difference, i.e., $100 \times A_{\rm N-D}({\rm SNO \; ES})$ defined in
eq.~(\ref{eq:daynightdefn}). Predictions are shown for the solar
neutrino oscillation scenarios allowed at $3\sigma$ and illustrated in
figure~\ref{fig:global3}. The three panels refer to results for
different analysis strategies described in section~\ref{sec:global}.
Figure~\ref{fig:snodaynight} was constructed using a recoil electron
kinetic energy threshold of $6.75$ MeV. Table~\ref{tab:esdaynight}
gives numerical results for $A_{\rm N-D}({\rm SNO \; ES})$ for kinetic energy
thresholds of $4.5$ MeV and $6.75$ MeV.
\label{fig:snoesdaynight}}}

Figure~\ref{fig:snoesdaynight} and table~\ref{tab:esdaynight} show the
predicted (percentage) night-day differences for $\nu + e$ scattering
in the SNO detector.  The calculated values for $A_{\rm N-D}({\rm SNO
~ ES})$ were obtained using the $3\sigma$ boundaries of the global
oscillation solutions shown in figure~\ref{fig:global3}.

\TABLE[!htb]{ \centering
\caption{\label{tab:esdaynight} {\bf The predicted $\nu + e$ Day-Night
Asymmetry in SNO.}  The table presents $A_{\rm N-D}({\rm SNO ~ ES})$ in
percent.  The results are tabulated for different neutrino oscillation
scenarios determined by our standard strategy and for two different
thresholds of the recoil electron kinetic energy, $4.5$ MeV (columns
two through four) and $6.75$ MeV (columns five through seven). The
ranges are obtained for the 3$\sigma$ regions for
figure~\ref{fig:global3}(a).}
\begin{tabular}{ccccccc}
\noalign{\bigskip}
\hline
\noalign{\smallskip}
 &\multicolumn{1}{c}{b.f.}&max&min&\multicolumn{1}{c}{b.f.}&max&min\\
Scenario & 4.5 MeV& 4.5 MeV& 4.5 MeV& 6.75 MeV& 6.75 MeV& 6.75 MeV\\
\noalign{\smallskip}
\hline
\noalign{\smallskip}
&A$_{\rm N-D}$&A$_{\rm N-D}$&A$_{\rm N-D}$&A$_{\rm N-D}$&A$_{\rm N-D}$&A$_{\rm N-D}$\\
\noalign{\smallskip}
\hline
\noalign{\smallskip}
 LMA& 4.1  & 10.1& 0.0   & 4.7   &  11.4   & 0.0  \\
 LOW& 3.3  & 7.8& 0.0   & 2.9   &  7.1    & 0.0  \\
 VAC& 0.0  & 0.1 & -0.1 &0.1 &  0.3  & -0.2  \\
\noalign{\smallskip}
\hline
\end{tabular}
}

The values of $A_{\rm N-D}({\rm ~ ES})$ that are predicted for
the Super-Kamiokande and the SNO locations are very similar.  For the
Super-Kamiokande location, the LMA solution, and a kinetic energy
threshold of $4.5$ MeV ( corresponding to the recent Super-Kamiokande
electron recoil energy threshold~\cite{superk}), the predicted range
of $A_{\rm N-D}({\rm SK \; ES})$ is between $0.0$\% and $10.4$\%, with a
best-fit value of $4.3$\% . Similarly for the LOW solution, the
predicted range is between $0.0$\% and $7.3$\%, with a best-fit value
of $3.1$\%. The predicted values of $A_{\rm N-D}({\rm ~ ES})$ at both
the SNO and the Super-Kamiokande locations are negligibly small for
the VAC solution. Comparing with table~\ref{tab:esdaynight}, we see
that the range of predicted values of $A_{\rm N-D}({\rm ~ ES})$
differs, between the SNO and Super-Kamiokande locations, by at most
$7$\% of the predicted range.

The predictions for SNO can be compared with the measured result
reported by the Super-Kamiokande collaboration~\cite{superk}

\begin{equation}
A_{\rm N-D}({\rm SK \;  ES}) ~=~0.033 \pm 0.022 ({\rm
stat.})^{+0.013}_{-0.012} ({\rm sys.}).
\label{eq:skescdaynight}
\end{equation}
Not surprisingly, the best-fit predictions for the $\nu + e$ day-night
 scattering difference in the SNO detector implied by the global solutions
 shown in figure~\ref{fig:global3}a are all rather close to the
 best-estimate value obtained by Super-Kamiokande. The SNO predictions
 are also bounded by the $3\sigma$ Super-Kamiokande limits, i.e., the
 predicted SNO values lie between $0$\% and
 $11$\%. Figure~\ref{fig:snoesdaynight} shows that the LMA solution
 allows the largest values for $A_{\rm N-D}({\rm SNO ~
 ES})$ and the VAC solution predicts no day-night difference in $\nu +
 e$ event rates.

The $1\sigma$ allowed range for $A_{\rm N-D}({\rm SNO ~ ES})$ is
$4.7^{+2.7}_{-3.1}\;$\% for $6.75$ MeV and $4.1^{+2.4}_{-2.8}\;$\% for
a $4.5$ MeV threshold, both for analysis strategy (a). The results for
strategies (b) and (c) are very similar.

The ``Before and After'' predictions for $A_{\rm N-D}({\rm SNO ~ ES})$
 are essentially the same, with only minor differences. That is to
 say, that the results do not depend significantly on the choice
 between the previously standard value of
 $S_{17}(0)$~\cite{adelberger} and the most recent and precise value
 of $S_{17}(0)$~\cite{junghans01}.

\subsubsection{Correlation between $A_{\rm N-D}({\rm SNO
~ CC})$ and $A_{\rm N-D}({\rm SNO
~ ES})$ }
\label{subsubsec:correlation}

Figure~\ref{fig:corrdaynight} shows the correlation between the
predicted values for the day-night effect of the CC rate, $A_{\rm
N-D}({\rm SNO ~ CC})$, and the day-night effect of the
neutrino-electron scattering rate, $A_{\rm N-D}({\rm SNO ~ ES})$. The
two day-night effects are essentially proportional to each other. The
essential features of the correlation shown in
figure~\ref{fig:corrdaynight} can be derived analytically, as we show
in the following discussion.

The survival probability of electron neutrinos can be written as
\begin{equation}
P_{ee} = P_D -  (1 - 2P_c) \cdot \cos2\theta_S \cdot f_{\rm reg},
\label{eq:probtot}
\end{equation}
where
\begin{equation}
P_D  = \frac{1}{2} + \frac{1}{2}(1 - 2P_c) \cdot \cos2\theta_S \cdot
\cos2\theta
\label{eq:Pday}
\end{equation}
is the survival probability in the absence of the Earth--matter effect,
i.e.,  during the day.
Here $\theta_S$ is the matter mixing angle at the production point
inside the Sun ($\theta_S=\frac{\pi}{2}$ when the production point occurs at
much higher densities than the resonant point)
, and $P_c$ is the level crossing probability which
describes the non-adiabaticity of the conversion inside the Sun.
The Earth regeneration factor, $f_{\rm reg}$, is defined
as:
\begin{equation}
f_{\rm reg} \equiv P_{2e}  - \sin^2 \theta~,
\label{eq:regen}
\end{equation}
where  $P_{2e}$ is the probability of the
$\nu_{2} \rightarrow \nu_e$ conversion inside the Earth.
In the absence of the Earth--matter effect we have
$P_{2e} = \sin^2 \theta$, so that $f_{\rm reg} = 0$.

For both LMA and most of LOW region the adiabaticity condition is
satisfied and we can take $P_c\simeq0$.
Furthermore in the LOW region and in the lower part of the LMA region
one has $ \cos2\theta_S=-1$. With these approximations the
probabilities take the form
\begin{eqnarray}
P_{ee}&=& \sin^2 \theta+ f_{\rm reg}\;,\\
P_D&=&\sin^2 \theta \;,
\end{eqnarray}
And the day-night asymmetry for CC events can be written as
\begin{equation}
A_{\rm N-D}({\rm SNO\; CC}) = 2\frac{\langle P_{ee}\rangle- \langle
P_{D}\rangle} {\langle P_{ee}\rangle} = 2 \frac{\langle f_{\rm
reg}\rangle }{\sin^2 \theta+ \langle f_{\rm reg}\rangle}
\label{eq:adncc}
\end{equation}
where we have denoted by $\langle \rangle$ the averaging over the
neutrino spectrum, the interaction cross section, and the energy
resolution.

\FIGURE[!ht]{
\centerline{\psfig{figure=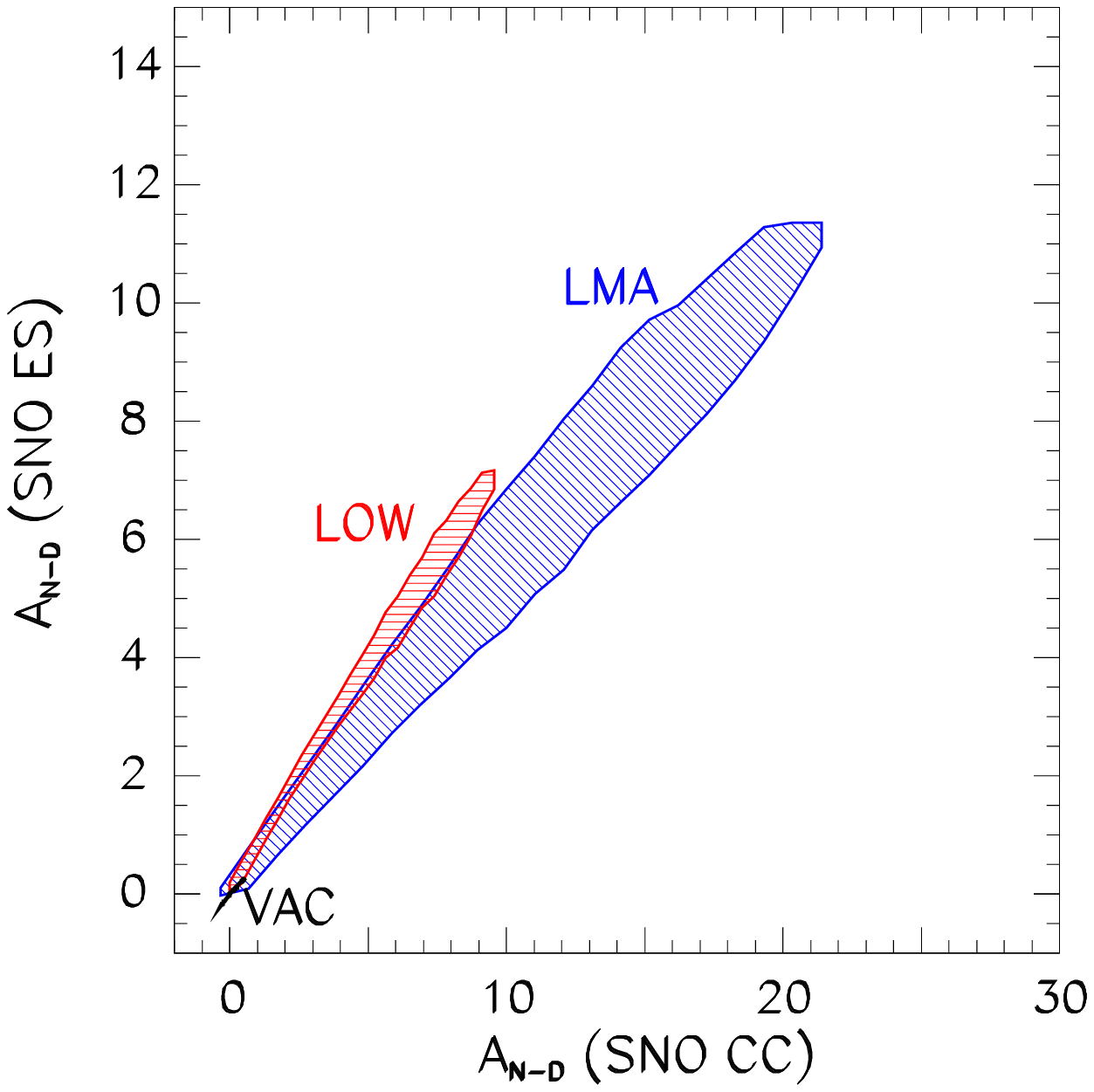,width=4.5in}}
\caption{{\bf The correlation between the day-night effects for $\nu +
e$ scattering and for CC events.} The day-night effects shown in
figure~\ref{fig:snodaynight} and figure~\ref{fig:snoesdaynight} are
correlated for the standard analysis strategy that resulted in
figure~\ref{fig:global3}a.  The predictions for the VAC solutions are
essentially a point at the position (0,0).
\label{fig:corrdaynight}}}

The [ES] event rate is [ES] $ \sim \langle P_{ee} +r (1- P_{ee})\rangle$
where $r\equiv \sigma_{\mu}/\sigma_{e}\simeq 0.15$ is the ratio of the
the $\nu_e - e$ and $\nu_{\mu} - e$ elastic scattering cross-sections.
Hence the day-night asymmetry for neutrino-electron scattering events
can be written as
\begin{equation}
A_{\rm N-D}({\rm SNO\; ES}) =
2 \frac{(1-r) \langle  f_{\rm reg}\rangle }
{\sin^2 \theta+ r\cos^2\theta+(1-r) \langle f_{\rm reg}\rangle}\;.
\label{eq:adnes}
\end{equation}
From eq.~(\ref{eq:adncc}) and eq.~(\ref{eq:adnes}), we see that the
slope in the $A_{\rm N-D}({\rm SNO\; CC})$ versus $A_{\rm N-D}({\rm
SNO\; ES})$ plot is given by
\begin{equation}
\frac{A_{\rm N-D}({\rm SNO\; ES})}{A_{\rm N-D}({\rm SNO\; CC})}=
(1-r)\frac{\sin^2 \theta+ \langle f_{\rm reg}\rangle}
{\sin^2 \theta+ r\cos^2\theta+(1-r) \langle f_{\rm reg}\rangle}
\rightarrow (1-r)\frac{\sin^2 \theta}{\sin^2 \theta+ r\cos^2\theta}
\label{eq:ratiodaynight}
\end{equation}
where the last implication applies in the region of small
asymmetries. Using this relation and the ranges of mixing angles in
the LOW and LMA solution we can reproduce the slope of the regions.
From eq.~(\ref{eq:ratiodaynight}), we see that the ratio of the
asymmetries increases with the mixing angle and is larger for mixing
angles on the dark side ($\theta > \pi/4$).  Therefore
eq.~(\ref{eq:ratiodaynight}) explains why the LOW solution has a
larger slope in figure~\ref{fig:corrdaynight} than the LMA solution.

\subsection{CC recoil energy spectrum: first and second moments}
\label{subsec:moments}

\FIGURE[!hb]{
\centerline{\psfig{figure=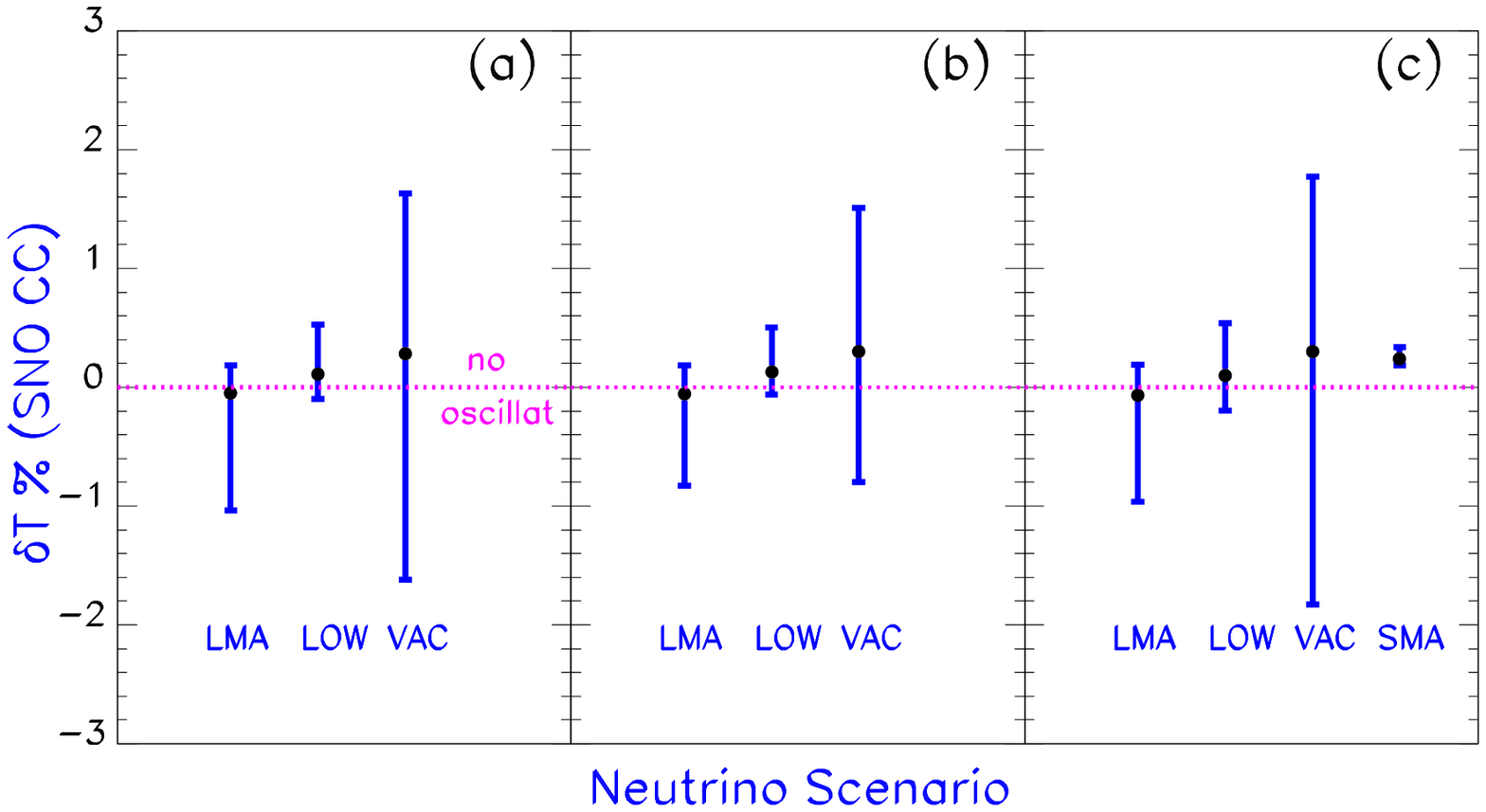,width=4.5in}}
\caption{{\bf The first moment: the fractional shift, \boldmath$\delta T$, in
percent for the average electron recoil energy.}  The
figure shows the fractional change  in percent for the average
electron recoil
energy, $\langle T\rangle$,  relative to what is
produced by an undistorted ${\rm ^8B}$ neutrino energy spectrum. The predictions are shown at $3\sigma$ C.L.
for the three different analysis strategies for neutrino oscillation
solutions that are illustrated in figure~\ref{fig:global3}. The
calculations were performed assuming a $6.75$ MeV recoil electron
kinetic energy threshold.
\label{fig:firstmoment}}}

The solar influence on the shape of the ${\rm ^8B}$ neutrino energy
spectrum is only about one part in
$10^5$~\cite{shapeindependence}. Therefore, in the absence of neutrino
oscillations or other new physics, the energy spectrum of solar
neutrinos incident on terrestrial detectors should be the same as the
energy spectrum of ${\rm ^8B}$ neutrinos produced in the
laboratory. The incident solar neutrino energy spectrum can be studied
well by observing the recoil energy spectrum of electrons produced by
neutrino absorption (CC) reactions on deuterium, as shown in
eq.~(\ref{eq:cc}). In the CC reaction, nearly all of the available kinetic
energy is taken by  the recoiling electron since the  other
particles in the final state are (massive) baryons.

The Super-Kamiokande experiment has already provided beautiful data on
the recoil energy spectrum from neutrino electron
scattering~\cite{superk}. In recent publications by the
Super-Kamiokande collaboration, these data are divided into $19$ separate
energy bins. A number of authors have questioned whether using so many
energy bins in a $\chi^2$ analysis of all the available data gives too
much weight to the measurement of the spectral parameters(see, e. g.,
ref.~\cite{creminelli} for a particularly insightful discussion of the
statistical questions).

\FIGURE[!t]{
\centerline{\psfig{figure=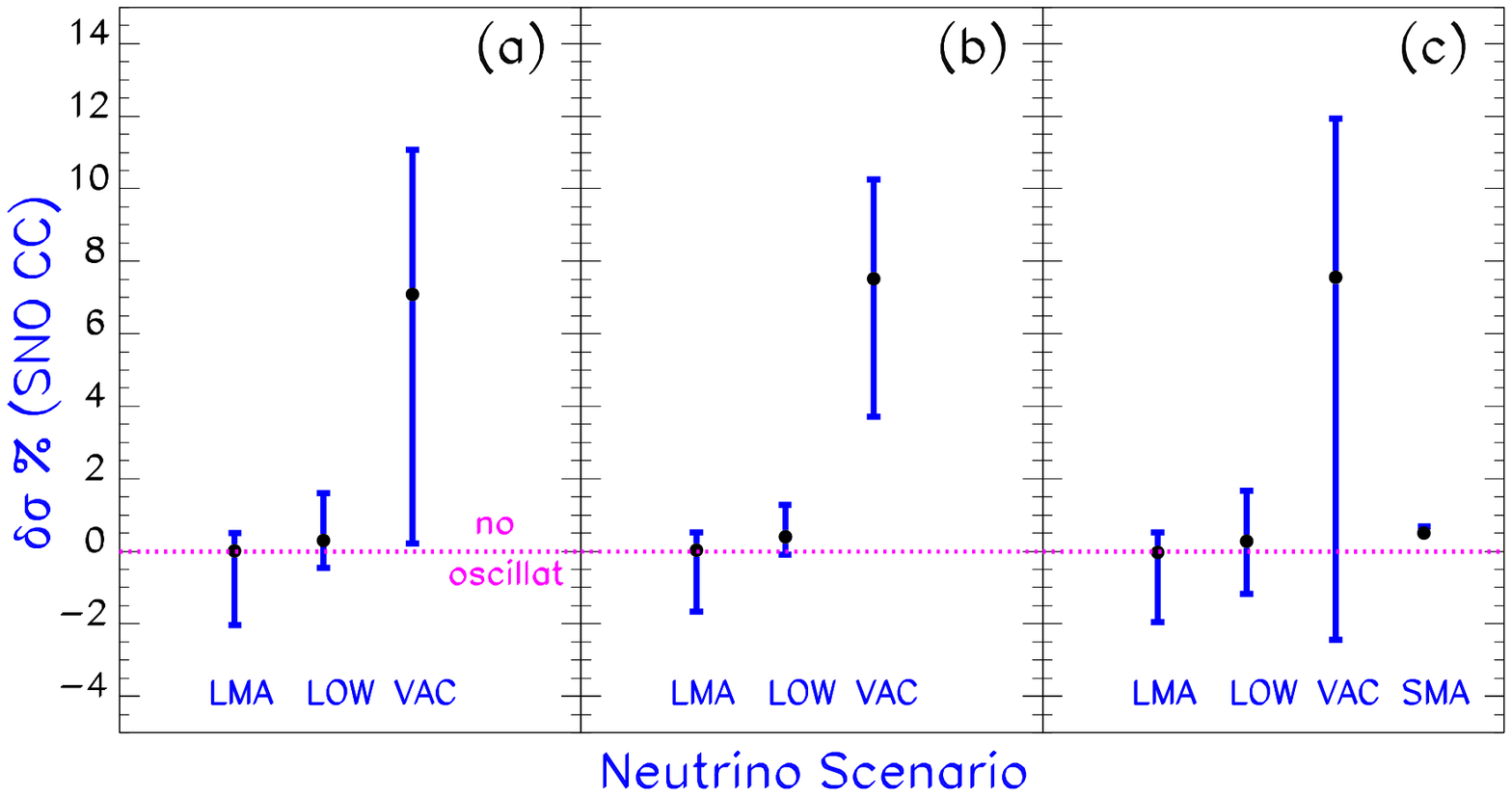,width=4.5in}}
\caption{{\bf The second moment: the fractional shift, \boldmath$\delta
\sigma$, in percent for the standard deviation of the electron recoil
energy spectrum.} The figure shows
the fractional change in the standard deviation of the electron recoil
energy spectrum relative to what is expected for an undistorted ${\rm ^8B}$
neutrino energy spectrum. The predictions are shown at $3\sigma$ C.L.
for the three different analysis strategies for neutrino oscillation
solutions that are illustrated in figure~\ref{fig:global3}. The
calculations were performed assuming a $6.75$ MeV recoil electron
kinetic energy threshold.
\label{fig:secondmoment}}}

In this subsection, we consider an alternative method of analysis of
the SNO recoil energy spectrum in terms of small number of the lowest
order moments of the distribution, a method that has been discussed in
connection with solar neutrino experiments in
refs.~\cite{tencommand,bl,bkl97,earlymoments}. We follow the notation
and analysis of refs.~\cite{tencommand,bkl97}.

Figure~\ref{fig:firstmoment} and figure~\ref{fig:secondmoment} show
the computed first and second moments of the recoil electron energy
spectrum from the CC reaction, eq.~(\ref{eq:cc}), for the three
different analysis strategies discussed in section~\ref{sec:global}
and illustrated in figure~\ref{fig:global3}.  These moments may well
represent the essential physical content of the electron energy spectrum
(cf.  refs.~\cite{tencommand,bkl97}) and could represent the results
of the spectral measurements in a global oscillation analysis.

For both the first and second moments, the VAC solution predicts
the largest deviation from the undistorted spectrum. Nevertheless,
the range of predicted values is only about $\pm
2$\% for the first moment and $\pm 7$\% for the second moment.
The non-statistical uncertainties  in measuring the first and second
moments in SNO have been estimated, prior to the operation of the
experiment,  in ref.~\cite{tencommand} and are,
respectively, about $1$\% and $2$\%.

Assuming the approximate validity of the prior estimate of the
measuring uncertainties given in ref.~\cite{tencommand},
it seems very unlikely that SNO will determine a $3\sigma$ deviation from the
undistorted spectrum if either the favored LMA or LOW oscillation solutions
shown in figure~\ref{fig:global3} is correct. Although this
is a negative prediction, it is nevertheless an important
prediction. Even for the VAC solution, it will be very difficult to
measure a significant deviation from an undistorted spectrum. The best
opportunity to see a deviation will be in the second moment.

\subsection{SNO: What if we are lucky?}
\label{subsec:lucky}

\FIGURE[!b]{
\centerline{\psfig{figure=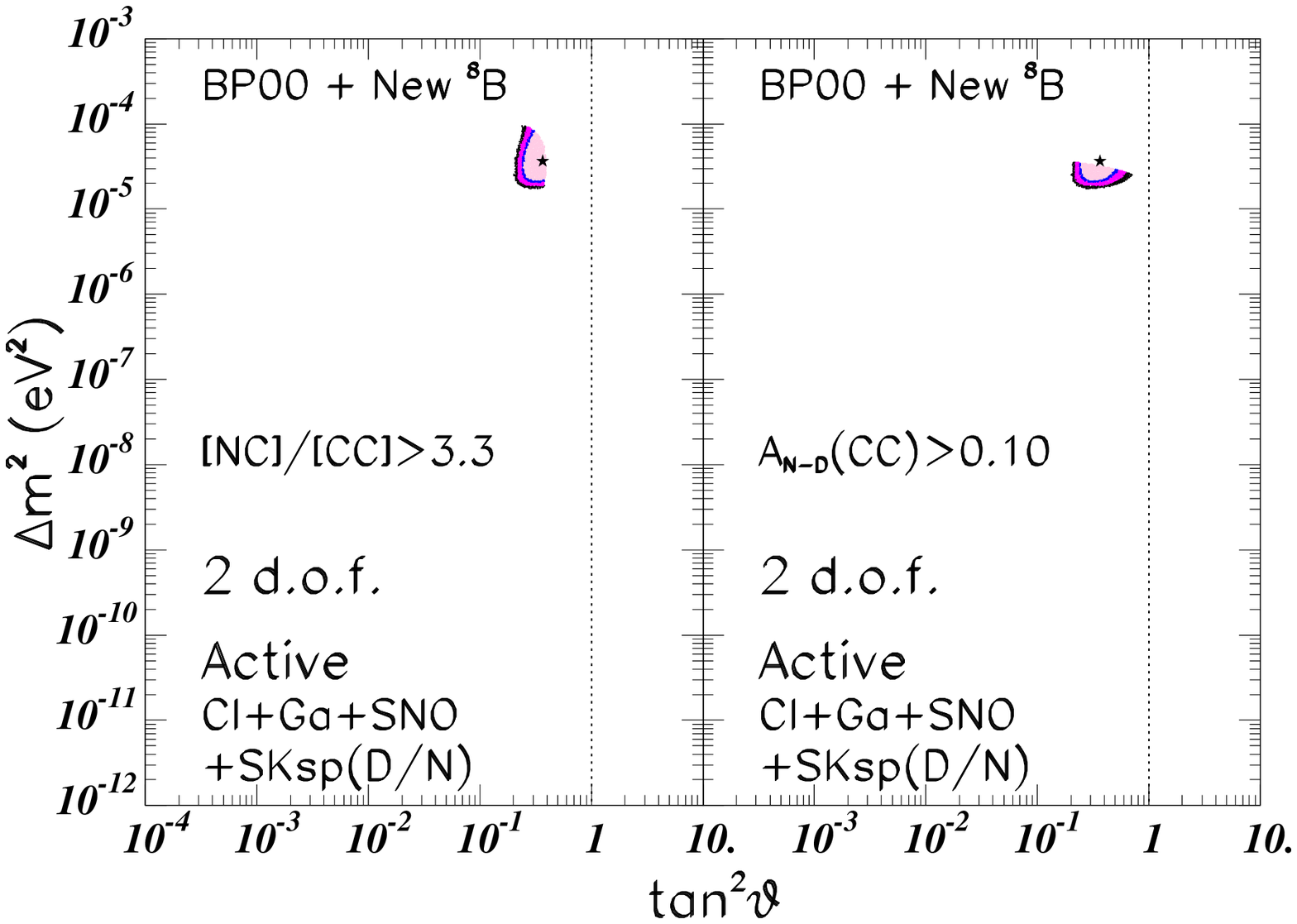,width=4.5in}}
\caption{{\bf If nature is kind.} The left panel shows the currently
allowed regions for the neutrino oscillation parameters that predict a
neutral current to charged current double ratio, [NC]/[CC], that is $>
3.3$ and the right panel shows the currently allowed region that
predicts a CC night-day difference, $A_{\rm N-D}({\rm SNO \; CC})$,
that is $ > 0.1$. The calculations were performed assuming a $6.75$
MeV recoil electron kinetic energy threshold. The regions shown in
figure~\ref{fig:lucky} will change after measurements become available
on either (or both) [NC]/[CC] or $A_{\rm N-D}({\rm SNO \; CC})$ and a
new global solution is calculated.  \label{fig:lucky}}}

Figure~\ref{fig:nccc}, figure~\ref{fig:snodaynight}, and
table~\ref{tab:nccc} show that if Nature has arranged things
favorably, then only a rather small range of neutrino oscillation
parameters can correctly predict the results.  If SNO measures either
a large value for the neutral current to charged current ratio, ${\rm
[NC]/[CC]} > 3.3$, or a large value for the night-day difference,
$A_{\rm N-D}({\rm SNO \; CC}) > 0.1$, then the $\Delta m^2$ and
$\tan^2 \theta$ will be rather well determined.

\FIGURE[!t]{
\centerline{\psfig{figure=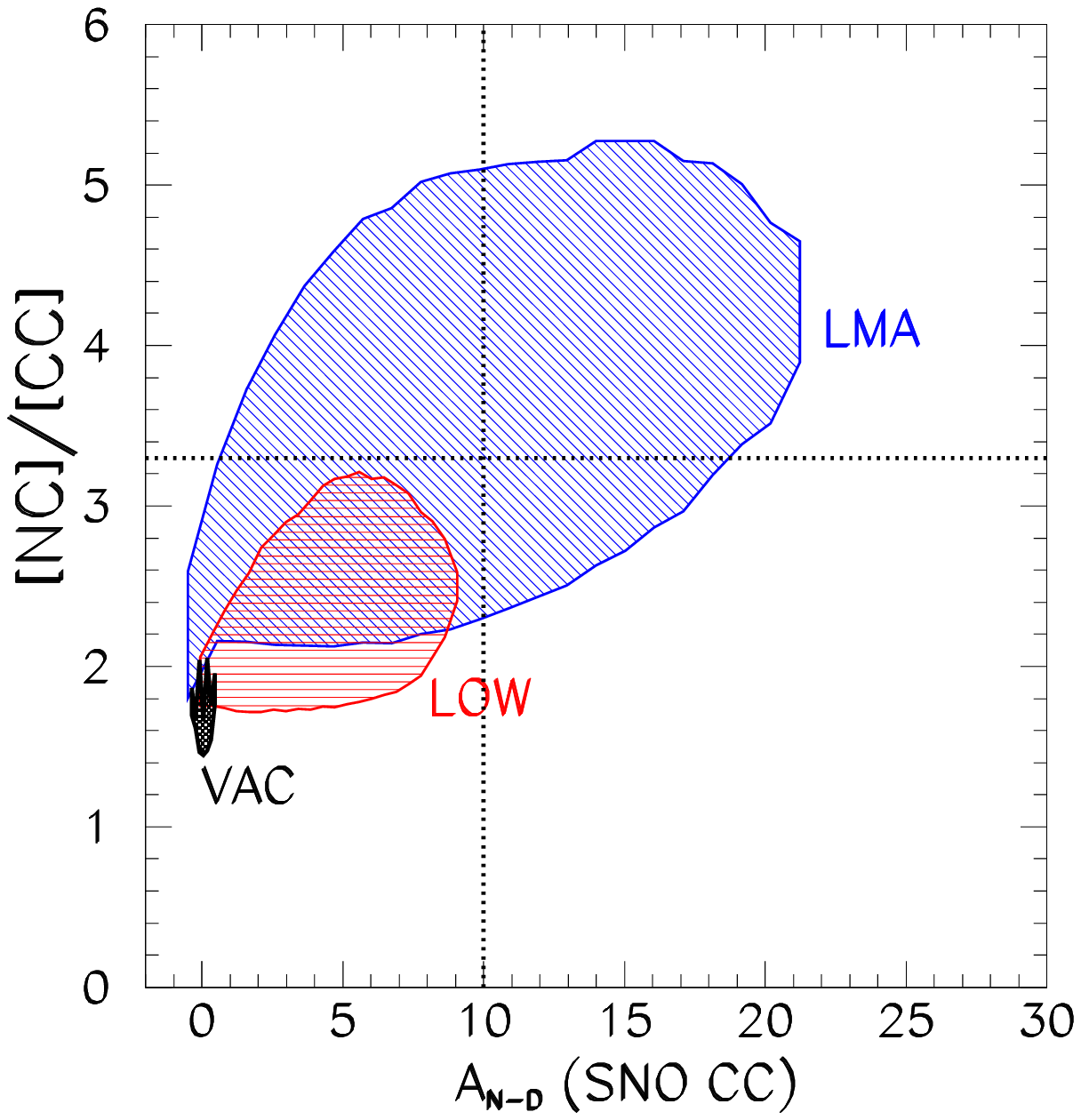,width=4.5in}}
\caption{{\bf The correlation between the neutral current to charged
current double ratio, [NC]/[CC], and the charged current day-night
effect $A_{\rm N-D}({\rm SNO \; CC})$.} The figure shows the allowed
values at $3\sigma$ of [NC]/[CC] versus $100 \times A_{\rm N-D}({\rm
SNO \; CC})$, using the analysis strategy (a) and a CC threshold of
$6.75$ MeV.  The dotted lines show the limiting values $A_{\rm
N-D}({\rm SNO \; CC}) = 10$\% and [NC]/[CC] $= 3.3$ used in
figure~\ref{fig:lucky}.\label{fig:corrncccdn}}}

Figure~\ref{fig:lucky} shows the small range of neutrino parameters
that, given the currently allowed regions defined by the currently
available experimental data, can correctly predict the results if SNO
shows that one of these inequalities is correct. In that fortunate
situation, SNO will narrow down the allowed neutrino parameters to a
small region in neutrino parameter space. With the currently available
data, the boundaries of the region corresponding to ${\rm [NC]/[CC]} >
3.3$ are for our standard analysis strategy, (a): $0.2 < \tan^2 \theta
< 0.4$ and $1.8\times 10^{-5}~{\rm eV^2} < \Delta m^2 < 1.0 \times
10^{-4}~{\rm eV^2} $.  For $A_{\rm N-D} > 0.1$, the corresponding
boundaries are $0.2 < \tan^2 \theta < 0.7$ and $1.8\times 10^{-5}~{\rm
eV^2} < \Delta m^2 < 3.5 \times 10^{-5}~{\rm eV^2} $.  The only change
in the above numbers if we use the free ${\rm ^8B}$ strategy
(cf. figure~\ref{fig:global3}c), rather than our standard analysis
strategy is that the $A_{\rm N-D}$ range is slightly affected and the
upper limit for $\Delta m^2$ decreases to $3.3 \times 10^{-5}~{\rm
eV^2} $.

The regions corresponding to ${\rm [NC]/[CC]} > 3.3$ and $A_{\rm
N-D}({\rm SNO \; CC}) > 0.1$ depend upon the data available when a
global analysis is made. The regions shown in figure~\ref{fig:lucky}
will change after measurements become available on either (or both)
[NC]/[CC] or $A_{\rm N-D}({\rm SNO \; CC})$ and a new global solution
is carried out.

If bi-maximal mixing~\cite{bimaximal} is correct, one would not expect
either ${\rm [NC]/[CC]} > 3.3$ (this would be a $10\sigma$ deviation
from predictions based upon existing experimental data,
cf. figure~\ref{fig:global3}) or $A_{\rm N-D}({\rm SNO \; CC}) > 0.1$
(a $3.8\sigma$ [$3.1\sigma$]deviation from the predictions of
currently allowed solutions for analysis (a) [(c)]).

Figure~\ref{fig:corrncccdn} shows the correlation between the
currently allowed values of [NC]/[CC] and $A_{\rm N-D}({\rm SNO \;
CC})$. There is a general tendency that larger values of $A_{\rm
N-D}\break({\rm SNO \; CC})$ are associated with larger values of [NC]/[CC],
although the correlation is rather broad. For a given value of
[NC]/[CC], a large range of values of$A_{\rm N-D}({\rm SNO \; CC})$ is
currently allowed and vice-versa.  Figure~\ref{fig:corrncccdn} also
shows clearly that for smaller values of [NC]/[CC] and $A_{\rm
N-D}({\rm SNO \; CC})$ discriminating among different oscillation scenarios
will not be easy. For example, for the LMA solution, [NC]/[CC] $ <2.1$
implies $A_{\rm N-D}({\rm SNO \; CC}) < 1$\% .  On the other hand, there is a
range of neutrino parameters characterizing the LOW solution for which
one expects ${\rm [NC]/[CC]} < 2.1$ while there exists a small but non
vanishing day-night asymmetry $1$\% $ < A_{\rm N-D}({\rm SNO \; CC}) < 8$\%.

\section{Predictions for ${\rm\bf ^7Be}$ rate and day-night effect}
\label{sec:be7}

In this section, we present the predictions of the globally
allowed solutions for the neutrino electron scattering rate and
the day-night effect in the BOREXINO~\cite{borexino} and
KamLAND~\cite{kamland} detectors. For specificity, we present
results for the location of BOREXINO, but comment in the text on
the differences in the predictions for KamLAND and BOREXINO. We
assume the same energy window for recoil electrons for both
detectors, so the only cause for the differences in the
predictions is the different latitude of the detector locations.
In all cases, the differences are smaller than the expected
measurement uncertainties.

We consider only electrons with recoil kinetic energies in the
experimentally preferred range of $0.25 ~{\rm MeV} < T_{\rm e} < 0.8
~{\rm MeV}$ (cf. ref.~\cite{borexino}). For simplicity in presentation
and following the general practice in the literature, we refer to
these events as ${\rm ^7Be}$ neutrino events. However, we include in
our discussion neutrinos from all solar neutrino sources that produce
recoil electrons in the indicated energy range. We make use of the
predicted neutrino fluxes and uncertainties given by the BP00 standard
solar model~\cite{bp2000}. (The ${\rm ^8B}$ neutrinos do not
contribute significantly in this energy range.) The ${\rm ^{13}N}$ and
${\rm ^{15}O}$ neutrinos are, after ${\rm ^7Be}$, the next most
important neutrino sources for producing electrons with recoil kinetic
energies between $0.25 ~{\rm MeV}$ and $0.8 ~{\rm MeV}$.

In order to indicate the insensitivity of our conclusions to the
poorly known CNO neutrino fluxes, table~\ref{tab:be7} gives results
for the scattering rate and the day-night difference both with and
without including CNO and $pep$ neutrinos. The predicted values are
insensitive to the non-${\rm ^7Be}$ neutrinos, if their fluxes are
comparable to the BP00 standard solar model fluxes.

We begin by discussing the $\nu - e$ scattering rate in
section~\ref{subsec:be7rate} and then discuss the day-night effect in
section~\ref{subsec:be7daynight}. The hypothetical accuracy with which
KamLAND and BOREXINO may, together with previous solar neutrino
experiments, determine neutrino oscillation parameters has been
discussed in ref.~\cite{vissani}.

\subsection{${\rm\bf  ^7Be}$ \boldmath$\nu - e$ scattering rate}
\label{subsec:be7rate}

Figure~\ref{fig:be7rate} shows, for the currently allowed oscillation
 solutions, the expected $3\sigma$ range in BOREXINO of the reduced
 neutrino-electron scattering rate, $[{\rm ^7Be}]$,

\TABLE[!t]{
\centering
\caption{\label{tab:be7}{\bf ${\rm\bf ^7Be}$: Neutrino-electron
scattering rate and Day-Night Asymmetry.}  The table presents the
reduced neutrino-electron scattering rate, [${\rm ^7Be}$]
(eq.~\ref{eq:defnratio}), and the night-day difference, $A_{\rm N-D}$
(see, eq.~\ref{eq:daynightdefn} but the result is presented here in
\%), for recoil electrons with kinetic energies in the range $0.25
~{\rm MeV} < T_{\rm e} < 0.8 ~{\rm MeV}$.  The numbers in parentheses
were calculated assuming that only ${\rm ^7Be}$ neutrinos contribute
while the entries not in parentheses were obtained assuming the
correctness of the BP00 standard solar model neutrino fluxes.  The
results are given for the three neutrino oscillation scenarios
described in figure~\ref{fig:global3}a.  The ranges shown in the table
correspond to the 3$\sigma$ allowed regions for
figure~\ref{fig:global3}a.}
\begin{tabular}{cccc|ccc}
\hline
\noalign{\smallskip}
 &\multicolumn{1}{c}{b.f.}&max&min&\multicolumn{1}{c}{b.f.}&max&min\\
Scenario &[R($^7$Be)]&[R($^7$Be)]&[R($^7$Be)]
&A$_{\rm N-D}$&A$_{\rm N-D}$&A$_{\rm N-D}$\\
\noalign{\smallskip}
\hline
\noalign{\smallskip}
 LMA& 0.65(0.65)   & 0.76(0.76) &  0.58(0.58) & 0.0(0.0) & 0.1(0.1) &0.0(0.0) \\
 LOW& 0.62(0.62)   & 0.74(0.74) &  0.54(0.54) & 27(27) & 42(42)& 0.0(0.0) \\
 VAC& 0.63(0.62)   & 0.79(0.78) &  0.53(0.51) & -3.5(-3.8) & -1.1(-1.3) & -4.8(-5.3)\\
\noalign{\smallskip}
\hline
\end{tabular}
}

\begin{equation}
[{\rm ^7Be}] ~\equiv~ \frac{\displaystyle
{\sum_i
\left[
\int\phi_i(\nu_e)\sigma_i(E,\nu_e) +
\int\phi_i(\nu_x)\sigma_i(E,\nu_x)
\right]}}
{\displaystyle \sum_i
{\rm
(Standard~Model~Value)_i}},~~0.25 ~{\rm MeV} < T_{\rm e} < 0.8 ~{\rm
MeV}.
\label{eq:defnratio}
\end{equation}
The neutrino-electron scattering rate, $[{\rm ^7Be}]$, is defined in
eq.~(\ref{eq:defnratio}) as the ratio of the number of recoil
electrons with kinetic energies in the range $0.25 ~{\rm MeV} < T_{\rm
e} < 0.8 ~{\rm MeV}$ divided by the number that is expected if the
BP00 model is correct and there are no neutrino oscillations. The
summation over the different neutrino sources $i$ includes all solar
neutrino fluxes, although the $[{\rm ^7Be}]$ neutrinos are expected to
dominate according to the BP00 solar model. The $\mu$ and $\tau$
neutrinos are denoted by $\nu_x$ in eq.~(\ref{eq:defnratio}).

Figure~\ref{fig:be7rate} shows that all three of the analysis
strategies described in section~\ref{sec:global} yield rather similar
predictions for the scattering rate.

\FIGURE[!ht]{
\centerline{\psfig{figure=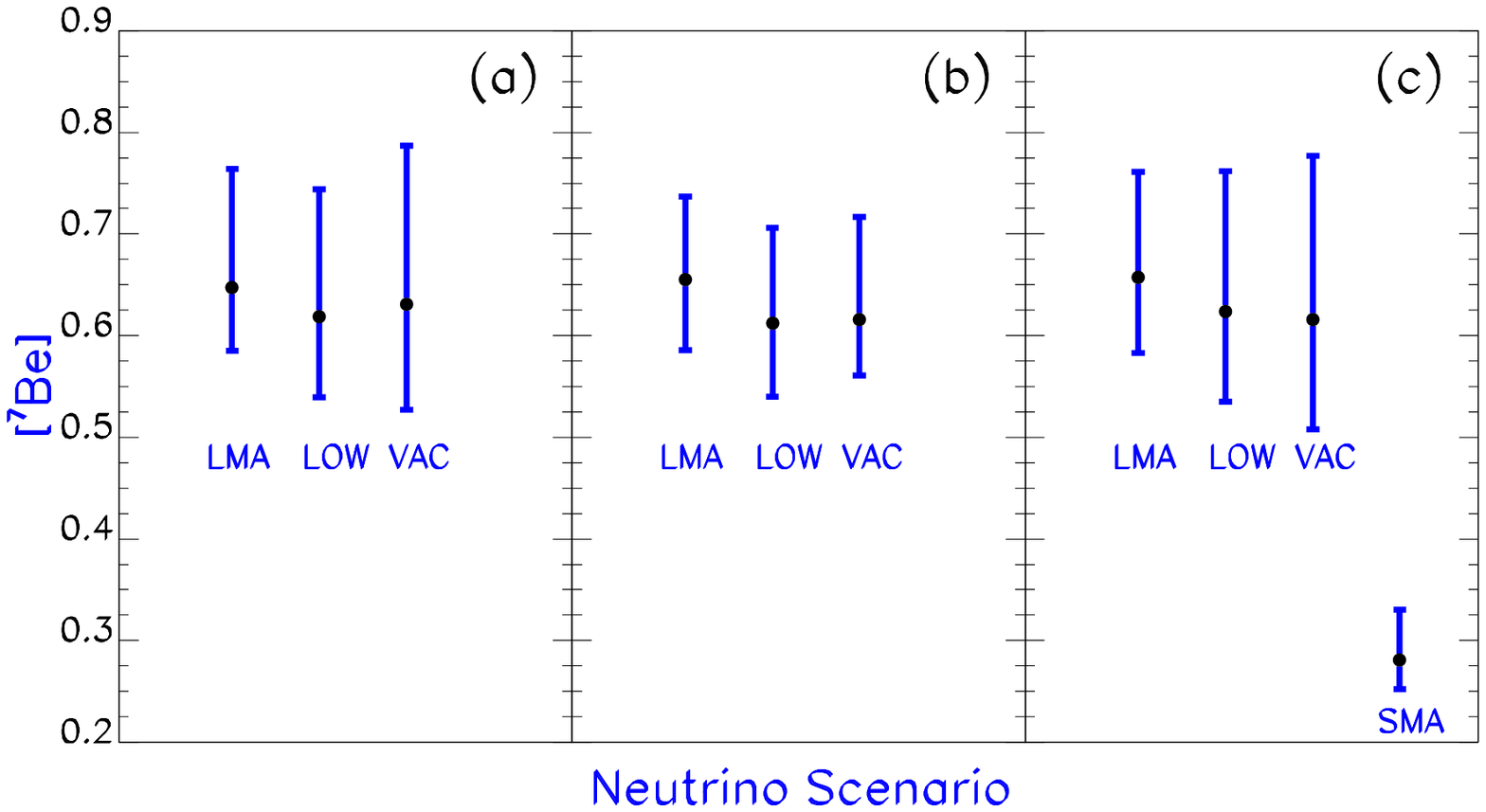,width=4.5in}}
\caption{{\bf The ${\rm\bf ^7Be}$ neutrino event rate.} The figure shows
[${\rm ^7Be}$], the calculated $\nu_e + e^-$ event rate for all recoil
electrons (including CNO neutrinos, see text) with kinetic energies in the range
$0.25 ~{\rm MeV} < T_{\rm e} < 0.8 ~{\rm MeV}$ relative to the rate
that is expected for the BP00 standard model neutrino flux and no
neutrino oscillations. The three panels of predictions in
figure~\ref{fig:be7rate} were derived from the three panels of global
solutions illustrated in figure~\ref{fig:global3} using the three
analysis strategies described in section~\ref{sec:global}.
\label{fig:be7rate}}}

The predicted $3\sigma$ range of the neutrino electron scattering rate
is (cf. table~\ref{tab:be7}) $[{\rm ^7Be}] = 0.65
^{+0.14}_{-0.12}$, where both the minimum value and the maximum value
are achieved by vacuum solutions and the quoted limits correspond to
our standard analysis strategy (cf. figure~\ref{fig:global3}a).  For
the favored LMA solution, the range is somewhat smaller: $[{\rm ^7Be}]
= 0.65 ^{+0.11}_{-0.07}$ (see also ref.~\cite{bilenky01}. The SMA solution,
which is allowed at $3\sigma$ only in the ``free ${\rm ^8B}$ analysis
strategy (cf. figure~\ref{fig:global3}c), predicts much smaller values
for $[{\rm ^7Be}]$, namely, $[{\rm ^7Be}] < 0.34$ .

The $1\sigma$ predicted range is $[{\rm ^7Be}] = 0.65^{+0.04}_{-0.03}$
for our standard global analysis strategy (a).  For the free ${\rm ^8B}$
analysis strategy (c), the spread in predicted values of $[{\rm
^7Be}]$ is somewhat larger, $[{\rm ^7Be}] = 0.66^{+0.05}_{-0.04}$.
Analysis strategy (a) and (b) predict essentially the same $1\sigma$
range.

For KamLAND, the predicted range of $[{\rm ^7Be}]$ is essentially
the same as for\break BOREXINO. For the LMA and VAC solutions, the
BOREXINO and KamLAND best-fit and maximum and minimum values are
the same to three significant figures. For the LOW solution, the
predicted range for KamLAND is about $1$\% lower than for
BOREXINO; the best-fit, maximum and minimum values are shifted
downward for KamLAND.

\subsection{${\rm\bf  ^7Be}$ day-night variations}
\label{subsec:be7daynight}

Figure~\ref{fig:be7daynight} shows the predicted percentage difference
between the night and the day rates that is expected for the BOREXINO
experiment. As a number of previous authors have discussed (see,
e. g., refs.~\cite{borexino,alexborexino,bkdaynight} and references
cited therein), the LOW solution is the only currently allowed
oscillation solution that predicts a large night-day difference in the
rates. The LMA solution predicts a negligible variation and therefore
the $1\sigma$ predicted range is effectively zero.

The small night-day difference for the VAC solution shown in
figure~\ref{fig:be7daynight} and table~\ref{tab:be7} is the first
calculation of this variation, with which we are familiar, for the low
energies to which BOREXINO and KamLAND are sensitive. The physical
origin of this variation has been described in section IXA of
ref.~\cite{tencommand}; it is due to the fact that the VAC survival
probability depends upon the earth-sun separation and the longest
nights occur in the Northern hemisphere when the earth-sun distance is
shortest.
\FIGURE[!ht]{
\centerline{\psfig{figure=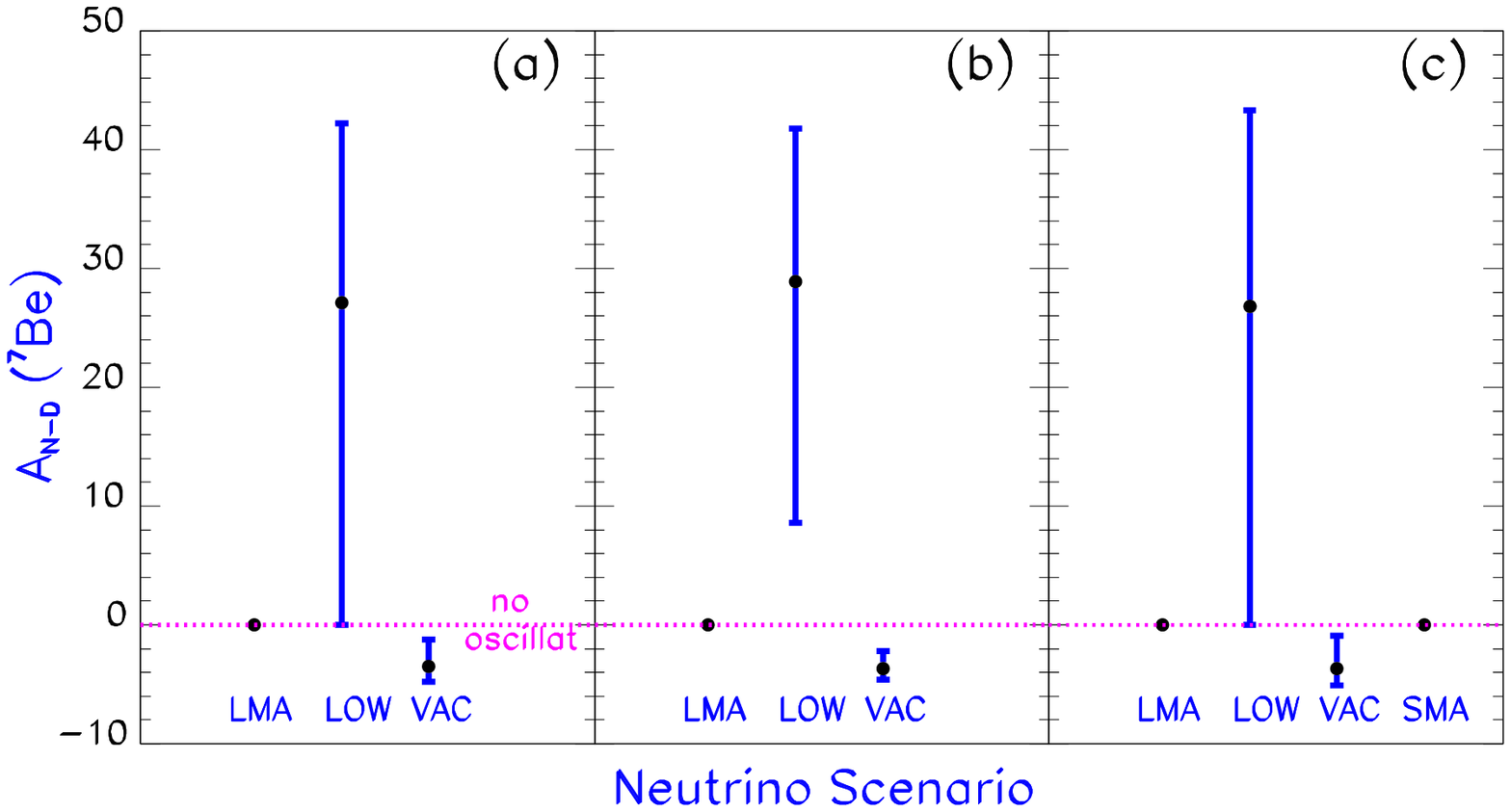,width=4.5in}}
\caption{ {\bf The percentage difference between the night and the day
rates for recoil electrons with kinetic energies in the range
\boldmath$0.25 ~{\rm\bf MeV} < T_{\rm e} < 0.8 ~{\rm\bf MeV}$. } The
figure shows the night-day percentage difference, i.e., $100 \times
A_{\rm N-D}$ defined in eq.~(\ref{eq:daynightdefn}). The predictions
are given for the solar neutrino oscillation scenarios allowed at
$3\sigma$ and illustrated in figure~\ref{fig:global3}. The three
panels refer to results for different analysis strategies described in
section~\ref{sec:global}.
\label{fig:be7daynight}}}

For KamLAND, the predicted range of values of $A_{\rm N-D}$ is  very
similar to the range predicted for BOREXINO. For the LMA solution,
the predicted range of $A_{\rm N-D}$ for the two experiments is  the same to
an accuracy of better than $0.1$\%. For the LOW solution, the maximum
value for KamLAND is about $2$\% less than for BOREXINO ($41$\%
compared to $42$\%) and for the VAC solution, the minimum value is
$-3.9$\% for KamLAND compared to $-4.8$\% for BOREXINO.

\section{Predictions for KamLAND reactor experiment}
\label{sec:kamland}

The KamLAND detector~\cite{kamland,KamLAND}, located in the site of
 the famous Kamiokande experiment~\cite{kamiokande}, consists of
 approximately one kton of liquid scintillator surrounded by
 photomultiplier tubes. KamLAND is sensitive to the $\bar{\nu}_e$
 flux,

 \begin{equation}
\label{eq:nubarabs}
\bar{\nu_e} + p \to n +  e^+\ ,
\end{equation}
from 17 reactors that are located reasonably close to the
 detector.  The distances from the different reactors to the
 experimental site vary from slightly more than 80~km to over 800~km,
 while the majority (roughly 80\%) of the neutrinos travel from 140~km
 to 215~km (see, {\it e.g.,}\/ \cite{KamLAND} for details).

KamLAND ``sees'' the antineutrinos by detecting the total energy
 deposited by recoil positrons, which are produced via
 reaction~(\ref{eq:nubarabs}).  The total visible energy, $E_{\rm
 visible}$, is
\begin{equation}
\label{eq:defnevis}
E_{\rm visible} ~=~ E_{e^+}~+~m_{e} ~=~E_{\bar{\nu}} +m_{\rm p}
-m_{\rm n} +m_{\rm e},
\end{equation}
 where $E_{e^+}$ is the total energy of the positron and $m_{\rm p},
 m_{\rm n}$,and $m_e$ are, respectively, the proton, neutron, and
 electron mass. The positron energy is related to the incoming
 antineutrino energy by $E_{e^+}=E_{\nu}-1.293$~MeV,  up to small
 corrections related to the recoil momentum of the daughter neutron.
 KamLAND is expected to measure the visible energy with a resolution
 better than $\sigma(E)/E=10\%/\sqrt{E}$, for $E$ in MeV
 \cite{KamLAND,Kam-KAM}.
\TABLE[!t]{
\centering
\caption{\label{tab:kamLAND} KamLAND event ratio and first and second
visible energy moments.
Here [CC] is the reduced charged current event ratio (defined below)
and $\delta E_{\rm visible}$ and $\delta \sigma$ are the first two
moments of the visible energy spectrum.
The results are tabulated for two different
thresholds of the visible energy, $1.22$ MeV (columns two through
four) and $2.72$ MeV (columns five through seven).
The ranges listed correspond to
the 1~$\sigma$ and 3~$\sigma$ allowed regions for our standard
LMA oscillation solution, (a).
}
\begin{tabular}{@{\extracolsep{-4pt}}ccccccc}
\noalign{\bigskip}
\hline
\noalign{\smallskip}
 &\multicolumn{1}{c}{b.f.}&max&min&\multicolumn{1}{c}{b.f.}&max&min\\
Scenario & 1.22 MeV& 1.22 MeV& 1.22 MeV& 2.72 MeV& 2.72 MeV& 2.72 MeV\\
\noalign{\smallskip}
\hline
\noalign{\smallskip}
&&1~$\sigma$&1~$\sigma$&&1~$\sigma$&1~$\sigma$\\
\noalign{\smallskip}
\hline
\noalign{\smallskip}
 [CC]        & 0.44 & 0.66   & 0.37   &0.39    & 0.68 &  0.33 \\
 $\delta E_{\rm visible}$ (\%)& -5 & +4   & -10   &+0.3   & +4 &  -8 \\
 $\delta \sigma$ (\%)& +8 & +12   & -15   &+5    & +6 & -13 \\
\noalign{\smallskip}
\hline
\noalign{\smallskip}
&&3~$\sigma$&3~$\sigma$&&3~$\sigma$&3~$\sigma$\\
\noalign{\smallskip}
\hline
\noalign{\smallskip}
 [CC]        & 0.44 & 0.73   & 0.27   &0.39    & 0.74 &  0.23 \\
 $\delta E_{\rm visible}$ (\%)& -5 & +9   & -14   &+0.3    & +7 &  -11 \\
 $\delta \sigma$ (\%)& +8 & +18   & -20   &+5    &+15 & -18 \\
\noalign{\smallskip}
\hline
\end{tabular}
}

In section~\ref{subsec:kamlandpredictions} we present the predictions
of the LMA solution (a) (cf. figure~\ref{fig:global3}a) for the
charged current capture rate (eq.~\ref{eq:nubarabs}) and for the
distortion of the visible energy spectrum. We characterize the
spectral distortion by the first and second moments of the energy
distribution.  No observable effect is predicted for the other
currently allowed oscillation solutions.  We describe in
section~\ref{subsec:kamlandcalculations} the calculational details of
how the predictions were obtained.

\FIGURE[!ht]{
\centerline{\psfig{figure=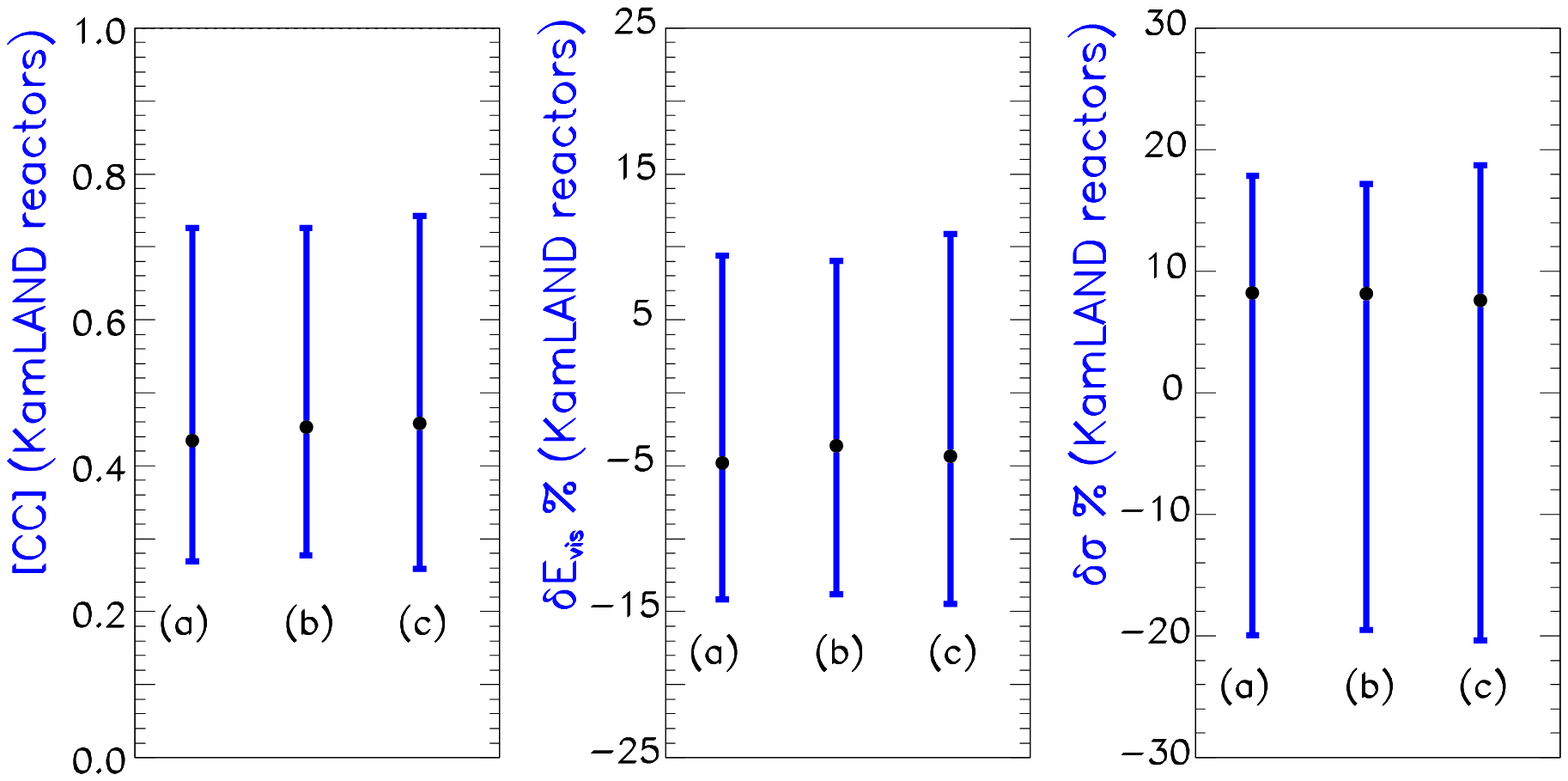,width=5in}}
\caption{ {\bf KamLAND reactor predictions. } The
figure shows the $3\sigma$ allowed predictions for the CC and the first
and second moments of the visible energy spectrum relative to the
expectations based upon the standard solar model and the standard
electroweak model (cf. Table~\ref{tab:kamLAND}). The results are
displayed for a visible energy threshold of $1.22$ MeV
(cf. eqs.~\ref{eq:cckamlandlow} and \ref{eq:cckamlandhigh}).
\label{fig:kamLAND}}}

\subsection{Predictions of global solution for charged current rate
and spectrum distortion}
\label{subsec:kamlandpredictions}

Table~\ref{tab:kamLAND} and Figure~\ref{fig:kamLAND} summarize our
principal predictions for the KamLAND reactor experiment.

The reduced charged current ratio, ${\rm [CC] }$, given in
Table~\ref{tab:kamLAND} is the ratio of the observed event rate for
eq.~(\ref{eq:nubarabs}) divided by the event rate predicted by the
standard solar model (BP00) and the standard electroweak theory (no
oscillations). Thus
\begin{equation}
{\rm [CC] } ~=~
{
{\rm (CC)_{Obs} \over (CC)_{SM} }
}.
\label{eq:defcckamland}
\end{equation}
If the predictions of the standard solar model and the standard
electroweak model are both correct, then ${\rm [CC]} ~=~ 1.0$.

We find that the allowed range of the charged current ratio computed
for a $1.2$ MeV total visible energy threshold and the LMA region
shown in figure~\ref{fig:global3}a is

\begin{equation}
\label{eq:cckamlandlow}
{\rm [CC]} ~=~0.44^{+0.22}_{-0.07} (1\sigma)~~[0.44^{+0.29}_{-0.17}
(3\sigma)],~~E_{\rm threshold} = 1.22 {\rm ~MeV}.
\end{equation}
For a somewhat higher visible energy threshold in which the background
is less problematic (cf. refs.~\cite{Kam-KAM,dgp,mpkamland}), we find
\begin{equation}
\label{eq:cckamlandhigh}
{\rm [CC]} ~=~0.39^{+0.29}_{-0.06} (1\sigma)~~[0.39^{+0.35}_{-0.16}
(3\sigma)],~~E_{\rm threshold} = 2.72 {\rm ~MeV}.
\end{equation}

Table~\ref{tab:kamLAND} and Figure~\ref{fig:kamLAND} represent the
distortion of the visible energy spectrum by the fractional deviation
from the undistorted spectrum of the first two moments of the energy
spectrum (cf. the discussion in section~\ref{subsec:moments}) In the
absence of oscillations, one expects $\langle E_{\rm
vis}\rangle_0=3.97$ MeV and $\langle\sigma\rangle_0=1.26$ MeV for
$E_{\rm threshold} = 1.22$ MeV ($\langle E_{vis}\rangle_0=4.33$ MeV
and $\langle\sigma\rangle_0=1.06$ MeV for $E_{\rm threshold} =1.72$
MeV).  The predicted distortion of the energy spectrum, which can be
as large as 20\%, may be more difficult to measure than the predicted
deviation from unity of the reduced charge current rate, [CC]. If a
significant distortion of the spectrum is observed, the magnitude of
the distortion will severely restrict the possible range of the
oscillation parameters.

The results given here are in general agreement with other
calculations that were made using previously determined allowed
regions~\cite{kamland,KamLAND,Kam-KAM,dgp,mpkamland,BS-KAM,vissani,gp,bargerkamland}. However,
our calculations are the first that we know of that present the
predicted distortions of the visible energy spectrum in terms of the
first and second moments and which give the predicted values of the
reduced charge current ratio for two separate energy thresholds (and
for $1\sigma$ and $3\sigma$ deviations).

\subsection{Calculational procedures and details}
\label{subsec:kamlandcalculations}

The antineutrino spectrum which is to be measured at KamLAND depends
on the power output and fuel composition of each reactor (both change
slightly as a function of time) and on the cross section for
reaction~(\ref{eq:nubarabs}). For the results presented here, we
follow the flux and the cross section calculations and the statistical
procedure described in ref.~\cite{dgp}.  We define one
``KamLAND-year'' as the amount of time it takes KamLAND to observe $800$
events with visible energy above $1.22$~MeV. This is roughly what is
expected after one year of running (assuming a fiducial volume of 1
kton), if all reactors run at (constant) $78\%$ of their maximal power
output \cite{KamLAND}. We assume a constant chemical composition for
the fuel of all reactors (explicitly, 53.8\% of $^{235}$U, 32.8\% of
$^{239}$Pu, 7.8\% of $^{238}$U, and 5.6\% of $^{241}$Pu, see
\cite{bargerkamland,chem-comp}).

The shape of energy spectrum of the incoming neutrinos can be
derived from a phenomenological
parametrization, obtained in \cite{shape-spectrum},
\begin{equation}
\frac{{\rm d}N_{\bar{\nu}_e}}{{\rm d}E_{\nu}}\propto
e^{a_0+a_1E_{\nu}+a_2E^2_{\nu}},
\label{eq:spectrum}
\end{equation}
where the coefficients $a_i$ depend on the parent nucleus.
The values of $a_i$ for the different isotopes we used are
tabulated in \cite{shape-spectrum,mpkamland}. These
expressions are  good approximations of the (measured) reactor
flux for values of $E_{\nu}\gtrsim 2$~MeV.

The cross section for $\bar{\nu}_e+p\rightarrow e^{+}+n$ has been
computed including corrections related to the recoil momentum of the
daughter neutron in \cite{cross-sec}.  We used the hydrogen/carbon
ratio, r=1.87, from the proposed chemical mixture (isoparaffin and
pseudocumene) \cite{KamLAND}.  It should be noted that the energy
spectrum of antineutrinos produced at nuclear reactors has been
measured accurately in previous reactor neutrino experiments (see
\cite{KamLAND} for references). For this reason, we will assume for
simplicity that the standard (without oscillations) antineutrino
energy spectrum is known precisely.  Some of the effects of
uncertainties in the incoming flux on the determination of oscillation
parameters have been studied in \cite{mpkamland} and are estimated to be
small.

The calculation is first done for visible energies $1.22<E_{vis}<7.22$
MeV and in the computation of the shifts on the visible energy moments
we assume 12 energy bins (binwidth is 0.5 MeV).  There still remains,
however, the possibility of irreducible backgrounds from geological
neutrinos in the lower energy bins ($E_{vis}\lesssim 2.6$ MeV)
\cite{Kam-KAM,dgp,mpkamland}. To verify how this possible background
may affect the results we have repeated the analysis discarding the
three lower energy bins i.e. considering only events with visible
energies $2.72<E_{vis}<7.22$ MeV.

\section{Discussion and summary}
\label{sec:discuss}
In this section, we summarize and discuss our principal results that
are presented in the previous sections.

We summarize in section~\ref{subsec:discussmodel} our conclusions
 regarding solar model predictions and present in
 section~\ref{subsec:discussglobal} the currently allowed solar
 neutrino oscillation solutions that are found with three different
 analysis strategies. We describe the predictions of these allowed
 solutions for the SNO experiment in section~\ref{subsec:discusssno}
 and for the BOREXINO and KamLAND experiments in
 section~\ref{subsec:discussbe7}. In section~\ref{subsec:whatborexino}
 we discuss what can be learned from the BOREXINO and KamLAND solar
 neutrino experiments, despite the fact that the favored LMA, LOW, and
 VAC experiments all predict similar event rates. We describe the
 unique sensitivity of the reactor KamLAND experiment in
 section~\ref{subsec:discusskamlandreactor}. Finally, we summarize in
 section~\ref{subsec:robust} the robustness of the neutrino
 oscillation predictions and the possibilities for uniquely
 identifying the oscillation solution.

We present predictions for both $1\sigma$ and $3\sigma$ ranges of the
currently allowed neutrino oscillation parameters.

We use the notation [Q] to represent the measured or predicted value
of the quantity Q divided by the value expected for Q if one assumes
the correctness of the standard solar model (BP00) and also assumes
that nothing happens to the neutrinos after they are created in the
sun.

\subsection{Uncertainties in model predictions of solar neutrino fluxes}
\label{subsec:discussmodel}

How does the ${\rm ^8B}$ neutrino flux predicted by the solar model
agree with the value inferred from the measurements by
SNO~\cite{sno2001} for the CC flux and by
Super-Kamiokande~\cite{superk} for the neutrino-electron scattering
rate?  Combining the SNO and Super-Kamiokande measurements, the SNO
collaboration infers~\cite{sno2001}(see
also~\cite{foglipostsno,conmpSKSNO}) a total active ${\rm ^8B}$
neutrino flux ($\nu_e + \nu_\mu + \nu_{\tau}$) of $\phi({\rm ^8B;
measured}) = (5.44 \pm 0.99 )10^6 {\rm ~ cm^{-2} s^{-1}}$. Comparing
the measured ${\rm ^8B}$ neutrino flux with the flux~\cite{bp2000}
calculated using the 1998 standard value for $S_{17}(0)$, $\phi({\rm
^8B;BP00}) = 5.05(1 ^{+0.20}_{-0.16})10^6 {\rm ~ cm^{-2} s^{-1}}$, one
finds

\begin{equation}
\frac{\phi({\rm SNO + Super-Kamiokande})}
{\phi({\rm BP00})}
= 1.08 \pm 0.20({\rm measurement}) \pm 0.18({\rm solar~ model}).
\label{eq:measuredoverbp}
\end{equation}
Thus the measured flux agrees with the BP00 solar model flux to within
$0.3\sigma$, combined experimental and theoretical errors.  If we use
instead the flux (see table~\ref{tab:rates}) obtained with the more
recent (Junghans et al.) value for $S_{17}(0)$, $5.93 \times
10^{-4}\left(1.00^{+0.14}_{-0.15}\right)$, we find

\begin{equation}
\frac{\phi({\rm SNO + Super-Kamiokande})}
{\phi({\rm BP00 + New ^8B})}
= 0.91 \pm 0.17({\rm measurement}) \pm 0.14({\rm solar~ model}).
\label{eq:measuredovernewbp}
\end{equation}
Therefore the measured flux agrees with the BP00 + ${\rm New
^8B}$ solar model flux to within $0.4\sigma$, combined
experimental and theoretical errors. In both cases, the measured
and the solar model fluxes agree to better than $1\sigma$. The
agreement between the predicted and measured ${\rm ^8B}$ solar
neutrino flux is a significant confirmation of the standard solar
model since the rare ${\rm ^8B}$ flux depends upon the $25$th power of
the central solar temperature~\cite{tempdependence}.

We present calculated results in this paper for two values of the
crucial low energy cross section factor, $S_{17}(0)$, namely,
$S_{17}(0) = 19^{+4}_{-2} {\rm eVb}$~\cite{adelberger} (the 1998
standard value) and $S_{17}(0) = (22.3 \pm 0.9) {\rm
eVb}$~\cite{junghans01} (the most recent and precise
measurement). These two values of $S_{17}(0)$ represent well the range
of measured cross section factors obtained in recent
experiments~\cite{b8refs}.  The reader can therefore see at a glance
the dependence of the calculated quantities upon the adopted value of
the cross section $S_{17}(0)$.

Essentially all global analyses of solar neutrino oscillations make
use of neutrino fluxes and their uncertainties obtained from a solar
model. If the eight most important solar neutrino
fluxes(cf. table~\ref{tab:rates} ) were allowed to vary without
constraints, there would be too many free parameters to make possible
an efficient study of solar neutrino oscillations. If the
BP00 model is adopted in the analysis, then the total $p-p$ flux is
assumed known to $\pm 1$\% and the total ${\rm ^7Be}$ flux is assumed
known to $\pm 10$\%.

 Guidance from a
solar model is required just to decide what are the most important
neutrino fluxes to include in the analysis. For example, all current
analyses of solar neutrino oscillations include a subset of the eight fluxes
calculated in the standard solar model and listed in Table~\ref{tab:rates}
but neglect the fluxes from the
reactions $e^- + {\rm ^3He} \rightarrow {\rm ^3H} + \nu_e$,\,
$e^- + {\rm ^8B} \rightarrow {\rm ^8Be} + \nu_e$,\, $e^- +
{\rm ^{13}N} \rightarrow {\rm ^{13}C} + \nu_e$, \,$e^- +
^{15}{\rm O} \rightarrow {\rm ^{15}N} + \nu_e$, and
\,$e^- + {\rm ^{17}F} \rightarrow {\rm ^{17}O} + \nu_e$.
One must use the parameters of
a solar model
to show that the neglected fluxes are
negligible~\cite{bahcalllines} and that the included fluxes are
important.

Even for  a Bayesian
analysis (see ref.~\cite{giuntimodelindependent}), one needs guidance
from a solar model to determine which fluxes should be included and which
should be neglected and,
e. g., to decide if the priors should be expressed
in terms of $10^9{\rm \, cm^2s^{-1}}$ for the $^7$Be flux and
$10^6{\rm \, cm^2s^{-1}}$ for the $^8$B flux or vice versa.
Nevertheless, the data for solar neutrino experiments is becoming
sufficiently precise and constraining that Bayesian analyses can
provide important and complimentary insights to results obtained with
other statistical techniques (see, e.g., figure 20 of
ref.~\cite{giuntimodelindependent}).

In a few cases, authors have carried out analyses in which the ${\rm
^8B}$ neutrino flux, but none of the other fluxes, is completely unconstrained by
the solar model predictions, which is our case (c) in
section~\ref{sec:global} and in figure~\ref{fig:global3}c.  The
experimental data from solar neutrino experiments are not yet
sufficient to permit a very restrictive analysis for unknown neutrino
oscillation parameters if neutrino fluxes other than the ${\rm ^8B}$
neutrino flux are allowed to be free variables.

The undesirable but currently unavoidable practice of using solar
models to help constrain neutrino parameters is, ironically, the
opposite practice to what was envisioned in the early days of solar
neutrino research. The original chlorine experiment was proposed to
test solar models using the assumed well-known properties of
neutrinos~\cite{johnray}.

In order to  document what we use in this paper and to facilitate
analyses of solar neutrino data by other groups, we summarize in
section~\ref{sec:newbp} the uncertainties and the best-estimates for
solar neutrino fluxes derived from the BP00 solar model. We describe
the results that are obtained when adopting
the precise Junghans et al. value of $S_{17}(0)$ and also the results
that are obtained using the 1998 standard value of
$S_{17}(0)$(see especially table~\ref{tab:rates} and
table~\ref{tab:uncertainties} and related comments in the text).

We stress the continued importance of precise measurements of the low
energy cross section factor, $S_{17}(0)$, for the ${\rm
^7Be}$(p,$\gamma$)${\rm ^8B}$ reaction. The neutrino fluxes measured in the
Super-Kamiokande, SNO, and ICARUS solar neutrino experiments all
depend linearly upon this cross section factor and the standard model
prediction of the event rate in the chlorine experiment is also
dominated by the ${\rm ^8B}$ neutrino flux.

The uncertainty in the low energy cross section factor, $S_{34}(0)$,
for the ${\rm ^3He}$(${\rm ^4He}$,$\gamma$) ${\rm ^7Be}$ reaction
dominates the uncertainty in the solar model calculation of the ${\rm
^7Be}$ solar neutrino flux. The total uncertainty in the solar model
calculation of the ${\rm ^7Be}$ neutrino flux is $9$\% and the ${\rm
^3He}$(${\rm ^4He}$,$\gamma$)${\rm ^7Be}$ reaction contributes $8$\%
to the total uncertainty that is computed by quadratically combining
uncertainties from different sources. The cross section factor
$S_{34}(0)$ is also the largest nuclear physics uncertainty in the
prediction of the ${\rm ^8B}$ solar neutrino flux if one adopts the
Junghans et al. value for the uncertainty in $S_{17}(0)$.

Now that there are precise measurements of $S_{17}(0)$ underway and
completed, we believe the highest priority nuclear astrophysics
measurement for the future is the precision determination of the low
energy cross section factor for the ${\rm ^3He}$(${\rm
^4He}$,$\gamma$)${\rm ^7Be}$ reaction. A measurement of $S_{34}(0)$ to
an accuracy of better than $4$\% is necessary in order  to
decrease the uncertainty in the reaction ${\rm ^3He}$(${\rm
^4He}$,$\gamma$)${\rm ^7Be}$ to where it is no longer the largest
uncertainty in the prediction of the ${\rm ^7Be}$ solar neutrino flux.

\subsection{Global neutrino oscillation solutions}
\label{subsec:discussglobal}

Using three different analysis strategies that span the range of
previously used strategies, we determine the globally allowed solar
neutrino oscillation solutions that are consistent with all the
available solar and reactor data.  The results are summarized in
figure~\ref{fig:global3}; the calculations on which this figure is
based used the new Junghans et al. value of $S_{17}(0)$.
Table~\ref{tab:globalbestbp00} gives the best-fit values of $\Delta
m^2$ and $\tan^2 \theta$, as well as the local value of $\chi^2_{\rm
min}$ for each oscillation solution; the results presented in
table~\ref{tab:globalbestbp00} were obtained using our standard
analysis strategy in which we take account of the BP00 + New ${\rm
^8B}$ predicted fluxes and estimated uncertainties and include the
Super-Kamiokande day and night energy spectrum but not the total rate.
The $3\sigma$ upper limit to the allowed value of $\Delta m^2$ lies
between $3.0$ and $7.5$, in units of $10^{-4}~{\rm eV^2}$, depending
upon what is assumed about $S_{17}(0)$ and the analysis strategy.

The LMA solution is favored, but the LOW and VAC solutions are
also allowed at a C.L. corresponding to $3\sigma$. Other
solutions such as oscillations into sterile neutrinos or the SMA
solution for active neutrinos are disfavored at $3\sigma$ if the
standard analysis strategy is adopted, but SMA is barely allowed
at $3\sigma$ if the ${\rm ^8B}$ neutrino flux is unconstrained by
the solar model predictions and uncertainties.

Figure~\ref{fig:beforeafter} is a ``Before and After'' comparison of
the effect of the ${\rm ^8B}$ production cross section on the global
oscillation solutions. The only difference between the ``Before and
After' panels in figure~\ref{fig:beforeafter} is the replacement in
the analysis of the 1998 standard value of $S_{17}(0) = 19^{+4}_{-2}
{\rm eVb}$~\cite{adelberger} by the value obtained by the recent
precise measurement, i.e., $S_{17}(0) = \left(22.3 \pm 0.9\right) {\rm
eVb}$~\cite{junghans01}. Thus the ``before'' (left) panel of
figure~\ref{fig:beforeafter} corresponds to the results shown in the
left panel of figure~9 of ref.~\cite{bgp}. The change in the value of
$S_{17}(0)$ is sufficient to drive the SMA and Just So$^2$ solutions
over the edge of the allowed region; they are allowed at $3\sigma$
with the older value of $S_{17}(0)$ but not with the newer value.

The global oscillation analysis yields a $3\sigma$ allowed range for
the inferred total ${\rm ^8B}$ solar neutrino flux expressed in terms
of the best-estimate flux predicted by the BP00 model with the
Junghans et al. value of $S_{17}(0)$. From table~\ref{tab:fb8}, we
find for active neutrinos
\begin{equation}
0.40 ~\leq~f_{\rm B} ~\leq~ 1.36,
\label{eq:fb8summary}
\end{equation}
for the case in which the ${\rm ^8B}$ neutrino flux is unconstrained
in the analysis. The best-fit value of $ f_{\rm B} = 1.0$ for this
unconstrained case.  The $3\sigma$ allowed range found here is
slightly smaller than determined directly for active neutrinos by the
SNO collaboration by comparing the SNO and Super-Kamiokande
results. The SNO collaboration found~\cite{sno2001} $ f_{\rm B} = 0.92
\pm 0.50$, $3\sigma$ range.  We show in section~\ref{subsec:fb8} that
if one considers an arbitrary admixture of sterile and active
neutrinos, the $3\sigma$ upper limit of $ f_{\rm B}$ increases to
$2.9$.

All of the global analyses of solar neutrino experiments that include
the important Super-Kamiokande data~\cite{superk} on the electron
recoil energy spectrum use the many energy bins provided by the
Super-Kamiokande group ($19$ energy bins in the last report). It would
be very instructive to be able to carry out global analyses while
representing the spectrum by only one or two measured quantities, the
first or the first and second moments. One could then determine the
change in the global solutions that result from giving the measurement
of the energy spectrum the same prominence in the analysis as one or
two measurements of the total rates\footnote{We already know from
previously published calculations(see
refs.~\cite{bks2001,bks98,krastev01} that are significant differences
in the allowed regions between the extreme cases in which only the
total rates of the solar neutrino experiments are considered and the
case in which all $19$ of the Super-Kamiokande energy bins are
included in addition to the total rates. In order to properly take
account of the characteristics of the detector, which can influence
the moments~\cite{bl,earlymoments}, the experimental collaboration
must publish both the measured moments and the values expected for the
moments if there is no distortion of the spectrum. }.

\TABLE[!ht]{
\centering
\caption{Neutrino oscillation predictions for the chlorine and gallium
radiochemical experiments.  The predictions are based upon the global
analysis strategy (a), described in section~\ref{sec:global}, and make
use of the neutrino fluxes given in table~\ref{tab:rates} for the BP00
solar model, the neutrino absorption cross sections given in
ref.~\cite{nuabs}, and the Junghans et al. value of
$S_{17}(0)$~\cite{junghans01}. The rates and $1\sigma$ uncertainties
are presented for the best-fit values of the allowed solutions listed
in table~\ref{tab:globalbestbp00}. The total rates should be compared
with the standard solar model values of table~\ref{tab:rates}, which
are, $8.65^{+1.2}_{-1.1}$ (chlorine) and $130^{+9}_{-7}$(gallium), and
the measured values, $2.56\pm 0.23$ (chlorine, see
ref.~\cite{chlorine})and $75.6\pm 4.8$(gallium, see
ref.~\cite{gallex,gno,sage}).  \protect\label{tab:radiochemical}}
\begin{tabular}{@{\extracolsep{-5pt}}lcccccc}
\hline
\noalign{\smallskip}
Source&Cl&Ga&Cl&Ga&Cl&Ga\\
&(SNU)&(SNU)&(SNU)&(SNU)&(SNU)&(SNU)\\
&LMA&LMA&LOW&LOW&VAC&VAC\\
\noalign{\smallskip}
\hline
\noalign{\smallskip}
pp            &  0   & 41.8 &  0   & 38.7 &  0   & 44.3\\
pep           & 0.12 & 1.49 & 0.11 & 1.35 & 0.16 & 1.95\\
hep           & 0.01 & 0.02 & 0.02 & 0.03 & 0.03 & 0.04\\
${\rm ^7Be}$  & 0.62 & 18.7 & 0.58 & 17.5 & 0.54 & 16.4\\
${\rm ^8B}$   & 2.05 & 4.27 & 2.94 & 6.13 & 3.95 & 8.33\\
${\rm ^{13}N}$& 0.05 & 1.80 & 0.04 & 1.69 & 0.05 & 2.01\\
${\rm ^{15}O}$& 0.17 & 2.77 & 0.16 & 2.64 & 0.20 & 3.29\\
${\rm ^{17}F}$& 0.00 & 0.03 & 0.00 & 0.03 & 0.00 & 0.04\\
\noalign{\medskip}
&\hrulefill&\hrulefill&\hrulefill&\hrulefill&\hrulefill
&\hrulefill
\\
Total         & $3.01 \pm 0.31$ & $71.0 \pm 2.6$ & $3.85 \pm 0.45$&
$68.1 \pm 2.8$& $4.93 \pm 0.60$ & $76.4 \pm 3.2$\\
\noalign{\smallskip}
\hline
\end{tabular}
}

We describe in Section~\ref{subsec:methods} the technical differences
between the three analysis strategies. This section is intended
primarily for neutrino analysis junkies.

Table~\ref{tab:radiochemical} presents the contributions of individual
sources and the total rates predicted by the favored neutrino
oscillation schemes, LMA, LOW, and VAC, for the chlorine and the
gallium radiochemical experiments. The predictions are given in the
table for our best-fit solutions obtained using the standard analysis
strategy (a). The two measured rates are also listed. The errors are
dominated by the solar model uncertainties for analysis strategy (a),
which is apparent from table~\ref{tab:radiochemical} by comparing the
calculated and measured values. The theoretical uncertainties
expressed in SNU for the chlorine and gallium rates are greatly
reduced for the oscillation solutions (table~\ref{tab:radiochemical})
compared to the no-oscillation values
(table~\ref{tab:rates}). However, the theoretical uncertainties
expressed as percentages of the total rates are comparable for the
oscillation and no-oscillation scenarios.

All three of the oscillation scenarios yield total event rates in good
agreement with the measured values for the gallium
experiments. However, the calculated rates for the LOW and the VAC
solutions are in poor agreement with the measured values
(discrepancies of $2.7$ and $3.8$ $\sigma$, respectively). The global
VAC solution is acceptable only because one can choose the ${\rm ^8B}$
normalization such that the predicted distortion is small for the
recoil energy spectrum measured by Super-Kamiokande. Of course, for
strategy (c), in which the ${\rm ^8B}$ neutrino flux is a free
parameter, one can obtain global allowed solutions that are much
better fits to the chlorine rate.

The predicted rates for the Super-Kamiokande experiment (in units of
$\times 10^{6}\break {\rm cm^{-2}s^{-1}}$) are: $2.39^{+0.33}_{-0.36}$
(LMA), $3.02^{+0.42}_{-0.45}$ (LOW), and $3.81^{+0.53}_{-0.57}$(VAC),
which should be compared with the measured rate of $(2.32\pm
0.03^{+0.08}_{-0.07})\times 10^{6} {\rm cm^{-2}s^{-1}}$~\cite{superk}.
For the SNO CC measurement, the predicted rates are (in units of
$\times 10^{6} {\rm cm^{-2}s^{-1}}$) : $1.72^{+0.24}_{-0.26}$
(LMA), $2.50^{+0.35}_{-0.38}$ (LOW), and $3.26^{+0.46}_{-0.49}$ (VAC),
which should be compared with the measured rate of
$(1.75\pm0.07^{+0.12}_{-0.11})\times 10^{6} {\rm
cm^{-2}s^{-1}}$~\cite{sno2001}.

\subsection{Predictions for SNO}
\label{subsec:discusssno}

All three analysis strategies yield essentially the same
$3\sigma$ range for the neutral current to charged current double
ratio(see figure~\ref{fig:nccc}) predicted by the favored LMA,
LOW, and VAC solutions, namely, $1.4 < {\rm [NC]/[CC]} < 5.3$(For
analysis strategy (c), with uses an unconstrained ${\rm ^8B}$ neutrino
flux, the upper limit extends to $6.2$.) The $1\sigma$ predicted range
is $3.5 \pm 0.6$ for our standard analysis strategy.

The predicted range of [NC]/[CC] for the favored oscillation
solutions is also rather insensitive (see
figure~\ref{fig:ncccbeforeafter}) to the choice of the ${\rm
^8B}$ production cross section within the range of the 1998
standard value, ref.~\cite{adelberger}, and the recent precision
determination, ref.~\cite{junghans01}. If the ${\rm ^8B}$ is
allowed to vary freely, our analysis strategy (c), then the SMA
solution for active neutrinos is allowed at the $3\sigma$ C.L.
and for this solution  $1.15 < {\rm [NC]/[CC]} < 1.31$.
Figure~\ref{fig:nccc} shows that all of the values of [NC]/[CC]
predicted by the allowed neutrino oscillation solutions are
separated from the no oscillation value of ${\rm [NC]/[CC]} =
1.0$ by more than the expected experimental uncertainty.

For the average difference in the CC day-night effect, $A_{\rm
N-D}({\rm SNO ~CC})$, all three analysis strategies also yield
essentially the same results, as is shown in
figure~\ref{fig:snodaynight}. Figure~\ref{fig:ccabeforeafter} shows
that the predictions for $A_{\rm N-D}({\rm SNO ~CC})$ are also robust
with respect to likely changes in the value of $S_{17}(0)$. The
$3\sigma$ range in percent for all the oscillation solutions is $0.0
\leq A_{\rm N-D}({\rm SNO~CC}) \leq 0.21$ (The $1\sigma$ range is
$A_{\rm N-D}({\rm SNO~CC}) = 8.15 \pm 5.15\;$\% .)

The predicted average difference in the day-night effect for
neutrino-electron scattering, $A_{\rm N-D}({\rm SNO ~ES})$, depends
upon which among the currently allowed global oscillation solutions
(LMA, LOW, VAC, or SMA) is assumed (see
figure~\ref{fig:snoesdaynight}). All of the results are bounded by the
$3\sigma$ allowed range for Super-Kamiokande, i. e., $A_{\rm N-D}({\rm
SNO ~ES})$ lies between $0$\% and
$11$\%. Figure~\ref{fig:corrdaynight} shows the predicted correlation
between the day-night effect for the CC and the day-night effect for
neutrino-electron scattering.  The predicted correlation between
$A_{\rm N-D}({\rm SNO ~ES})$ and $A_{\rm N-D}({\rm SNO ~CC})$
constitutes an important consistency check for the oscillation
solution.

The first and second moments of the electron recoil energy spectrum
from CC reactions summarize most of the measurable information about
the energy spectrum of ${\rm ^8B}$ neutrinos that are observed by SNO
(and by Super-Kamiokande). The low order moments may represent a more
appropriate way to characterize the relatively undistorted energy
spectrum rather than to provide event rates in a large number of
separate spectral energy bins(cf. ref~\cite{superk}). SNO has a
significant advantage over Super-Kamiokande in measuring potential
distortions of the energy spectrum due to neutrino oscillations,
because the electron is the only light particle (hence it takes most
of the energy) in the CC reaction, eq.~(\ref{eq:cc}).

The current set of allowed oscillation solutions do not predict
distortions that are large enough to be detected by SNO at a high
level of significance.  Figure~\ref{fig:firstmoment} and
figure~\ref{fig:secondmoment}, when combined with the discussion in
section~\ref{subsec:moments} of the likely experimental uncertainties,
establish this pessimistic prediction.  Nevertheless, it is important
to measure accurately the first two moments of the recoil energy
distribution. The pessimistic prediction summarized here could be
wrong if the correct oscillation solution is one not favored by the
initial pioneering set of solar neutrino experiments (see further
discussion of this point in section~\ref{subsec:robust}).

If we are lucky, if Nature is kind, then one measurement may define
rather well the solar neutrino oscillation
parameters. Figure~\ref{fig:lucky} shows that only relatively small
regions in neutrino oscillation parameter space predict that SNO will
measure either ${\rm [NC]/[CC]} > 3.3$ or $A_{\rm N-D}({\rm SNO \; CC}) >
0.1$, If either of these inequalities is established experimentally,
then the oscillation parameters $\Delta m^2$ and $\tan^2 \theta$ will
be strongly constrained.

\subsection{Solar neutrino predictions for BOREXINO and KamLAND}
\label{subsec:discussbe7}

Table~\ref{tab:be7} and figure~\ref{fig:be7rate} show that the
currently LMA, LOW, and VAC oscillation solutions predict similar
values for the neutrino-electron scattering rate for ${\rm ^7Be}$
neutrinos, $[{\rm ^7Be}] = 0.655 \pm 0.035, 1\sigma(0.66 \pm 0.13, 3\sigma)$.
The SMA solution, which is allowed at $3\sigma$ only in the ``free
${\rm ^8B}$ analysis strategy (cf. figure~\ref{fig:global3}c),
predicts a much smaller value, ${[\rm ^7Be]} < 0.34$ .

Figure~\ref{fig:be7daynight} and table~\ref{tab:be7} show that the LOW
solution is the only currently allowed (at $3\sigma$) neutrino
oscillation solution that predicts a significant day-night variability
in BOREXINO or KamLAND. If a difference between day and night rates is
actually observed, this will very strong evidence in favor of the LOW
solution.

We have tested the robustness of the predictions for the BOREXINO and
KamLAND experiment in several ways: i) we computed the predicted
results for all three of the analysis strategies described in
section~\ref{sec:global}; ii) we compared the results at the
terrestrial locations of BOREXINO and KamLAND; and iii) we compared
the predictions with and without including events from neutrino
sources other than ${\rm ^7Be}$. In all cases, the variations are
small (and are given quantitatively in section~\ref{sec:be7}).

How well can we predict $[{\rm ^7Be}]$ if SNO measures either ${\rm
[NC]/[CC]} > 3.3$ or $A_{\rm N-D}({\rm SNO \: CC}) > 0.1$?  Unfortunately,
[${\rm ^7Be}$] is not a unique function of the neutrino mixing
parameters, $\Delta m^2$ and $\tan^2 \theta$. Instead, [${\rm ^7Be}$]
can take on the same value for many different pairs of ($\Delta m^2$,
$\tan^2 \theta$). The predicted range of [${\rm ^7Be}$] over the full
LMA region allowed at $3\sigma$ is from $0.585$ to $0.76$.  For the
parameters corresponding to ${\rm [NC]/[CC]} > 3.3$, the range is
almost as large: $0.585$ to $0.74$. For $A_{\rm N-D}({\rm CC}) > 0.1$,
the predicted range of [${\rm ^7Be}$] is $0.585$ to $0.72$.  Thus a
hypothetical (and optimistic) experimental determination of the
behavior of the survival probability, represented by
figure~\ref{fig:lucky}, in the $6$ MeV to $10$ MeV region of neutrino
energies most effectively probed by SNO would not be sufficient to
allow a precise prediction of the survival probability at $0.86$ MeV,
which corresponds to the energy of the ${\rm ^7Be}$ neutrino line.

\subsection{What will we learn from BOREXINO and KamLAND solar
neutrino experiments?}
\label{subsec:whatborexino}

At first impression, it may seem somewhat discouraging that the
currently most favored oscillation solutions, LMA, LOW, and VAC, all
predict very similar event rates for neutrino-electron scattering in
BOREXINO and KamLAND (see figure~\ref{fig:be7rate}). However, this is
really an advantage.

If the currently favored oscillation solutions and the standard solar
model are correct, then BOREXINO or KamLAND must confirm that $[{\rm
^7Be}] = 0.66 \pm 0.13$. Thus the fact that the predicted event rates
for BOREXINO and KamLAND are similar for the LMA, LOW, and VAC
solutions means that a measurement of the neutrino electron scattering
rate is a critical test of the validity of the standard solar model
prediction and the favored oscillation solutions. In addition, we may
be somewhat surprised and the measurements may show that $[{\rm
^7Be}] < 0.34$, favoring an SMA solution.

Also, a large measured value for the day-night effect in BOREXINO
or KamLAND would imply the correctness of the LOW oscillation
solution, which is currently allowed but is less favored than the LMA
solution (cf. Table~\ref{tab:globalbestbp00}).

Experiments like BOREXINO and KamLAND that detect neutrinos with
energies less than $1$ MeV are necessary to test the validity of solar
neutrino oscillation solutions. The solutions explored in this paper,
and in related papers by other authors, are primarily constrained by
data that refer to the relatively high energies($> 5$ MeV) for solar
neutrinos to which the chlorine, Kamiokande, Super-Kamiokande and SNO
experiments are primarily or exclusively sensitive. Of all the solar
neutrino experiments performed so far, only the gallium experiments,
SAGE and GALLEX + GNO, have a large sensitivity to low energy
neutrinos.  The gallium experiments are radiochemical experiments and
therefore do not measure neutrino energies.

BOREXINO and KamLAND will also tell us something new and critical
about the Sun.  We have very little direct observational information
about the important ${\rm ^7Be}$ solar neutrino flux. If one supposes
that ${\rm ^7Be}$ is the only source contributing to the observed
rates in the chlorine and gallium experiments, then the the $3\sigma$
upper limit implied by the chlorine measurements~\cite{chlorine} is
$2.8$ times the BP00 predicted flux. The upper limit implied by the
gallium experiments~\cite{sage,gallex,gno} is $2.6$ times the BP00
predicted flux. All of the existing experiments are consistent with a
${\rm ^7Be}$ flux that is identically zero. Thus the direct
experimental constraints only require that

\begin{equation}
0.0 ~\leq ~\frac{\phi(^7{\rm Be})}{\phi(^7{\rm Be})_{\rm BP00}}~ \leq ~2.6 ,
  ~~3\sigma.
\label{eq:be7experimental}
\end{equation}
It is important to test experimentally whether the solar model
prediction for the ${\rm ^7Be}$ neutrino flux is correct. The branch
of the $p-p$ fusion chain that is represented by ${\rm ^7Be}$
neutrinos occurs in $15$\% of the completions of the chain, according
to the BP00 solar model.

In order to test the solar model prediction of the ${\rm ^7Be}$
total neutrino flux, we must perform at least one additional experiment
beyond the BOREXINO and KamLAND $\nu-e$ scattering experiments, which
measure a linear combination of CC and NC events. Either a CC
measurement or a NC measurement of the ${\rm ^7Be}$ line is essential
to test the solar model in a way that is completely free of all
influence of solar models.

\subsection{Predictions for the KamLAND reactor experiment}
\label{subsec:discusskamlandreactor}

The KamLAND reactor experiment may provide definitive evidence for or
against the LMA oscillation solution. The reduced charged current rate
is predicted by the LMA solution to satisfy ${\rm [CC]}
~=~0.44^{+0.22}_{-0.07} (1\sigma)~~[0.44^{+0.29}_{-0.17}
(3\sigma)],~~E_{\rm threshold} = 1.22 {\rm ~MeV}$, with a somewhat
larger allowed range for a higher energy
threshold. Table~\ref{tab:kamLAND} summarizes the predicted distortion
of the energy spectrum (in terms of the first and second moment) as
well as the rate of the CC reaction for two plausible energy
thresholds. If the LMA solution is correct, then the KamLAND
experiment should observe a measurable deficit in the charged current
rate. The spectral distortion, if measured, will determine the solar
neutrino oscillation parameters with unprecedented precision.

\subsection{Robustness and Uniqueness}
\label{subsec:robust}

The neutrino oscillation solutions that describe all the available
solar and reactor data are, with one exception, robust with respect to
variations among the three analysis strategies we have used in this
paper. The SMA solution is the sole exception,  and only
just barely an exception. This solution exists at
the $3\sigma$ C.L. if the ${\rm ^8B}$ neutrino flux is treated as a
free parameter (cf. figure~\ref{fig:global3}), but the SMA is absent
for all three analysis strategies at $99$\% C.L.

The predictions for the quantities that are expected to be most easily
measured by the SNO, BOREXINO, and KamLAND experiments are also robust
with respect to analysis strategies.

If we are lucky, one of these experiments may determine uniquely the
solar neutrino oscillation parameters. SNO could potentially identify
the LMA solution as correct by observing a large value for either the
neutral to charged current ratio or the day-night difference.  KamLAND
could also establish the LMA solution by finding an appreciable
deficit of reactor antineutrinos in the first phase of the
experiment.  On the other hand, BOREXINO and KamLAND could
definitively select the LOW solution if a large day-night effect is
observed.

All of the above comments depend upon the validity of the global
theoretical analysis which fits oscillation solutions to the reported
results of the first six solar neutrino experiments: chlorine,
Kamiokande, SAGE, GALLEX + GNO, Super-Kamiokande, and SNO.  Because
the analyses depend upon experimentally untested constraints implied
by the standard solar model and because there is insufficient
redundancy in the experiments performed so far, there may be startling
surprises when solar neutrino oscillations are probed in the future in
different experimental ways.

\subsection{Late breaking news from Super-Kamiokande}
\label{subsec:late}

The Super-Kamiokande collaboration has discussed~\cite{smy}
preliminary data for $1496$ days of observation, a $19$\% increase in
the length of data-taking over
the previously reported  data set of $1258$ days of observation. We
are grateful to M. Smy for making the preliminary data available to us
so that we could access the robustness of the predictions made in this
paper to a modest amount of additional data.

Figure~\ref{fig:lateglobal} shows the global solutions for the same
three analysis strategies and input data as were used in producing
figure~\ref{fig:global3}, except that figure~\ref{fig:lateglobal}
makes use of $1496$ days of Super-Kamiokande data. At first glance,
figure~\ref{fig:global3} and figure~\ref{fig:lateglobal} are very
similar to each other. With the new data, there is a small reduction
in the lower-mass region of the LMA solution and in the upper-mass
region of the LOW solution. The primary reasons for the changes that do occur are
the somewhat smaller differences between the day and the night data
and the slightly flatter (with respect to the undistorted standard
spectrum) recoil energy spectrum.
We spare the reader the details of the
best-fit points (which are shown in figure~\ref{fig:lateglobal}) since
these values bounce around within the allowed region as new data
are obtained.
\FIGURE[!ht]{
\centerline{\psfig{figure=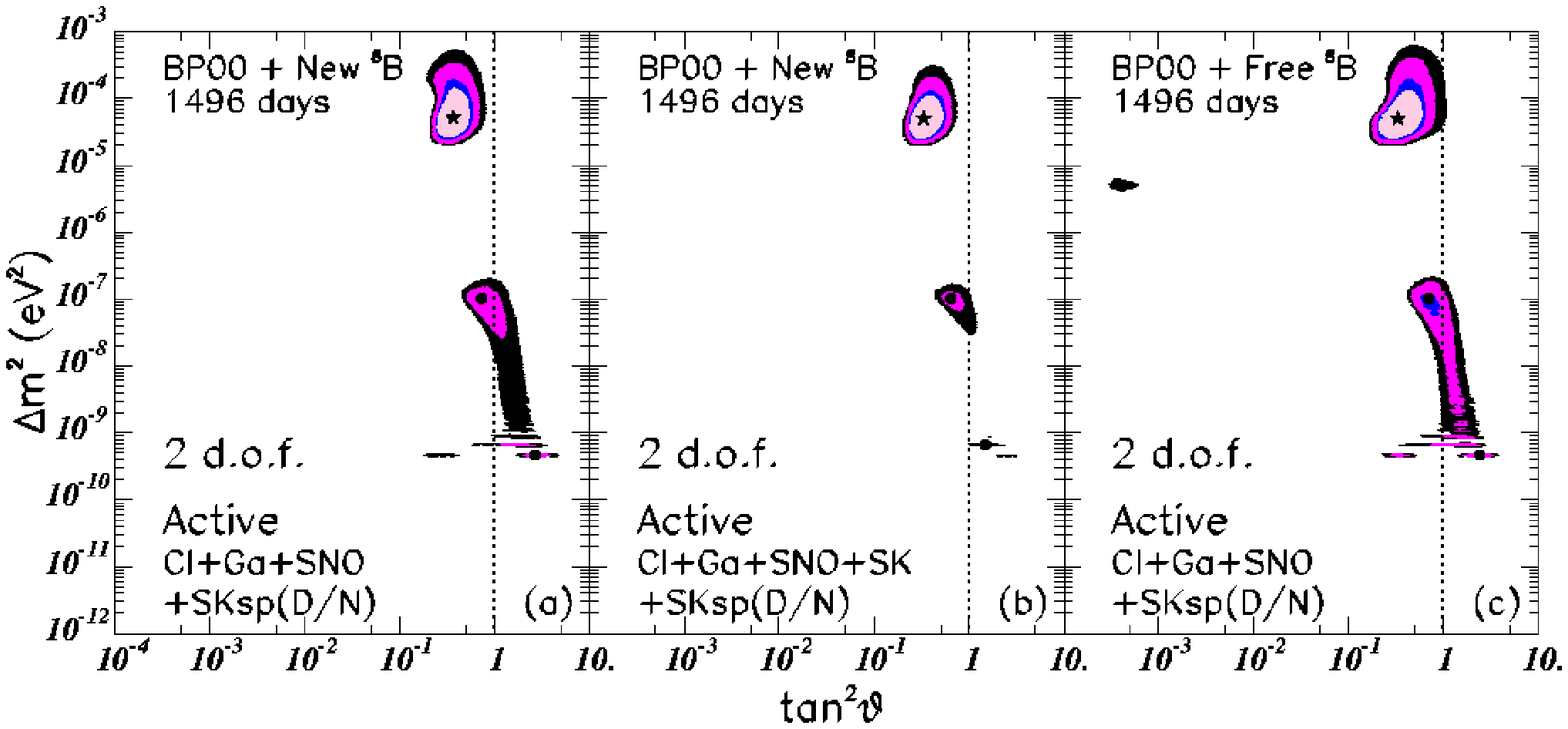,width=5in}}
\caption{ {\bf Global oscillation solutions including $1496$ days of
Super-Kamiokande observations. } Figure~\ref{fig:lateglobal}
is the same as
figure~\ref{fig:global3} except that data from $1496$ days of
Super-Kamiokande observations were used in constructing
figure~\ref{fig:lateglobal} whereas $1258$ days of Super-Kamiokande
observations were used in constructing figure~\ref{fig:global3}.
\label{fig:lateglobal}}}

At what CL is LMA the unique solution to the solar neutrino data? In
the global solution (strategy a) represented in
figure~\ref{fig:global3}, which includes 1258 days of Super-Kamiokande
data, LMA is the only solution at $96$\% ($2\sigma$) CL and in
figure~\ref{fig:lateglobal}, for $1496$ days of Super-Kamiokande
observations, LMA is the only solution at $97$\% ($2.1\sigma$)
CL. Even before the availability of the SNO data, LMA was the only
solution at $86$\% ($1.5\sigma$) CL \cite{gpbeforesno}.  These results
are in good agreement with those described by Smy~\cite{smy} (who
analyzed the Super-Kamiokande data with more zenith-angle bins and
fewer spectral energy bins than we do). We conclude that LMA is the
preferred solution with a CL that has been increasing slowly with time.

Table~\ref{tab:update} presents the $1\sigma$ and $3\sigma$ ranges for
the principal solar neutrino observables that were calculated in
sections~\ref{sec:global}-\ref{sec:kamland}. In constructing
table~\ref{tab:update}, we again use analysis strategy (a) and the
same input data as in the previous calculations except that we now
include $1496$ days of Super-Kamiokande data.It is instructive to
compare the entries in table~\ref{tab:update} with the previously
tabulated results. The changes due to the new data are smaller than or
comparable to the differences among the predictions of the three
analysis strategies for the total event rates and the energy shifts,
i. e., for the quantities [NC]/[CC], $\delta$T(SNO CC), [$^7$Be], [CC]
(KamLAND), and $\delta E_{\rm vis}$ (KamLAND). We conclude, once
again, that these predictions are robust.

The predicted maximum value at $3\sigma$
of the day-night asymmetry  is reduced by about
$15$\% for the CC measurement by SNO
(from $21$\% to $18$\%) and by about
$10$\% for BOREXINO (from $41$\% to $38$\%). These changes reflect the
difficulty in measuring day-night asymmetries, which are determined by
relatively small differences between two large numbers.

\TABLE[!ht]{ \centering \caption{{\bf Predictions including $1496$
days of Super-Kamiokande data.}  This table presents the $1\sigma$ and
$3\sigma$ ranges for the measurable quantities calculated in
sections~\ref{sec:global}-\ref{sec:kamland}. We use here analysis strategy
(a) and the same input data as in the previous calculations, except
that $1496$ days of Super-Kamiokande data were included. Results are
presented for measurables in the SNO, BOREXINO, and KamLAND experiments. The
threshold of the recoil electron kinetic energy used in computing the
SNO observables for this table is $6.75$ MeV and for the KamLAND reactor
observables,  $E_{\rm th}=1.22$
MeV. The recoil energy range for the BOREXINO experiment is the same
as adopted in section~\ref{sec:be7}.}  \protect\label{tab:update}
\begin{tabular}{lcc}
\noalign{\bigskip}
\hline
\noalign{\smallskip}
Observable&  b.f. $\pm 1\sigma$  &b.f. $\pm 3\sigma$ \\
\noalign{\smallskip}
\hline
\noalign{\smallskip}
$[$NC$]/[$CC$]$
&$3.41^{+0.64}_{-0.50}$& $3.41^{+1.64}_{-2.00}$\\
\noalign{\smallskip}
\noalign{\smallskip}
A$_{\rm N-D}$(SNO CC)
& $5.1^{+5.5}_{-3.1}$&$5.1^{+12.9}_{-5.4}$\\
\noalign{\smallskip}
\noalign{\smallskip}
A$_{\rm N-D}$(SNO ES)
& $2.9^{+2.9}_{-1.7} $&$2.9^{+6.7}_{-3.1}$ \\
\noalign{\smallskip}
\noalign{\smallskip}
$\delta$T\% (SNO CC)
&  $0.00^{+0.02}_{-0.36}$&$0.00^{+1.53}_{-1.33}$\\
\noalign{\smallskip}
\noalign{\smallskip}
$[$R($^7$Be)$]$
& $0.66\pm0.03$& $0.66^{+0.13}_{-0.11}$\\
\noalign{\smallskip}
\noalign{\smallskip}
A$_{\rm N-D}$($^7$Be)
& --- & $ 23^{+15}_{-28}$\\
\noalign{\smallskip}
\noalign{\smallskip}
$[$CC$]$(KamLAND)
& $0.56^{+0.11}_{-0.16}$ &$0.56^{+0.17}_{-0.27}$\\
\noalign{\smallskip}
\noalign{\smallskip}
$\delta E_{\rm visible}$(KamLAND) (\%)
 & $-7^{+11}_{-3}$ & $-7^{+15}_{-7}$\\
\noalign{\smallskip}
\hline
\end{tabular}
}

\clearpage

 We are grateful to E. G. Adelberger and K. A. Snover for discussions
 of their important new measurement of the ${\rm ^7Be + p}$ cross
 section and to V. Barger, S. Glashow, D. Marfatia, A. McDonald, M. Smy,
 R. Svoboda, and J. Wilkerson for valuable comments and
 discussions. A. Piepke made the very useful suggestion that we
 include predictions for the KamLAND reactor experiment and
 M. Chen and A. McDonald drew our attention to the relation between
 the day-night effects in SNO of the CC and the neutrino-electron
 scattering. JNB acknowledges support from NSF grant No. PHY0070928.
 MCG-G is supported by the European Union Marie-Curie fellowship
 HPMF-CT-2000-00516.  This work was also supported by the Spanish
 DGICYT under grants PB98-0693 and PB97-1261, by the Generalitat
 Valenciana under grant GV99-3-1-01, and by the ESF network 86.

\end{document}